\documentclass[amsmath,amssymb,nobibnotes,aps,prx,onecolumn,twocolumn,reprint]{revtex4-2}

\usepackage[utf8]{inputenc}
\usepackage[T1]{fontenc}
\usepackage{color}
\usepackage{graphicx}
\usepackage{epsfig}
\usepackage{dcolumn}
\usepackage{float}
\usepackage{comment}
\usepackage{multibib}
\usepackage{pstricks}
\usepackage{verbatim}
\usepackage[percent]{overpic}
\usepackage{geometry}
\usepackage{hyperref}
\hypersetup{
    colorlinks=true,
    linkcolor=blue,
    citecolor=blue,
    urlcolor=blue
}
\usepackage{cleveref}

\newcites{supp}{Supplementary References}
\newcommand{\SSG}[1]{\textcolor{black}{{#1}}}
\newcommand{\AV}[1]{\textcolor{black}{#1}}
\begin{document}

\title{Thermal Fluctuation Driven Structural Relaxation in Undeformed Glasses: Unraveling the Evolution of Mechanical Stability }
\author{Avinash Kumar Jha}
\affiliation{Department of Physics, Indian Institute of Technology Roorkee, Roorkee 247667, Uttarakhand, India}

\author{Shiladitya Sengupta}
\affiliation{Department of Physics, Indian Institute of Technology Roorkee, Roorkee 247667, Uttarakhand, India}
\email{shiladityasg@ph.iitr.ac.in}


\begin{abstract}
\SSG{Glasses are mechanically rigid, still undergo structural relaxation which changes their properties and affects potential technological applications. Understanding the underlying physical processes is a problem of broad theoretical and practical interest. We investigate intermittent structural relaxation events or ``avalanches'' occurring inside glassy regime. Contrary to the more well-known avalanches due to shear, here they are induced by thermal fluctuations in undeformed glass. By analyzing changes in structural, mechanical, dynamical, topological and vibrational properties of the system, we provide a multi-faceted characterization of avalanches. Overall we find that the system softens due to avalanches. Further, we develop a formalism to extract local measures of non-Affine displacement and tensorial strain for thermal amorphous solids in absence of any external deformation. Our analysis highlights a key difference between two types of driving: while the shear deformation response is dominated by volume preserving deviatoric strain, changes in local density must be considered to model response of undeformed glass under thermal noise.} The observations \SSG{suggest} the idea of Generalized Strain Transformation Zones (GSTZ), where coupled shear and volume‐changing deformations govern thermally-mediated plasticity.  \SSG{Our work paves the way for a unified description of elasto-plastic response of (athermal) mechanically deformed and thermally driven undeformed glasses.} 
\end{abstract}
\maketitle

\section{Introduction}\label{sec:intro}
\SSG{Glasses are ubiquitous in our everyday life. However, understanding their nature - whether they are `liquids that are frozen' or `solids that flow' \cite{dyre2024solid} - has remained elusive \cite{angell2000relaxation, lubchenko2007theory, ngai2010nature, berthier2011theoretical}. Nevertheless, it is evident that even below the experimental glass transition temperature ($T_g$), glassy materials do indeed undergo structural relaxation, manifested as aging and de-vitrification. These phenomena changes properties of the materials that affect their technological applications \cite{2003Vestel, 2014sansinena, alhalaweh2015physical,  yanagishima2017common, jafary2018critical, 2019Salunkhe}. Thus, it is desirable to have a theoretical description of the structural relaxation process in glass if we are to design functional devices from glassy materials \cite{wondraczek2022advancing}.}

\SSG{Recent computer simulations and experiments in colloidal glasses \cite{2011SanzPRL, keys2011excitations, sanz2014avalanches, yanagishima2017common} have shown that deep inside the glassy regime {\it i.e.} well below $T_g$ the structural relaxation takes place {\it via} sudden irreversible particle rearrangement events in-between long quiescent states,} manifesting as abrupt jumps in the mean square displacement (MSD) \cite{2011SanzPRL, sanz2014avalanches, yanagishima2017common, yanagishima2021towards}. \SSG{These events, which are induced by thermal noise, are termed ``plastic events'' or ``avalanches'' as they are reminiscent of intermittent dynamics in various driven systems \cite{sethna2001crackling}.}

\SSG{There are formidable challenges towards a comprehensive description of the underlying physical processes. First, the timescales involved are extremely long, being always at the limits of available experimental and computational capabilities; second, owing to disorder there is no unambiguous definition of structural defects in glasses, unlike dislocations in crystals; third and perhaps most importantly, general principles such as equilibrium statistical physics and thermodynamics are lacking for non-equilibrium systems.}
\SSG{In such a scenario, one approach that has yielded deep insights is to focus on mechanical or elasto-plastic response for which thermodynamic equilibrium is not necessary. Besides, stress, strain and elastic moduli can be defined even in absence of long range structural order, at least in a time-dependent way that is sufficient for most practical applications \cite{saw2016rigidity, tanguy2002continuum, hentschel2011athermal, nath2018existence}.} 

Understanding plasticity in disordered materials \SSG{such as glasses } \cite{eshelby1957determination, spaepen1977microscopic, picard2004elastic, maloney2006amorphous, schuh2007mechanical, hunter2012physics, cubuk2015identifying, chu2015colloidal, nicolas2018deformation, ye2013dynamic, richard2020predicting} has traditionally focused on strain-mediated processes, where external stresses drive irreversible deformations in systems like colloidal, metallic and structural glasses \cite{angell1995formation, 2019Parmar, 2025Karmakar, pingua2025cascade}. Plastic rearrangements mediated by thermal fluctuations \SSG{without any external force} is a \SSG{relatively} less explored phenomenon despite its profound implications for the stability of amorphous materials. \SSG{This is the topic of the present work. Our main results are as follows.}

\SSG{First, previous works on thermal-mediated plasticity have mainly focused on the MSD to detect avalanches \cite{zaccarelli2009colloidal, sanz2014avalanches, yanagishima2017common, yanagishima2021towards}.} To unravel the intricate changes during an avalanche event, in the present work we employ a multifaceted analysis using a range of structural, mechanical, dynamical, topological and vibrational metrics to detect avalanche events. \SSG{We also examine the correlations among the different categories of observables. Thus we greatly extend the scope of characterization of thermal-mediated plasticity. Overall, our analysis shows a softening of the system due to an thermal avalanche event. }

\SSG{Second, shear deformation response of glasses are typically described in terms of measures of non-Affine displacement field such as $D^2_{min}$ and tensorial strain,  stress and elastic moduli \cite{falk1998dynamics, chikkadi2011spatial, falk2011deformation, kondic2012microstructure, shang2014evolution, pinney2016structure, kim2017molecular, wang2018nanometer, jana2019correlations, yang2022hidden, karimi2022shear}. We develop a formalism to compute local $D^2_{min}$ and local tensorial strain fields for thermal amorphous solids in absence of external deformation. It enables one to treat thermal mediated plastic response on the same footing as shear induced response. Thus we are able to comment on the similarity and difference of the glass response under the two different situations. On one hand we find the signature Eshelby-like strain fields similar to shear induced avalanche events. On the other hand,} strain-mediated plasticity is \SSG{typically considered to be driven by volume preserving (deviatoric) strain:} external stress leads to the formation of shear transformation zones (STZs) where localized shear strain accumulates, ultimately driving plastic deformation \cite{langer2004dynamics, falk2004thermal, falk2005toward, bouchbinder2007athermal, hermundstad2010energetics, langer2012shear, langer2015shear, richard2021simple}.  \SSG{Remarkably}, our analysis shows a key difference between strain-driven and thermally activated plasticity: while the former primarily involves localized shear strain, the latter introduces an additional volumetric component, altering the overall stress distribution and mechanical response of the system. This fundamental difference challenges conventional views of plasticity in amorphous materials and underscores the need to consider both shear and volumetric strain contributions when analyzing thermally induced structural relaxations \cite{zhang2021interplay}.

The organization of the paper is as follows: Sec \ref{sec:simuDet} presents the model system and computational details. We present several indicators of an avalanche event manifested in changes in mechanical, vibrational and topological properties of the system in Sec. \ref{sec:defAval}. We present a method to extract non-Affine contribution to particle displacement field in terms of local $D^2_{min}$ for undeformed glasses in Sec. \ref{sec:nonAff}. We also analyze particle mobility in real space as well as along the potential energy landscape for an thermal avalanche event. In Sec. \ref{sec:strain}, we present our method to extract local elastic strain measures in absence of external deformation and use it to determine the relative importance of volumetric and deviatoric strains in thermally-mediated avalanches. We directly demonstrate softening of the system by analyzing the structural order parameter called ``softness'' in Sec. \ref{sec:soft}. The relationship among the different types of observables are quantified in Sec. \ref{sec:relat}. Finally we summarize the results and present the conclusions in Sec. \ref{sec:disc}.

\section{Simulation Details}\label{sec:simuDet}
We study a two-dimensional polydisperse system of soft particles interacting via the Weeks-Chandler-Andersen (WCA) potential \cite{weeks1971role}:
\begin{align}
    U(r_{jk})&=4\epsilon\left[\left(\frac{\sigma_{jk}}{r_{jk}}\right)^{12}-\left(\frac{\sigma_{jk}}{r_{jk}}\right)^{6} +\frac{1}{4}\right],\;\; \frac{r_{jk}}{\sigma_{jk}} < 2^{\frac{1}{6}} \nonumber\\
    &=0 \;\; \mbox{otherwise.}\label{eqn:WCA}
\end{align}
\SSG{The system has 11\% size poly-dispersity with Gaussian distribution \cite{tanaka2010critical}. The mean particle diameter $\langle \sigma\rangle=1.0$ is used as the unit of length.} $\epsilon$ denotes the energy scale, \SSG{and the Boltzmann constant $k_B=1$.} Temperature is expressed in units $\frac{\epsilon}{k_B}$.

\paragraph*{Sample preparation protocol:} We perform NVT molecular dynamics (MD) simulation using an isokinetic thermostat implemented by Brown and Clarke \cite{brown1984comparison} \SSG{at the system size $N=2000$ and the density $\rho = 0.900$ over a broad range of temperatures both in and out of equilibrium} \cite{Jha_unpub}. \SSG{At each temperature in equilibrium,  we generate \AV{12 independent, well-equilibrated MD trajectories  with runlength $\sim 100 \, \tau_\alpha$ or longer}, 
where $\tau_\alpha$ denotes the $\alpha$-relaxation time. The simulation glass transition temperature $T_g$ is defined by the condition $\tau_\alpha (T_g) = 10^6$ (reduced units). To generate trajectories out of equilibrium, the initial configurations taken from equilibrated trajectories at a parent temperature $T_p=\AV {0.500}$ are quenched to different target low temperatures. Typically, a waiting time $t_w = 9.45 \times 10^5$ (in reduced units) is allowed for the system to age before analysis [$t_w \approx \tau_\alpha (T_g)$].}
\SSG{Inherent structure (IS) trajectories are generated from MD trajectories at same density and temperatures by minimizing the potential energy using the conjugate gradient method.}  

\paragraph*{Frame to time conversion:}
All \SSG{MD} trajectories in this work are plotted against an integer frame index $f$.  
\SSG{The MD step $S(f)$ and the time $t(f)$ in reduced units corresponding to frame \(f\) are given  by the conversion formula}
\begin{align}
   S(f) &= i\,B + 2^j \nonumber\\
   t(f) &= S(f)\,\delta t.
   \label{eqn:frm2time}
\end{align}
\SSG{Here $B$ denotes the blocksize in MD steps, $j$ is an integer such that $0\le j < N_{\rm cfg}$ with $N_{\rm cfg}$ being the number of stored configurations per block, and $\delta t$ is the incremental time step in reduced units corresponding to one MD step. $B=1050000$, $N_{\rm cfg}=21$, and $\delta t=0.003$ unless otherwise specified. The IS frame index is same as the corresponding MD frame index.}

\section{Mechanical, Vibrational and Topological Indicators of a Thermally Mediated Avalanche}\label{sec:defAval}

\begin{figure*}[htbp!]
    \centering
    \includegraphics[width=0.32\textwidth]{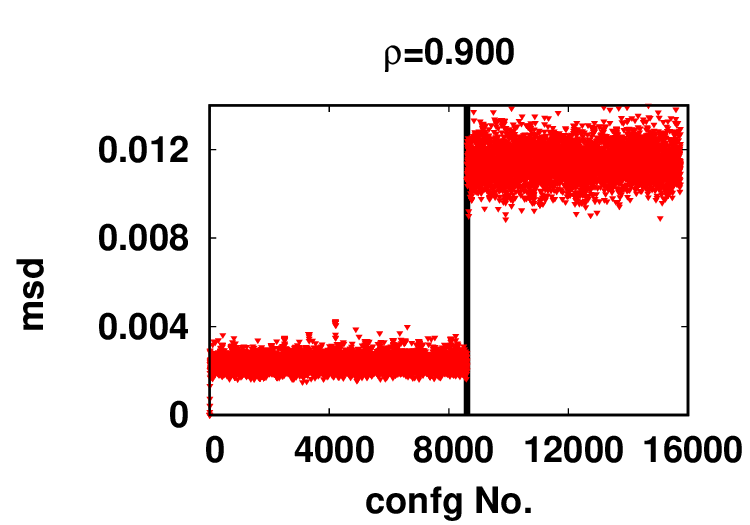}
    \includegraphics[width=0.32\textwidth]{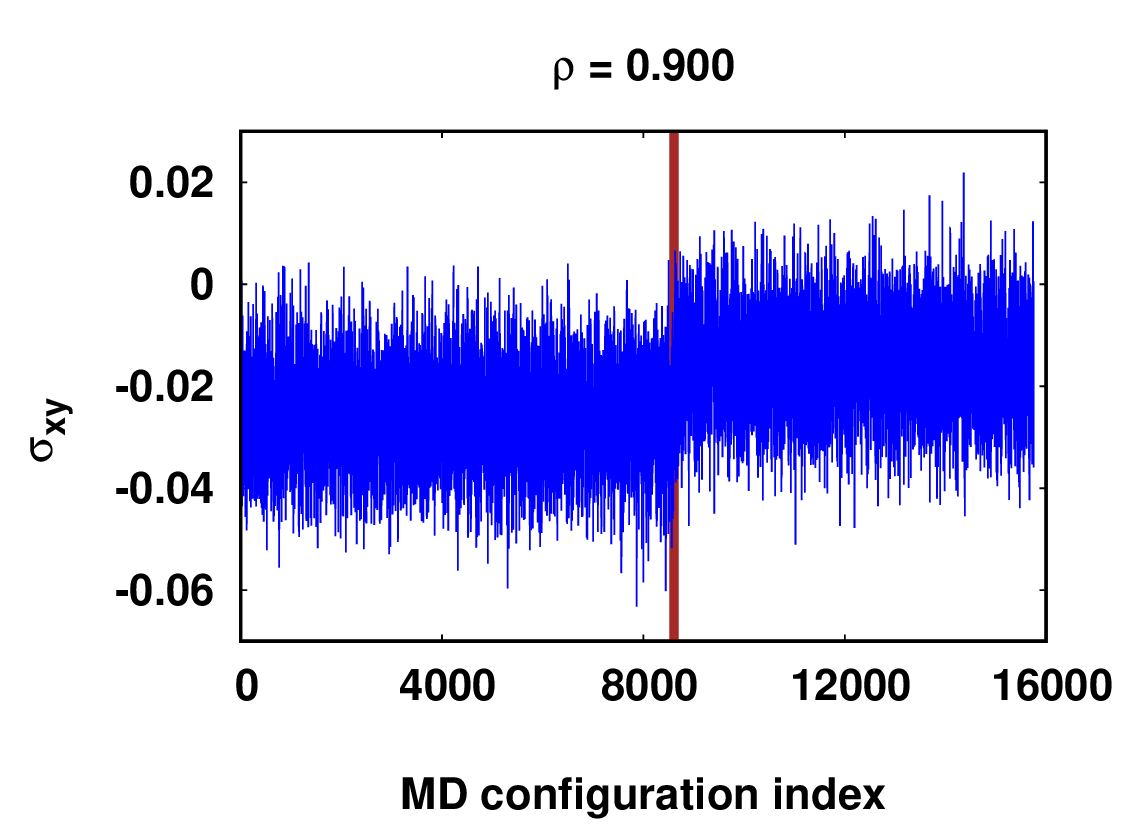}
    \includegraphics[width=0.32\textwidth]{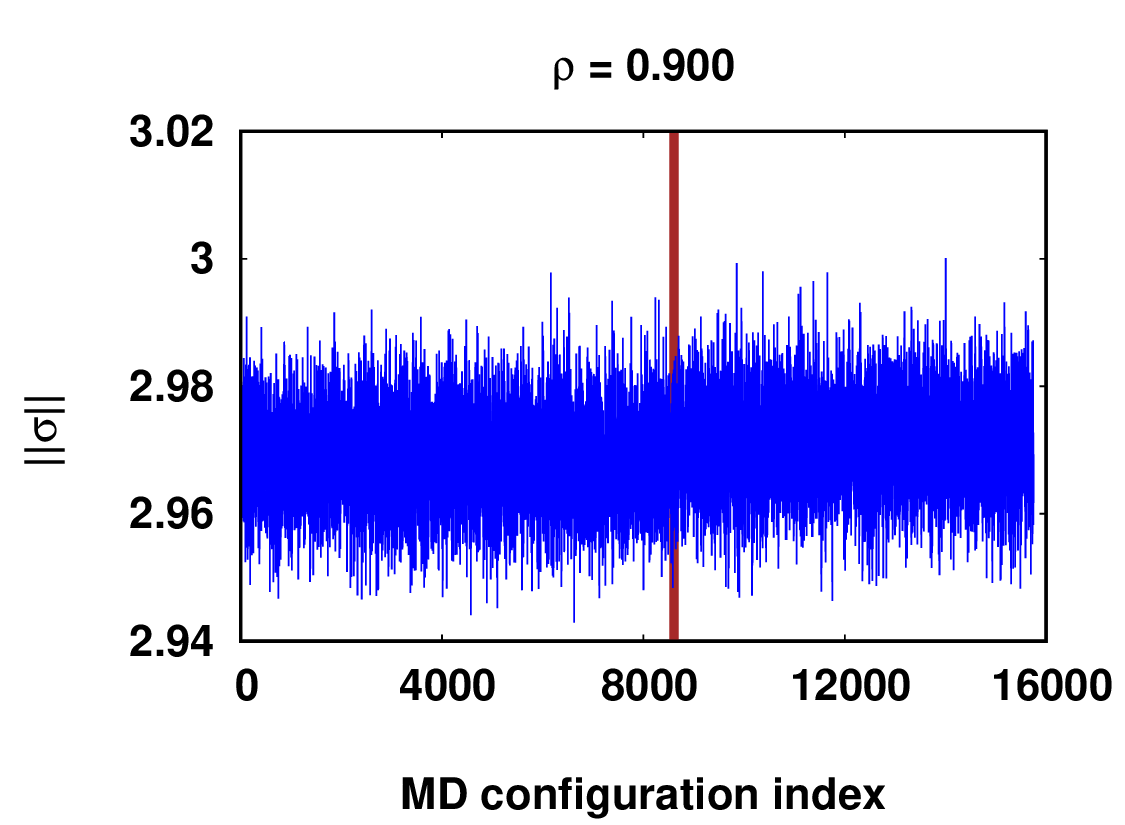}
    \put(-360,65){\textbf{(a)}}
    \put(-230,70){\textbf{(b)}}
    \put(-100,70){\textbf{(c)}}\\
    \includegraphics[width=0.32\textwidth]{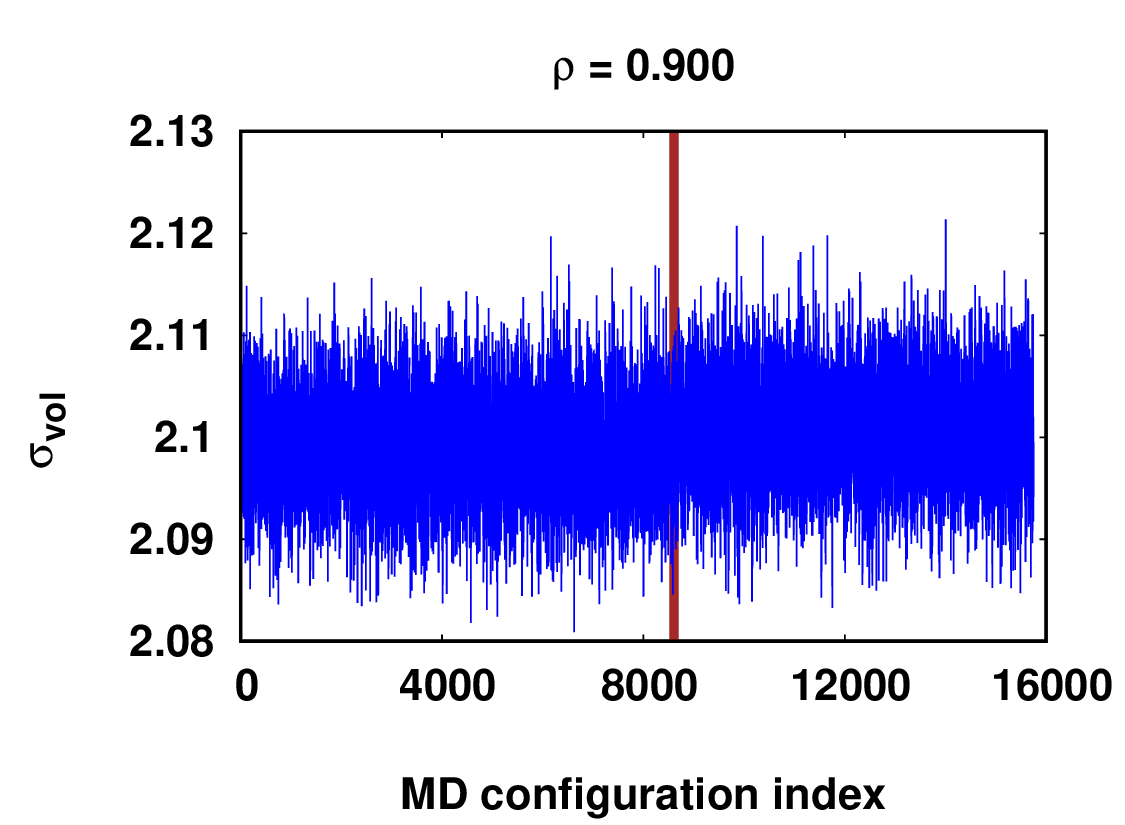}
    \includegraphics[width=0.32\textwidth]{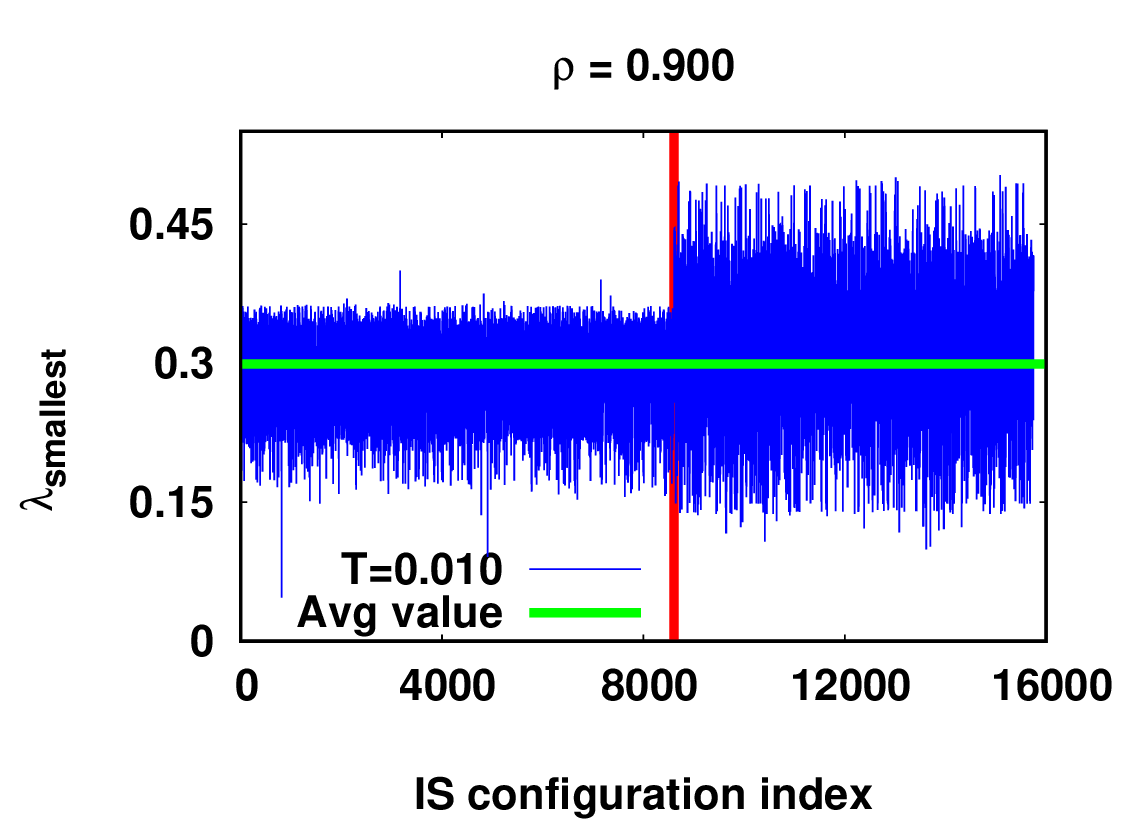}
    \includegraphics[width=0.32\textwidth]{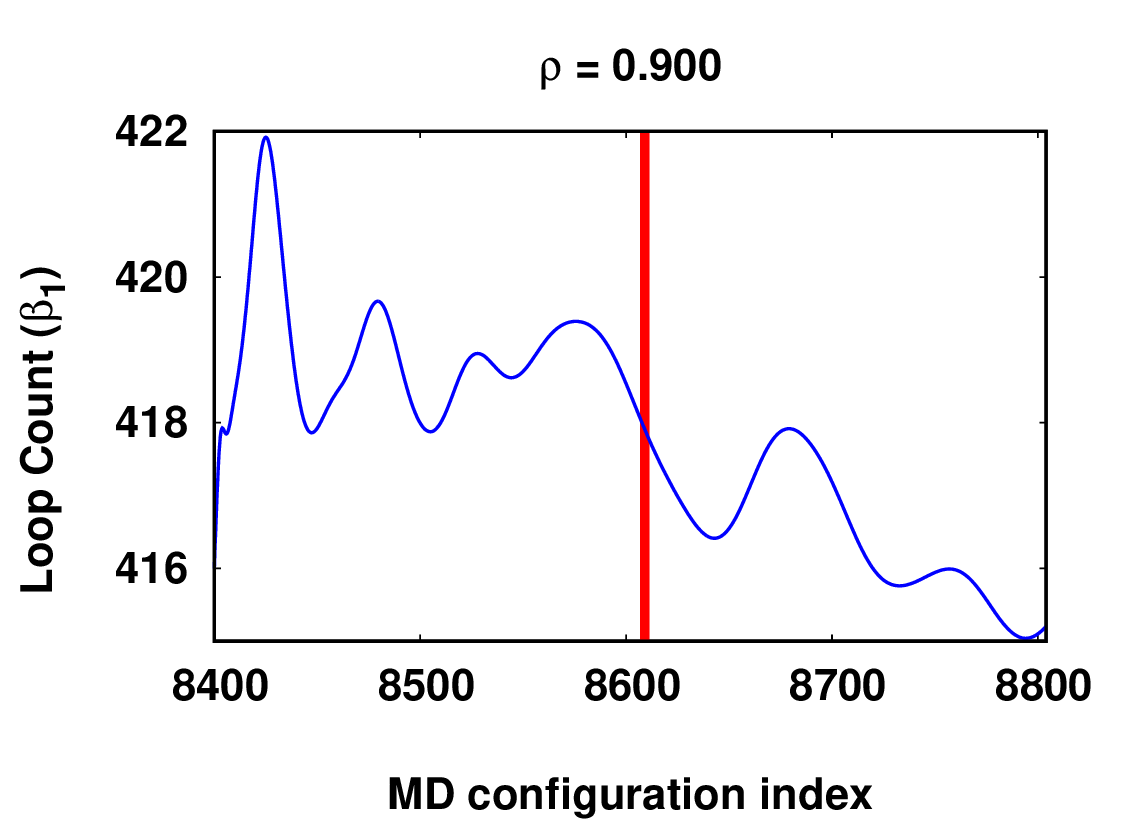}
    \put(-370,68){\textbf{(d)}}
    \put(-230,70){\textbf{(e)}}
    \put(-70,70){\textbf{(f)}}
    \caption{ \SSG{
    \textbf{(a):} \emph{Defining an avalanche event} by sudden jump in the mean square displacement (MSD) plotted against the configuration index of a representative MD trajectory. The vertical line indicates the configuration at which a thermally mediated plastic event (avalanche) occurs. In panels (b)-(f), vertical lines denote this configuration, either in the MD or in the corresponding inherent structure (IS) trajectory. \textbf{(b)-(d)):} \emph{Signature of avalanche in the stress fluctuation} by monitoring along the MD trajectory (b) shear and (c) volumetric stress as well as (d) norm of the stress tensor, measured at a finite temperature, see Eqns. ~\eqref{stress26d}, \eqref{stress26e}, and \eqref{stress26f}. Volumetric stress and the norm of the stress tensor remain largely unchanged across the avalanche, indicating that isotropic stress components are not markedly affected. In contrast, shear stress exhibits discontinuous change at the avalanche. \textbf{(e):} \emph{Indication of avalanche in the vibrational spectrum.} Evolution of smallest eigenvalues $\lambda_{smallest}$ (Eqn. \ref{small_egn}) of the Hessian matrix of the corresponding inherent structures (IS) shows prominent discontinuity at the avalanche event. The width of the fluctuation before and after the avalanche are clearly different, reflecting a significant change in the vibrational spectrum. A green horizontal line indicates the average value of the smallest eigenvalue over the entire trajectory. \textbf{(f):} \emph{Topological change at avalanche.} The Betti number ($\beta_1$) {\it i.e.} the total number of 1D holes, computed via a Vietoris–Rips complex, changes sharply across the avalanche, indicating significant restructuring of the underlying network due to the avalanche. 
    }}
    \label{fig:av1}
\end{figure*}

Figure \ref{fig:av1} captures the evolution of structural, mechanical, and topological properties \SSG{along a representative MD trajectory showing an avalanche event triggered by thermal noise. Avalanche in undeformed glass at finite temperature has been discussed elsewhere \cite{zaccarelli2009colloidal, sanz2014avalanches, yanagishima2017common, yanagishima2021towards}. Following these works, we identify avalanche from the mean square displacement (MSD) \cite{sanz2014avalanches, yanagishima2017common}.} The abrupt change in MSD in Fig. \ref{fig:av1}(a) indicates a structural relaxation event or “avalanche”, during which the system transitions from one quiescent state to another. 

\paragraph{Stress response:} Since thermally mediated particle rearrangements redistribute internal forces, we next examine \SSG{the time evolution} of the full stress tensor to identify how the avalanche perturbs stress \SSG{fluctuations}. For a two‐dimensional system of \SSG{$N$ particles in an} area \(A\) and number density \(\rho\) \SSG{at a temperature $T$}, components of the stress tensor $\boldsymbol{\sigma}$ are given by the finite‐temperature virial expression \cite{irving1950statistical,lutsko1988stress}:
\begin{align}
    \sigma_{xx} \;=\; \rho\,k_{B}T \;-\; \frac{1}{2A} \sum_{i<j}^{N} \left(\frac{\partial U}{\partial r_{ij}}\right)\frac{r_{x}^{ij}\,r_{x}^{ij}}{r^{ij}} \nonumber\\
    \sigma_{yy} \;=\; \rho\,k_{B}T \;-\; \frac{1}{2A} \sum_{i<j}^{N} \left(\frac{\partial U}{\partial r_{ij}}\right)\frac{r_{y}^{ij}\,r_{y}^{ij}}{r^{ij}} \nonumber\\
    \sigma_{xy} \;=\; -\,\frac{1}{2A} \sum_{i<j}^{N} \left(\frac{\partial U}{\partial r_{ij}}\right)\frac{r_{x}^{ij}\,r_{y}^{ij}}{r^{ij}}  \nonumber\\
    \sigma_{yx} \;=\; -\,\frac{1}{2A} \sum_{i<j}^{N} \left(\frac{\partial U}{\partial r_{ij}}\right)\frac{r_{y}^{ij}\,r_{x}^{ij}}{r^{ij}} 
    \label{stress26d}
  \end{align}
\SSG{where $U = \sum\limits_{i<j} U_{ij}$ is the potential energy and $x,y$ represent Cartesian components}. $r_{ij}$ is the distance between particles $i$ and $j$. The terms $r^{ij}_{x}$ and $r^{ij}_{y}$ are the x and y components, respectively, of $\mathbf{r}_{ij}$. $\rho k_{B}T$ is the thermal average of the kinetic contribution to stress, which includes velocity effects. 
\SSG{We also compute the} volumetric stress defined as
\begin{align}
    \sigma_{\mathrm{vol}} \;=\; \frac{1}{2}\,\bigl(\sigma_{xx} + \sigma_{yy}\bigr)
    \label{stress26e}
\end{align}
and the norm of the stress tensor (``Norm Stress''):
\begin{align}
    \bigl\lVert \boldsymbol{\sigma} \bigr\rVert 
    \;=\; \sqrt{\boldsymbol{\sigma}:\boldsymbol{\sigma}}
    \;=\; \sqrt{\sigma_{xx}^{2} \;+\; \sigma_{yy}^{2} \;+\; \sigma_{xy}^{2} \;+\; \sigma_{yx}^{2}}\,.
    \label{stress26f}
\end{align}

\SSG{Fig. \ref{fig:av1}(b)-(d) shows the time evolution of the stress fluctuations along a representative MD trajectory at a finite low temperature much below $T_g$.} The volumetric stress $\sigma_{\mathrm{vol}}$ and the ``norm stress'' $\bigl\lVert \boldsymbol{\sigma} \bigr\rVert$  do not exhibit any pronounced jump at the avalanche event, suggesting that the avalanche event does not significantly influence the isotropic stress components. In contrast, the shear stress fluctuation, reveals a \SSG{clear discontinuity} at the avalanche event. \SSG{See also supplementary information (SI) Figs. \ref{fig:av14nw2}-\ref{fig:av14nw4} for corresponding data for four other avalanche events.} Thus our results indicate that shear stress plays a critical role in triggering the thermally mediated avalanche, in agreement with the notion that shear-driven instabilities are central to plastic events in glasses \cite{maloney2006amorphous}.

\paragraph{Vibrational signature:} \SSG{An avalanche event represents a transition from one potential energy minimum to another. Hence, we analyze the low energy excitation modes for signatures of mechanical instability events.} In particular, Fig. \ref{fig:av1}(e) shows the evolution of $\lambda_{\text{smallest}}$ - the average of the first $m$ smallest positive eigenvalues of the Hessian matrix computed for each energy-minimized configuration, also known as inherent structures \cite{wyart2005geometric, zylberg2017local}:
\begin{align}
    \lambda_{\text{smallest}} = \frac{1}{m} \sum_{i=1}^{m} W(i).
    \label{small_egn}
\end{align}
Here $W(i)$ denote the \SSG{positive} Hessian eigenvalues in ascending order. \SSG{Following Ref. \cite{zylberg2017local}, we choose $m=30$.}  Eqn. \ref{small_egn} serves as a sensitive indicator of low energy excitations of the system and enhances our ability to detect subtle changes in mechanical stability during thermally mediated plastic events. \SSG{Similar to the MSD, $\lambda_{\text{smallest}}$ shows two distinct plateaus clearly separated by discontinuity at the avalanche event along with a significant increase in fluctuations post-avalanche. To assess the statistical significance of these observations, four other independent avalanche events are analyzed in SI Figs. \ref{fig:av14nw1} and the same trends are observed.} Thus our results highlight that \emph{vibrational spectrum is highly sensitive to structural rearrangements and significant changes in it can be triggered by thermal excitations.
}

\paragraph{Topological change:} 
\SSG{In real space the avalanches are triggered from localized region which are thought to be ``structural defects'' \cite{maloney2004subextensive, manning2011vibrational, richard2020predicting, pingua2025cascade}. Recent studies have pointed to the interesting possibility that such putative defects are topological in nature \cite{charan2023anomalous, bera2024clustering, desmarchelier2024topological}. In the context of thermal-mediated plasticity, previous studies have shown that an avalanche event changes the local topology \cite{yanagishima2017common,Jha_unpub}}. 

Here we analyze the topology of the configurations using a topological invariant called the ``Betti number'' ($\beta_1$) which counts the total number of 1D holes (rings) \cite{edelsbrunner2010computational, maria2014gudhi}. The details of the computation are described in the Appendix \ref{sec:appTopo}. In Fig. \ref{fig:av1}(f) we present the evolution of $\beta_1$ along the MD trajectory with the same for four other independent trajectories are shown in SI Fig. \ref{fig:av14nw5}. Interestingly in each case, there is a \emph{marked decrease in $\beta_1$ at the avalanche}. In 3 out of 5 events, two distinct steady states before and after the avalanche with a discontinuity at the avalanche can be clearly discerned, revealing significant topological changes in the configuration due to the avalanche event.
The substantial reduction in \SSG{the number of} 1D holes post-avalanche suggests a more ordered and more homogeneous local structure emerging after the event.

\section{Displacement fields for thermal avalanche event}\label{sec:nonAff}
\subsection{$D^2_{min}$ for thermal avalanche event}
\SSG{Sec \ref{sec:defAval} provides a comprehensive picture of a thermally mediated avalanche event by revealing changes in the mechanical, vibrational, and topological characteristics. Now we show the real-space aspects of the event by analyzing the non-Affine (NA) displacement fields and particle mobility. Non-affine displacement fields under shear deformation are commonly characterized by the measure $D^2_{min}$ \cite{falk1998dynamics, shiba2010plastic, chikkadi2011long}. However, the present study does not involve external shear and the plastic events are induced purely by thermal fluctuations. Thus, we first develop an appropriate formalism to measure local $D^2_{min}$ for undeformed glass. The details are presented in Appendix \ref{sec:appStrain} and here we focus only on the main results.}

\begin{figure*}[htbp!]
    \centering
    \includegraphics[width=0.32\textwidth]{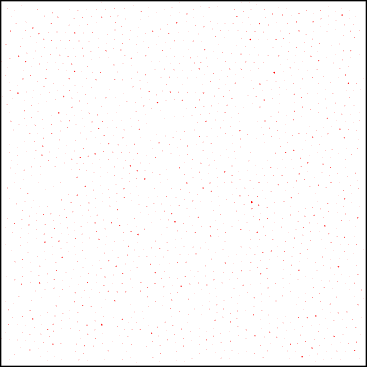}
    \includegraphics[width=0.32\textwidth]{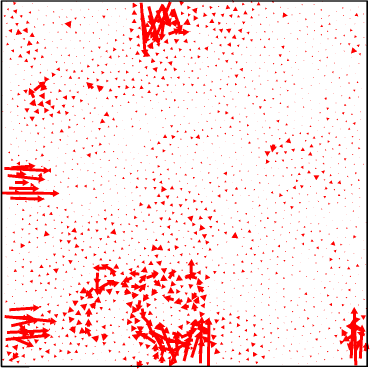}
    \includegraphics[width=0.32\textwidth]{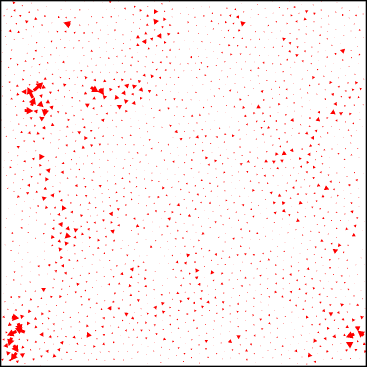}
    \put(-415,110){\large{\textbf{(a) \small{before}}}}
    \put(-180,110){\large{\textbf{(b) \small{at}}}}
    \put(-70,110){\large{\textbf{(c) \small{after}}}}\\
    \vspace{5mm}
    \includegraphics[width=0.32\textwidth]{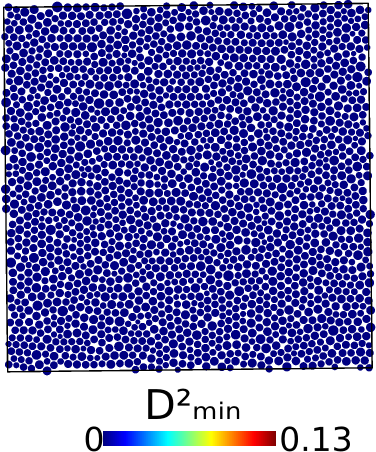}
    \includegraphics[width=0.32\textwidth]{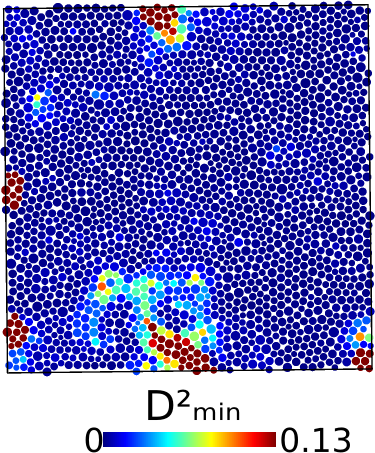}
    \includegraphics[width=0.32\textwidth]{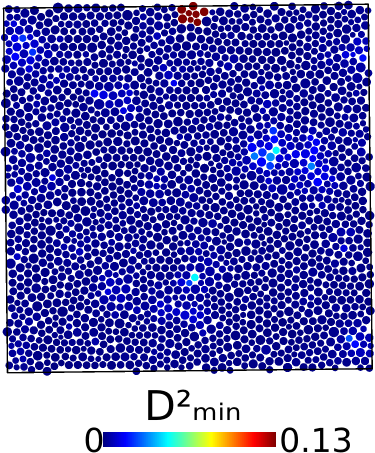}
    \put(-420,18){\textbf{(d) before}}
    \put(-278,18){\textbf{(e) at}}
    \put(-138,18){\textbf{(f) after}}\\
    \vspace{5mm}
    \includegraphics[width=8cm,height=5.7cm]{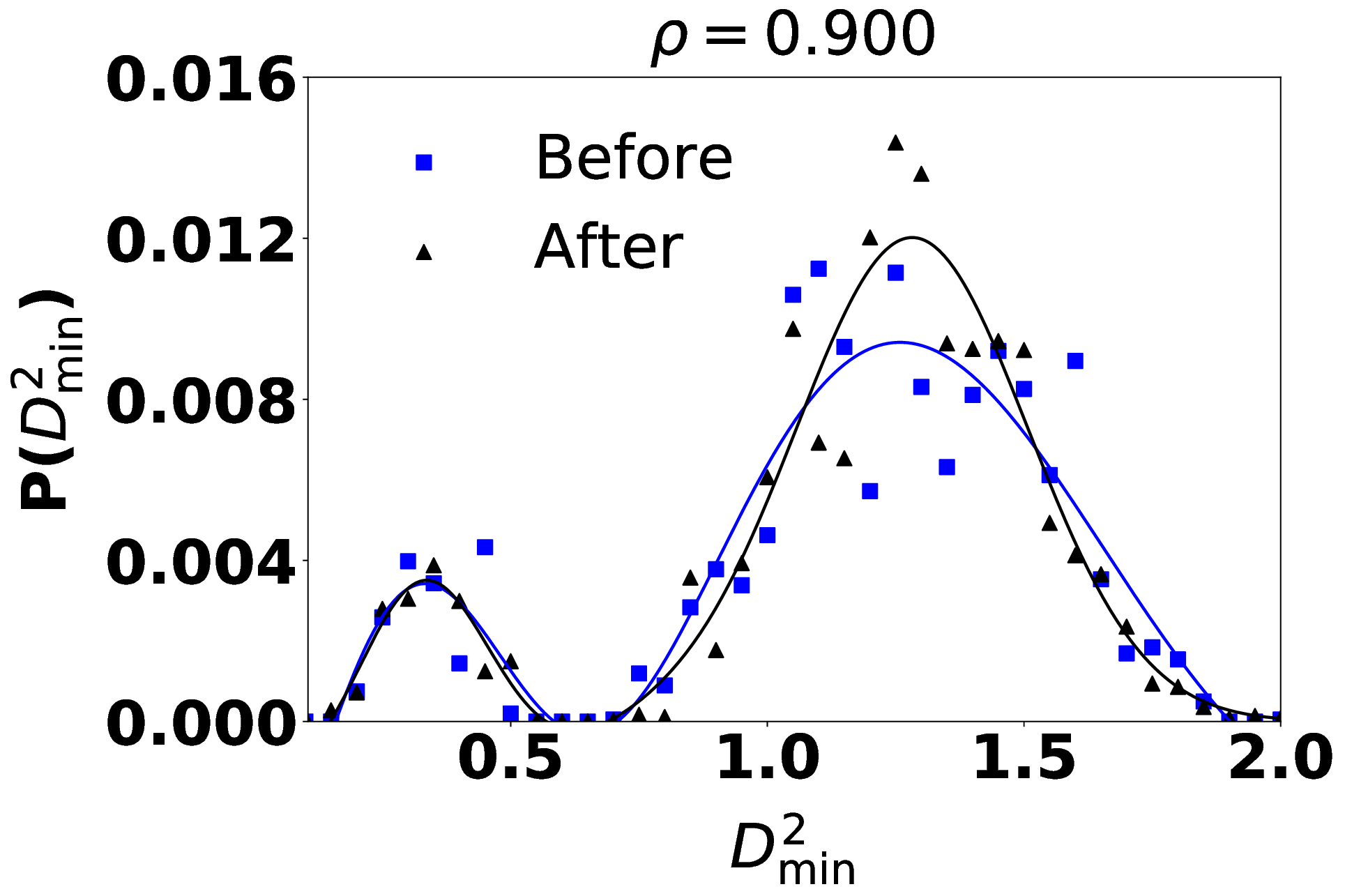}
    \includegraphics[width=0.46\textwidth]{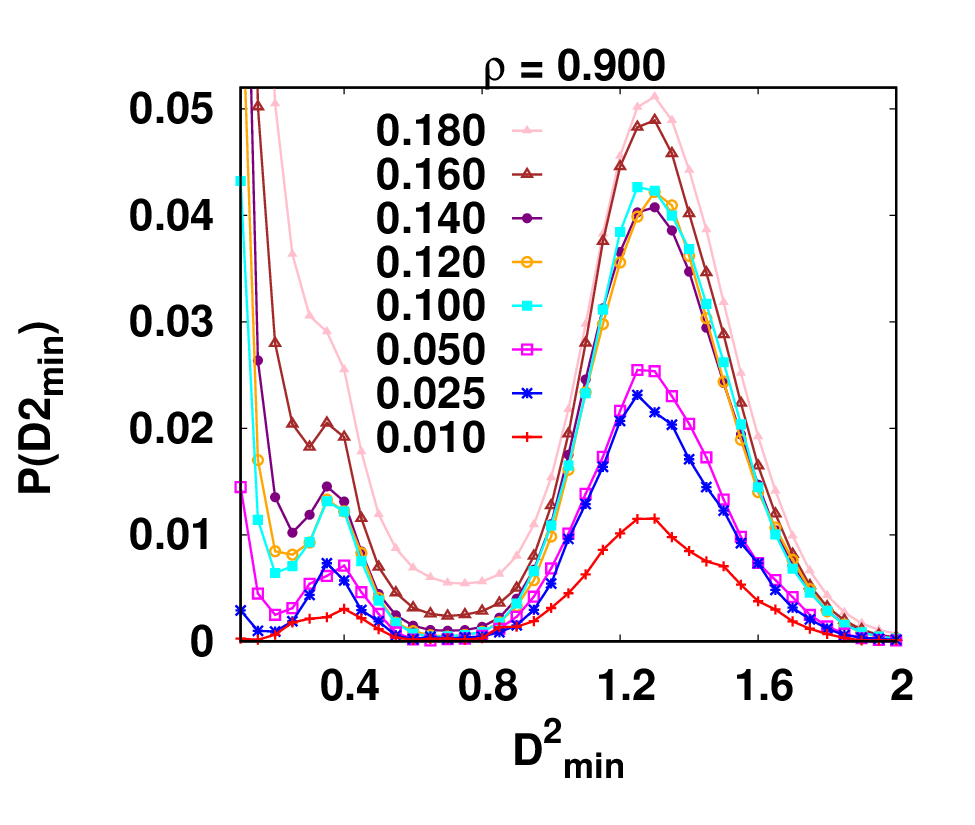}
    \put(-250,125){\textbf{(g)}}
    \put(-50,125){\textbf{(h)}}
    \caption{\SSG{
    \textbf{(a)-(c):} \emph{Visualization of the vector non-affine displacement vector field} for representative snapshots from a MD trajectory. Arrows represent x and y components of non-affine (NA) displacements of particles, describing both the magnitude and direction of local rearrangements. They reveal details such as swirl-like patterns, and rotational clusters. Notably, in the post-avalanche state, the displacement vectors are much more pronounced, with clearly visible, extended arrows (or “tails”) indicating regions of higher non-affinity.
    \textbf{(d)-(f):} \emph{Visualization of the scalar $D^2_{min}$ field}. Snapshots of the system color-coded by the scalar $D^2_{min}$ field of particles (d) before, (e) at, and (f) after the avalanche.
    \textbf{(g)} Comparison of $D^2_{min}$ distributions before and after a thermally mediated avalanche. The post-avalanche peak shifts to higher values, signaling a pronounced increase in non-affine displacements and a decrease in mechanical stability. 
    \textbf{(h)} Distribution of $D^2_{min}$ at various temperatures, each averaged over multiple independent trajectories. Two distinct peaks emerge, shifting to lower $D^2_{min}$ values with decreasing temperature, indicating a reduction in non-affine activity. 
    }}
    \label{fig:av2}
\end{figure*}

Let ${\mathbf{X}_i(t)}$ and ${\mathbf{x}_i(t+\Delta t)}$ denote the position vectors of a particle $i$ at times $t$ and $t+\Delta t$ respectively. We define separation vectors between particles $i,j$ at times $t$ and $t+\Delta t$ as 
\begin{align}
 \Delta \mathbf{X}_{ij}(t)&=\mathbf{X}_j(t)-\mathbf{X}_i(t) \nonumber\\
 \Delta\mathbf{x}_{ij}(t+\Delta t)&=\mathbf{x}_j(t+\Delta t)-\mathbf{x}_i(t+\Delta t)
 \label{main1}
\end{align}
$\Delta t$  denotes the time separation between two successive frames, which are, however, separated nonlinearly in actual time. The frame-to-time conversion is explained in Sec. \ref{sec:simuDet}. 

\SSG{The first key idea of our formalism is that an \emph{Affine} mapping from a configuration at time $t$ to that at time $t+\Delta t$ is expressed by a second rank tensor $\mathbf{H}$ termed the ``deformation gradient tensor'', such that $\mathbf{H}\cdot\Delta \mathbf{X}_{ij}(t)$ would be the separation vector at $t+\Delta t$ if the dynamics were purely Affine. During actual MD,} atoms of course do not move according to affine transformation. \SSG{Upto first order approximation, the non-affine contribution to the displacement of particle $i$ in a time interval $\Delta t$ can be written as}
\begin{align}
    \Delta x^{\text{NA}}_{i} &= \frac{\sum\limits_{j \in N_i} \Bigl[ \Delta x_{ij} - \Bigl( H_{xx}\,\Delta X_{ij} + H_{xy}\,\Delta Y_{ij} \Bigr) \Bigr]}{N_i} \nonumber\\
    \Delta y^{\text{NA}}_{i} &= \frac{\sum\limits_{j \in N_i} \Bigl[ \Delta y_{ij} - \Bigl( H_{yx}\,\Delta X_{ij} + H_{yy}\,\Delta Y_{ij} \Bigr) \Bigr]}{N_i}
    \label{d2xy}
\end{align}
where $x,y$ represent Cartesian coordinates and $H_{xx}$, $H_{xy}$, $H_{yx}$, and $H_{yy}$ denote the components of $\mathbf{H}$ and $N_i$ is the number of neighbors of particle $i$ upto second nearest neighbors, calculated from the pair correlation function $g(r)$. \SSG{In the context of strain-mediated plasticity, the ``error''} between the actual particle displacement and the Affine mapping \SSG{is typically described by a simpler, scalar metric:}
\begin{align}
    D^2_{ij}&=(\Delta\mathbf{x}_{ij}-\mathbf{H}_i\cdot\Delta\mathbf{X}_{ij})^T\cdot(\Delta\mathbf{x}_{ij}-\mathbf{H}_i\cdot\Delta\mathbf{X}_{ij})\nonumber\\
    D^2_{i}&=\frac{1}{N_i}\sum_{j \in N_i} D^2_{ij}
    \label{main4}
\end{align}
\SSG{The second key idea of the present formalism is that there exists an ``optimal''  $\mathbf{H}$ which minimizes the above loss function. This is obtained by minimizing $D^2_i$ with respect to $\mathbf{H}$, see Eqn. \ref{d2xy_sup3a}. The loss function for this optimal $\mathbf{H}$ will be denoted as $D^2_{min}$.}

Equations \ref{d2xy} and \ref{d2xy_sup3a} allow us to visualize the \SSG{vector non-affine displacement field. In Fig. \ref{fig:av2}(a)-(c), snapshots from a representative MD trajectory are shown before, at and after an avalanche event respectively.} Before the avalanche, the arrows are sparse and short, reflecting minimal rearrangements.  In contrast, large, well-organized vectors emerge at the onset of the avalanche. The direction of the local vectors reveal distinct clusters of arrows with patterns that are reminiscent of Eshelby-like lobes and vortex-like swirls  - indicative of coordinated rotational or swirling motion. These observations demonstrate that the avalanche involves a complex interplay of \SSG{non-Affine motion} reminiscent of Eshelby-type plasticity, \emph{even without an externally applied shear.}
In Fig. \ref{fig:av2}(d)-(f), the scalar $D^2_{min}$ field is visualized for the same configurations. The scalar nature highlights the magnitude of the displacement. Prior to the avalanche, most particles exhibited relatively low non-affinity suggesting a quiescent, stable arrangement, while at the avalanche, non-affinity increases drastically \SSG{revealing a localized event.} In the post-avalanche frame, red regions of high $D^2_{min}$ persist in a scattered pattern, highlighting that the avalanche have led to a distinct new configuration. 
The difference between configurations before and after avalanche is quantified in Fig. \ref{fig:av2}(g) by computing  the distribution of $D^2_{min}$. Two prominent peaks emerge in each histogram: one at low $D^2_{min}$ representing particles undergoing minimal displacement, and a second one at larger $D^2_{min}$ reflecting more significant, thermal-mediated structural rearrangements. \SSG{Data for four more independent avalanche events are shown in SI Fig. \ref{fig:av14nw7}. In all cases, the height of the second peak increases after avalanche.}

\SSG{To assess the influence of temperature in thermal-mediated plasticity, in Fig. \ref{fig:av2}(h) we analyze the average $D^2_{min}$ distribution, averaged over 12 independent MD trajectories, over a broad range of temperatures far out of equilibrium to about the simulation $T_g$ (where the available computing power fails to equilibrate the system). Two peaks are apparent at all temperatures. As the temperature decreases, both peaks shift toward lower $D^2_{min}$ values, indicating a general reduction in non-Affine activity, a natural consequence of diminished thermal agitation.}

\subsection{Particle Mobility}

\begin{figure*}[htbp!]
    \centering
    \includegraphics[width=0.49\textwidth]{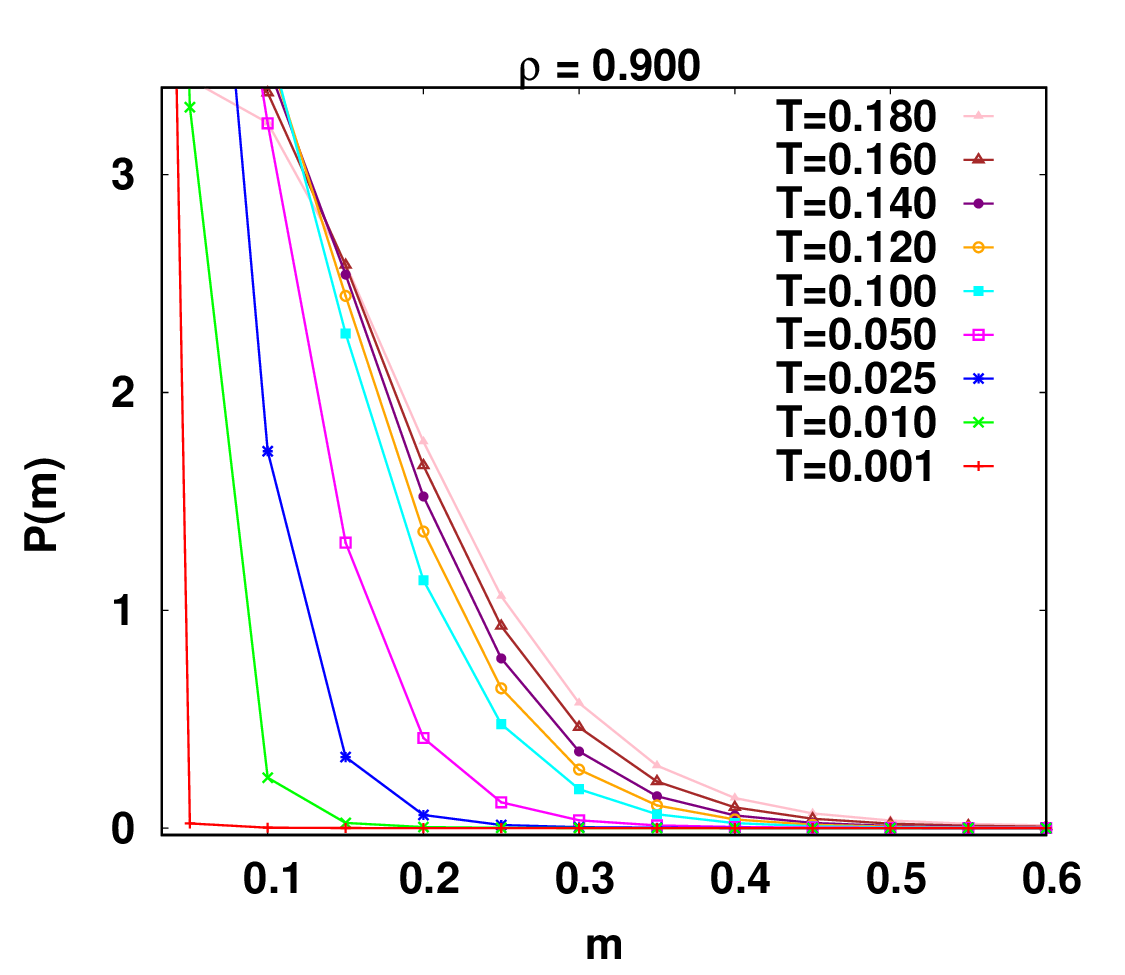}
    \includegraphics[width=0.49\textwidth]{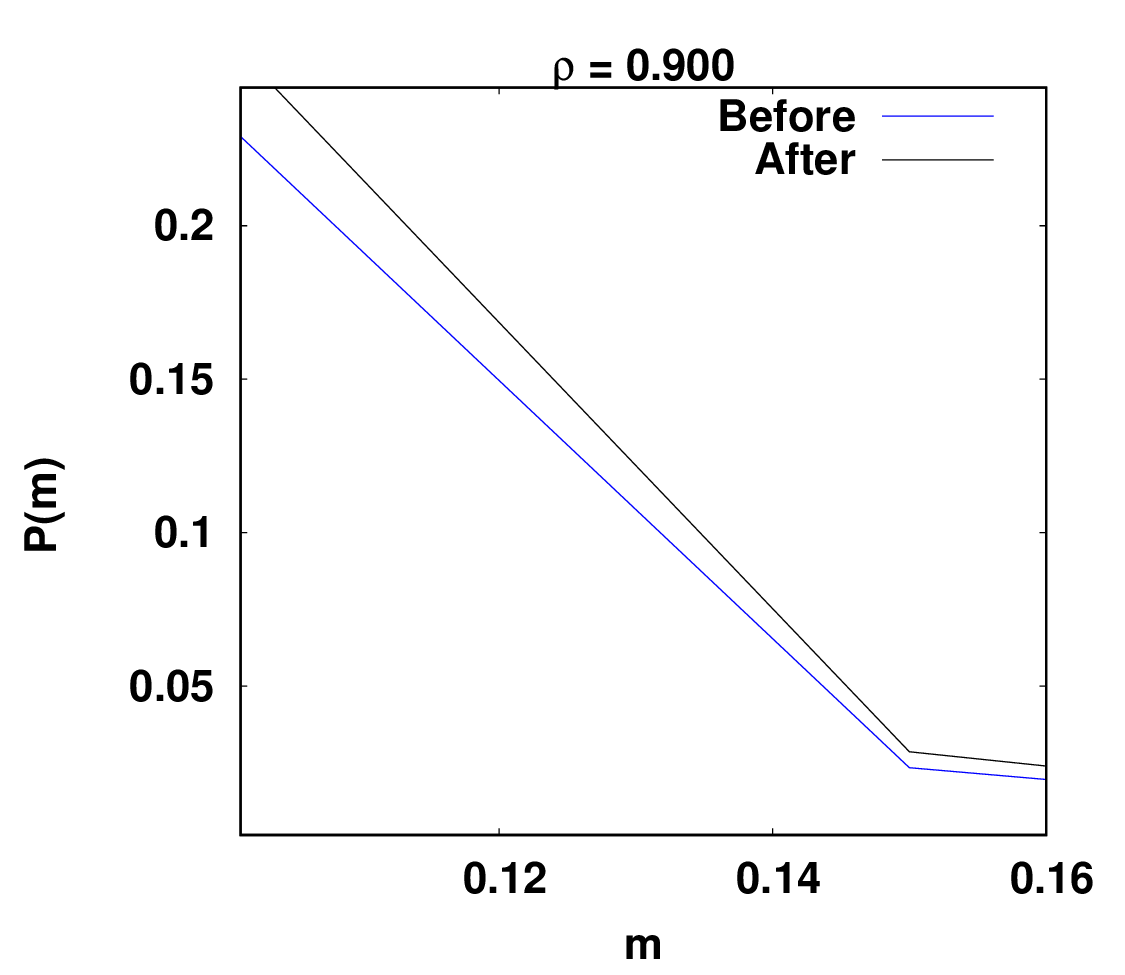}
    \put(-300,80){\textbf{(a)}}
    \put(-125,80){\textbf{(b)}}
    \\
    \vspace{5mm}
    \includegraphics[width=0.32\textwidth]{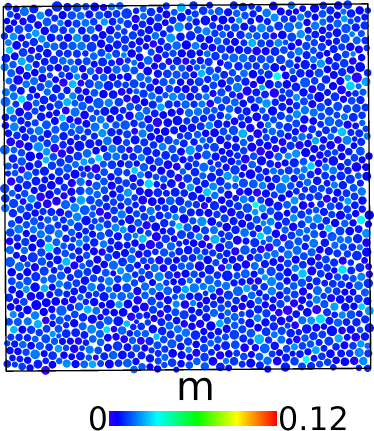}
    \includegraphics[width=0.32\textwidth]{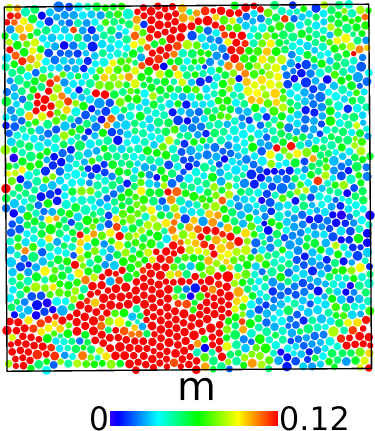}
    \includegraphics[width=0.32\textwidth]{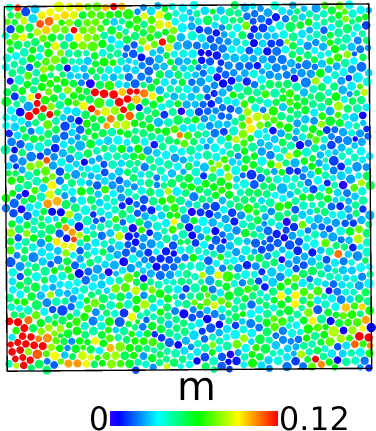}
    \put(-420,13){\textbf{(c) before}}
    \put(-280,13){\textbf{(d) at}}
    \put(-140,13){\textbf{(e) after}}
    \\
    \vspace{5mm}
    \includegraphics[width=0.32\textwidth]{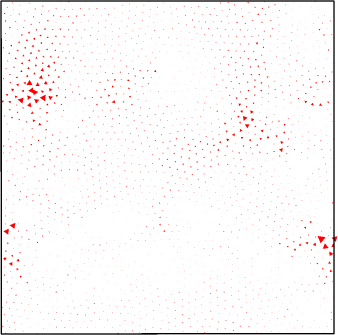}
    \includegraphics[width=0.32\textwidth]{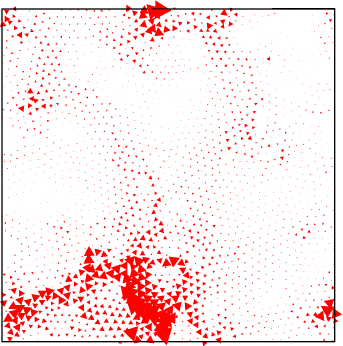}
    \includegraphics[width=0.32\textwidth]{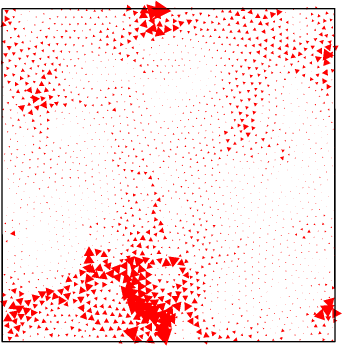}
    \put(-375,110){\textbf{(f) \small{before}}}
    \put(-245,110){\textbf{(g) \small{at}}}
    \put(-105,110){\textbf{(h) \small{after}}}\\
    \caption{ 
    \textbf{(a):} \SSG{Temperature dependence of ensemble-averaged mobility} distribution As temperature decreases, the distributions shift toward lower mobility. 
    \textbf{(b):} Dynamic mobility before (blue) and after (black) a thermally mediated avalanche, \SSG{highlighting the difference between the pre- and post-avalanche configurations.} 
    \textbf{(c)-(e):} \SSG{Snapshots of mobility field} (c) before, (d) at, and (e) after the avalanche.
    \textbf{(f)-(h):} Particle displacement field $\mathbf{u}^i_{IS(j)}$ [Eqn. \ref{main10}] measuring distances between potential energy minima [``inherent structures'' or IS] explored by the system (f) before, (g) during and (h) after a thermal-mediated avalanche. Before the avalanche \SSG{the displacement field shows uniform spatial distribution} with small displacements. The displacement field at the onset of the avalanche shows spatial heterogeneity with localized regions of large displacement indicating the core of the event. The post-avalanche displacement field is clearly distinct from the pre-avalanche one.
    }
    \label{fig:av6}
\end{figure*}

\SSG{To further investigate the mechanism underlying thermal-mediated avalanche events, in this section we analyze the particle mobility field using two different measures.}

\paragraph{Mobility $m$ from MD:} The magnitude of the displacement \SSG{for any particle $i$ between two successive configurations} is defined as
\begin{align}
    m(i) &= \sqrt{\left(x_i - x_{i,0}\right)^2 + \left(y_i - y_{i,0}\right)^2}
    \label{main8b}
\end{align}
where $x_i$ and $y_i$ are the current coordinates, and $x_{i,0}$ and $y_{i,0}$ are the coordinates from the previous time step. \SSG{Note that,} unlike fixed reference methods, which calculate displacements relative to an initial configuration, \SSG{in the present analysis, } the reference frame is updated immediately after computing the displacement:
\begin{align}
 x_{i,0} \leftarrow x_{i}, \; y_{i,0} \leftarrow y_{i}
\label{main9}
\end{align}  

Fig. \ref{fig:av6}\SSG{(a) shows the temperature dependence of the distribution of ensemble-averaged mobility.} We see that \SSG{particle} mobility decreases systematically \SSG{with the distribution getting narrower} with \SSG{decreasing} temperature, \SSG{as expected}. \SSG{In Fig. \ref{fig:av6}(b)-(e) we analyze the evolution of mobility across a thermal‐mediated avalanche event.} Fig. \ref{fig:av6}(b) compares the dynamic mobility distribution before and after an avalanche. The post‐avalanche curve (black) is shifted toward higher mobility values, indicating that the system becomes \SSG{softer and thus more susceptible to local rearrangements} after the avalanche. \SSG{The same conclusion is reached for four other events, see SI Fig. \ref{fig:av14nw6}.} Fig. \ref{fig:av6}(c)-(e) shows \SSG{the real-space distribution of the particle mobility using representative configurations (c) immediately before, (d) at and (e) after an avalanche event. The mobility field is relatively homogeneous with pre-dominantly low values before the avalanche. It shows significant heterogeneity with patches of high mobility appearing in the region with large $D^2_{min}$. Note also that the mobility field demonstrates that the post-avalanche state is distinct from the pre-avalanche one.} 

\paragraph{Displacement along the potential energy landscape:}
\SSG{Next, we directly examine the evolution of the system in the potential energy landscape by computing the particle displacement between the potential energy minima (``inherent structures'' or IS) explored by the system.}  We designate the first energy-minimized configuration as the fixed reference inherent structure $IS(0)$ to filter out transient thermal fluctuations. For every subsequent inherent structure $IS(j)$, the displacement vector for the $i^{th}$  particle is defined as
\begin{align}
    \mathbf{u}^i_{IS(j)}=\mathbf{X}_{IS(j)}^i-\mathbf{X}_{IS(0)}^i
    \label{main10}
\end{align}
Where $\mathbf{X}^i_{IS(j)}$ denotes the position of the $i^{th}$ particle in the $j^{th}$ inherent structure and $\mathbf{X}^i_{IS(0)}$ denotes the position of the $i^{th}$ particle in the reference inherent configuration.
\SSG{Fig. \ref{fig:av6}(f)-(h) displays the displacement vector field $\mathbf{u}^i_{IS(j)}$ for representative snapshots (f) before, (g) at and (h) after an avalanche event from the corresponding IS trajectory. In particular,} Fig. \ref{fig:av6}(g) reveals pronounced clusters of large displacements in the same region where $D^2_{min}$ reach elevated values [compare also with Fig. \ref{fig:av5}]. This shows that the avalanche event is highly localized. 
\emph{Surrounding these clusters are characteristic ``swirl'' pattern reminiscent of Eshelby-like quadrupolar displacement field in shear induced plasticity.} Despite the absence of externally imposed shear, \SSG{such similarity suggests that quadrupolar displacement field is also the lowest energy excitation in thermal-mediated plasticity.} 

\section{Relative significance of volumetric {\it vs.} deviatoric strains} \label{sec:strain}
\SSG{After analyzing the non-Affine displacement field in sec. \ref{sec:nonAff} we next focus on understanding the evolution of local strain field during thermal-mediated avalanche events. In shear deformation response it is typically assumed that strain is dominated by shear \cite{falk1998dynamics, maloney2006amorphous, picard2004elastic} {\it i.e.} the local density remains unchanged by avalanches, however recent studies have noted the importance of volumetric strain \cite{zhang2021interplay}. It will provide valuable insight to explicitly test the thermal-mediated case. Thus we analyze different contributions to (local) strain to determine if volumetric strain plays any significant role.}

\paragraph{Local strain measures:}
\SSG{Upto linear order approximation, the strain tensor $\mathbf{\epsilon }$ is related to the (optimal) deformation gradient tensor $\mathbf{H}$ {\it via}} 
\begin{align}
 \mathbf{H}&=\mathbf{I}+\mathbf{\epsilon }   
 \label{main7}
\end{align}
where $\mathbf{I}$ is the $2\times2$ identity tensor. From the theory of elasticity, \SSG{the total strain tensor $\mathbf{\epsilon}$ can be decomposed into volume-preserving, deviatoric strain tensor $\mathbf{e}$ and the volumetric strain tensor $\epsilon_{vol}$} which captures local compression or dilation within the material. These strains can be used to understand the degree of deformation locally during a thermally mediated avalanche. \SSG{In particular, we use the norm $\Vert\mathbf{e}\Vert$ [Eqn. \ref{22}] to measure the ``magnitude'' of the shear strain field, and the norm $\Vert\epsilon_{vol}\Vert$ [Eqn. \ref{23}]} to quantify the magnitude of the local volume changes regardless of their direction. An alternative approach to characterize local volume changes is through the determinant $\det \mathbf{H}$ of the \emph{optimal} deformation gradient tensor [Eqn. \ref{d2xy_sup3a}]. Unlike the norm $\Vert\epsilon_{vol}\Vert$, which does not distinguish between compression and dilation, $\det \mathbf{H}$ retains the directional information about the deformation. \SSG{In a local region subjected to a volume (area in 2D) changing transformation,} the transformed area $A'$ is given by the transformation rule:
\begin{align}
A'=(\det \mathbf{H})\, A
\label{main8a}
\end{align}
Thus a given region is ``dilating'' if $\det \mathbf{H} > 1$, ``compressing'' if $\det \mathbf{H} < 1$ and \SSG{undergoes no local volume change} if $\det \mathbf{H} = 1$, see Appendix \ref{sec:appStrain} for further details. \SSG{For the sake of completeness, we also analyze the full strain field using the norm $\Vert \mathbf{\epsilon} \Vert$ of the total strain tensor [Eqn. \ref{23ad1}].}

\begin{figure*}[ht!]
    \centering
    \includegraphics[width=0.45\textwidth]{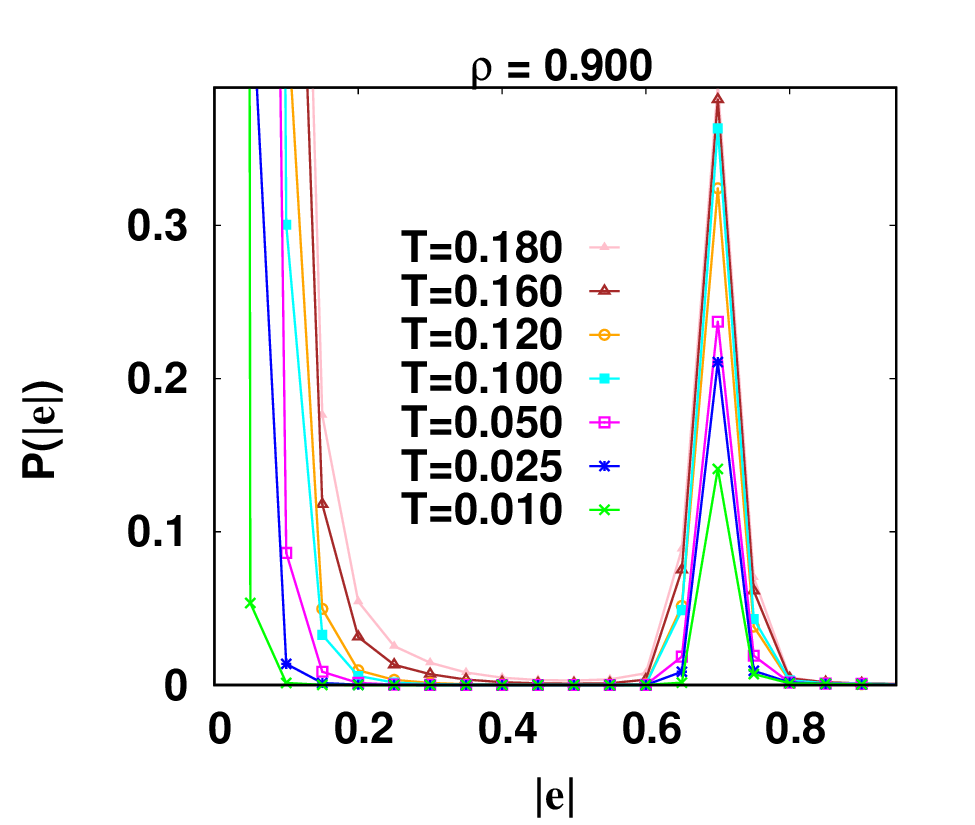}
    \includegraphics[width=0.45\textwidth]{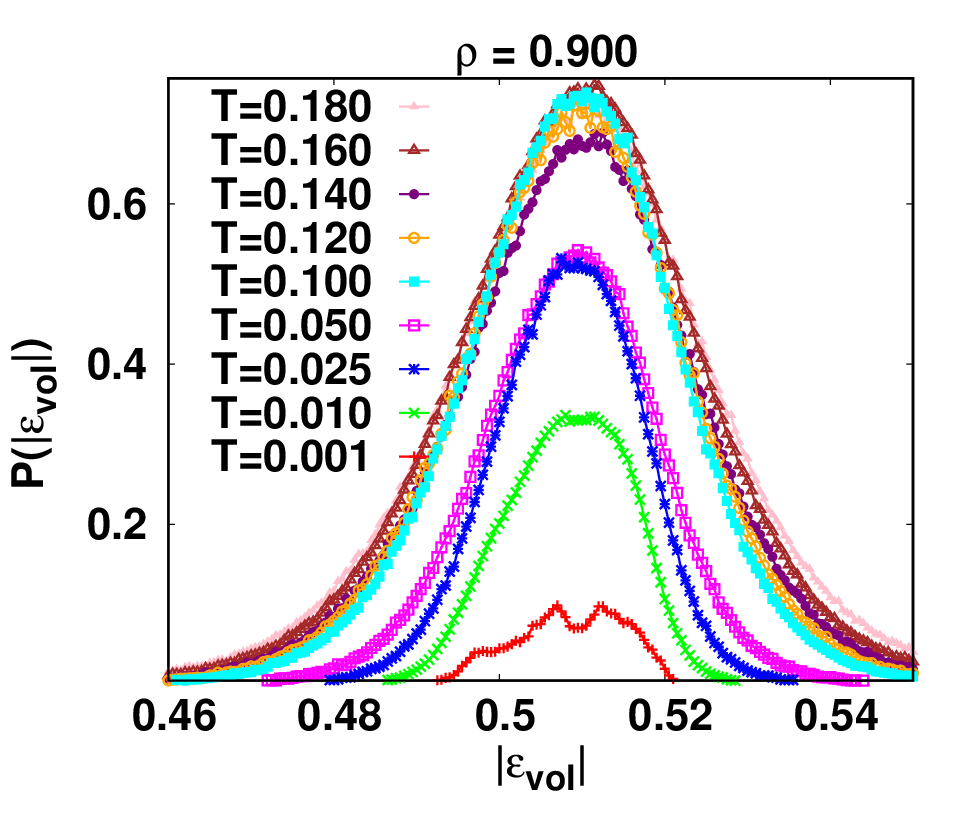}
    \put(-300,125){\textbf{(a)}}
    \put(-50,125){\textbf{(b)}}\\
    \includegraphics[width=0.45\textwidth]{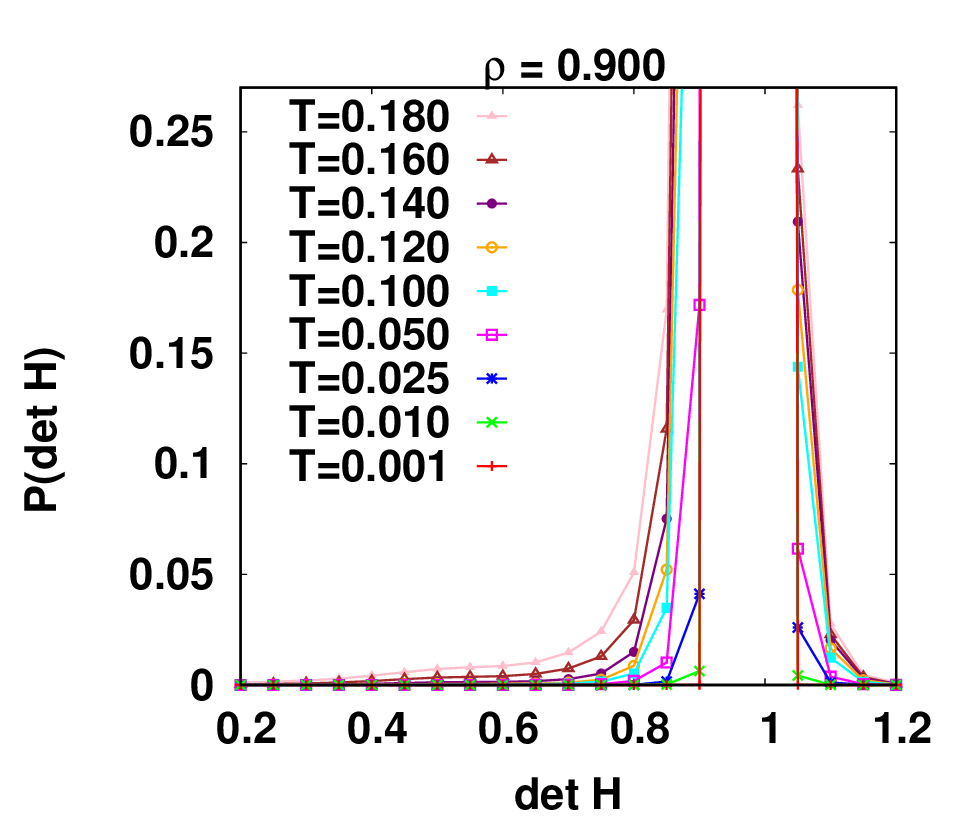}
    \includegraphics[width=0.45\textwidth]{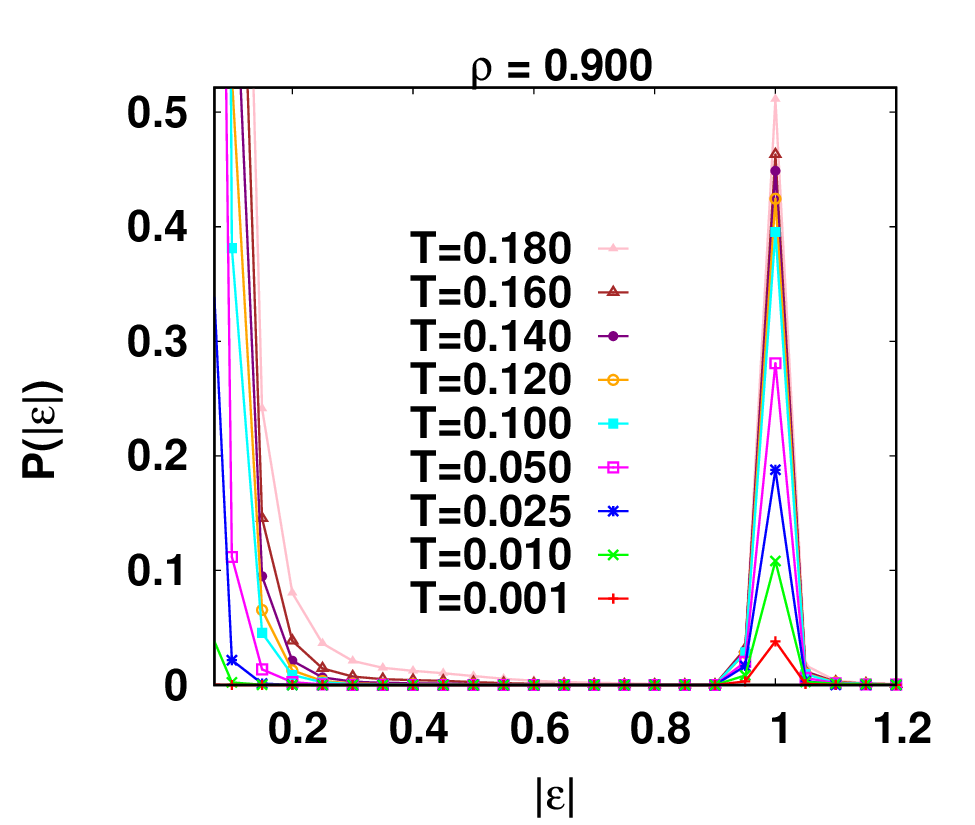}
    \put(-300,55){\textbf{(c)}}
    \put(-100,125){\textbf{(d)}}
    \caption{\SSG{
    \emph{Temperature ($T$) dependence of ensemble-averaged local measures of strain.} 
    Distributions of \textbf{(a)} norm $\Vert\mathbf{e}\Vert$ of local deviatoric strain tensor, \textbf{(b)} norm $\Vert \epsilon_{vol}\Vert$ of local volumetric strain tensor,  \textbf{(c)} $\det \mathbf{H}$ and \textbf{(d)} norm $\Vert \mathbf{\epsilon} \Vert$  of total local strain over a broad range of low temperatures, each averaged over 12 independent trajectories. As temperature decreases, both the height of the peak and the width of the respective distributions decreases for all strain measures ($\Vert\mathbf{e}\Vert$, $\Vert \epsilon_{vol}\Vert$, $\Vert \mathbf{\epsilon} \Vert$). This indicates that lowering the temperature leads to a more quiescent, less heterogeneous local strain environment. Further, although both local dilation $\det \mathbf{H}<1$ and local compression $\det \mathbf{H}>1$ are present, compression events, having longer tail, dominate across all temperatures.
    }}
    \label{fig:av3}
\end{figure*}

\begin{figure*}
    \centering
    \includegraphics[width=0.45\textwidth]{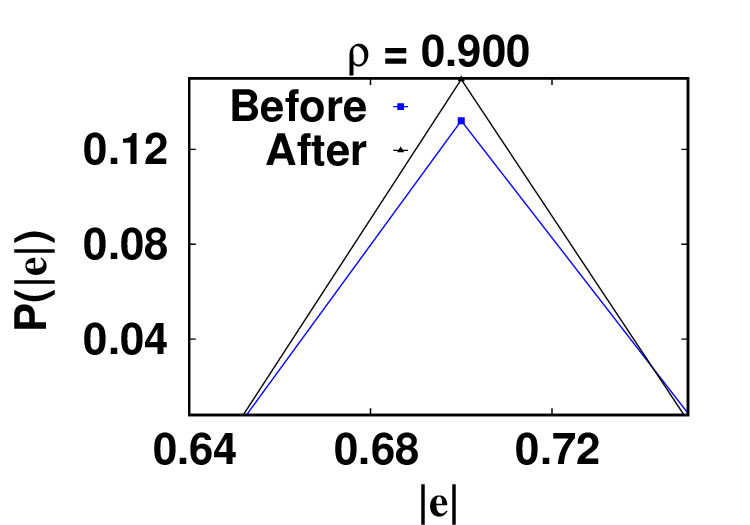}
    \includegraphics[width=6.5cm, height=4.5cm]{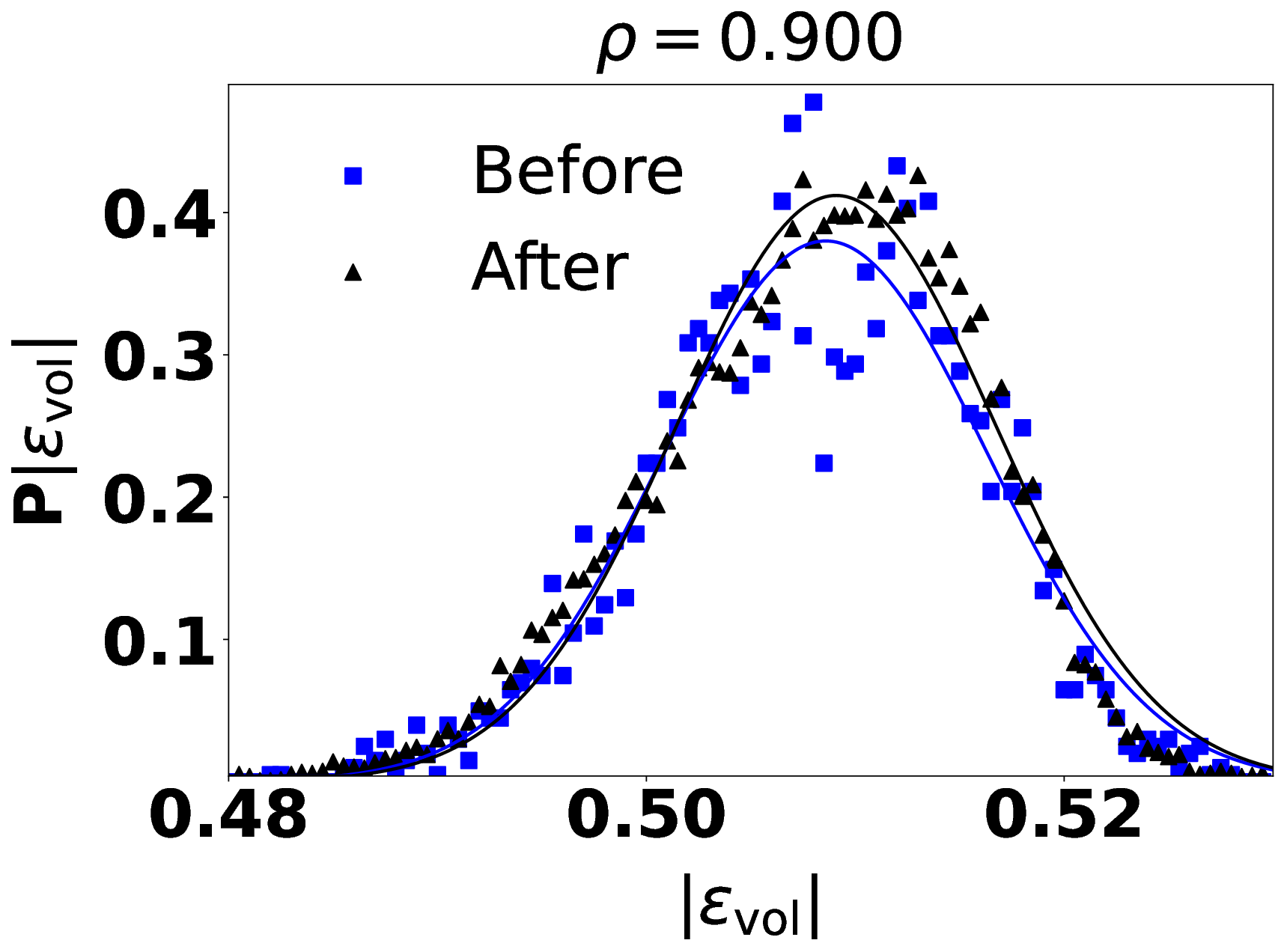}
    \put(-330,90){\textbf{(a)}}
    \put(-140,80){\textbf{(b)}}\\
    \includegraphics[width=0.4\textwidth]{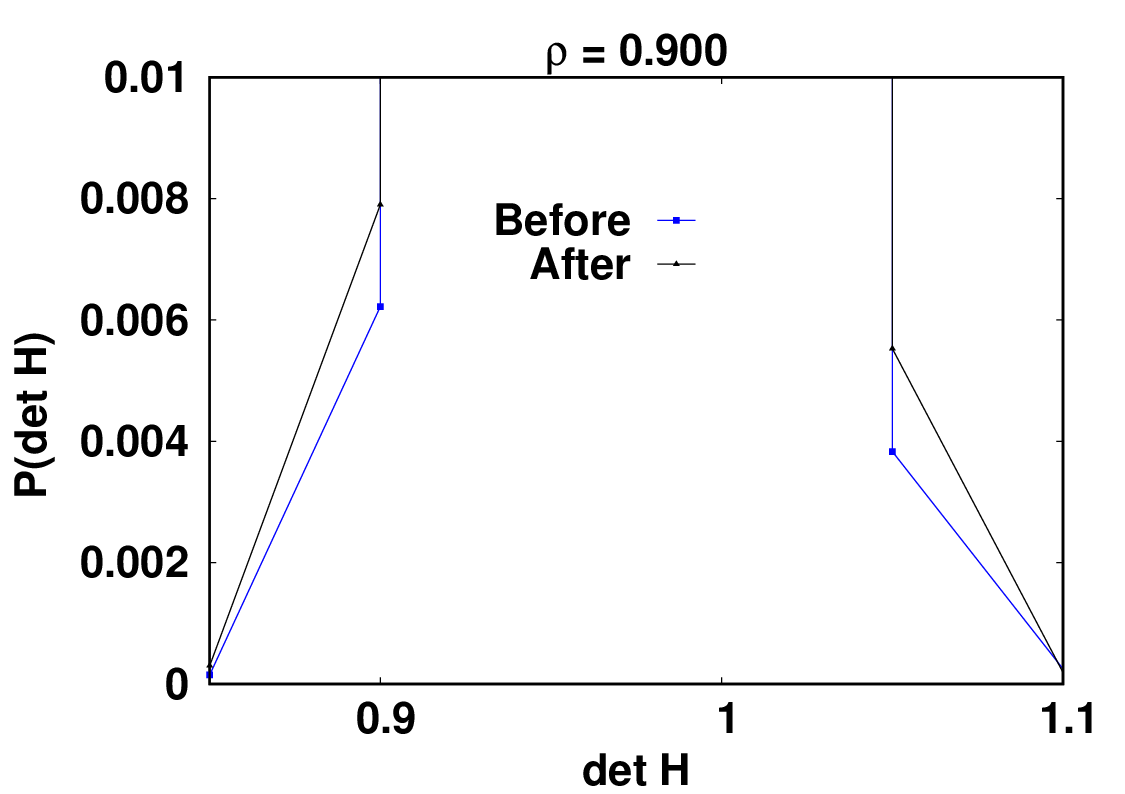}
    \includegraphics[width=6.5cm, height=5cm]{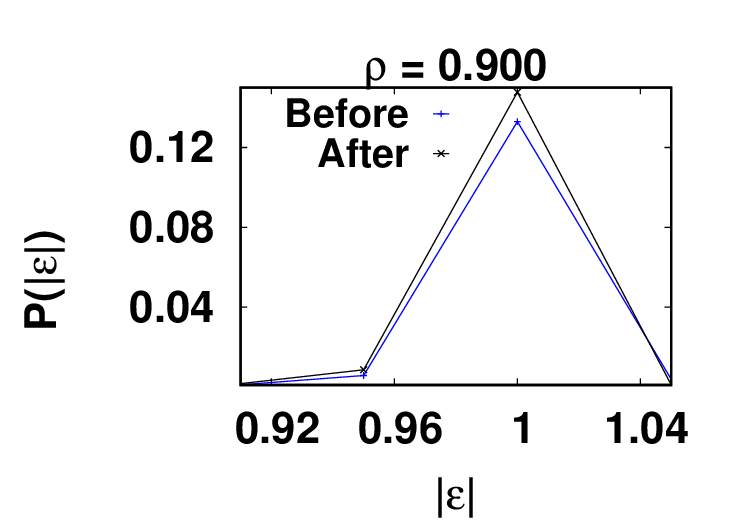}
    \put(-320,80){\textbf{(c)}}
    \put(-120,80){\textbf{(d)}}
    \caption{\SSG{
    \emph{Comparison of distributions of local strain measures immediately before (blue) and after (black) a thermally mediated avalanche event.}
    Distributions for representative configurations of \textbf{(a)} Norm $\Vert\mathbf{e}\Vert$ of local deviatoric strain, \textbf{(b)} norm $\Vert \epsilon_{vol}\Vert$ of local volumetric strain, \textbf{(c)} $\det \mathbf{H}$ and \textbf{(d)} norm $\Vert \epsilon \Vert$ of total local strain tensor show the difference in the configurations before and after an avalanche event. The post‐avalanche regime exhibits higher peak in both deviatoric and volumetric strains as well as in the total strain. This suggests an increased propensity plasticity due to an avalanche event. Also $\det \mathbf{H}$ shows an enhanced high‐compression tail after avalanche indicating that local compression becomes more pronounced following structural relaxation. See also SI Figs. \ref{fig:av14nw10}, \ref{fig:av14nw8}. \ref{fig:av14nw9} and \ref{fig:av14nw11} for additional data for four other independent avalanche events.
    }}
    \label{fig:av4}
\end{figure*}

\begin{figure*}
    \centering
    \includegraphics[width=0.23\textwidth]{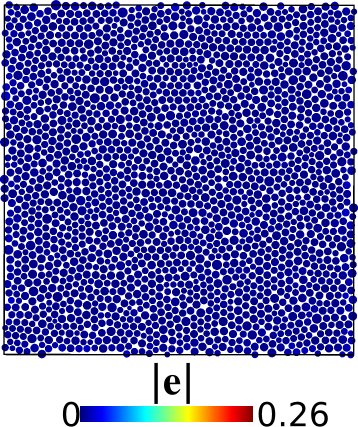}\hspace{10mm}
    \includegraphics[width=0.23\textwidth]{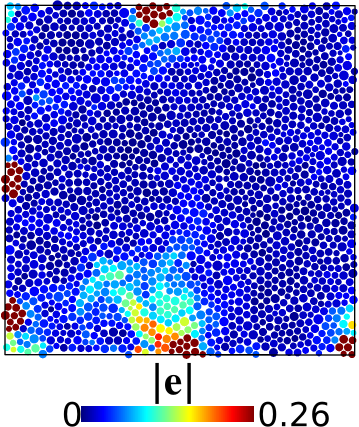}\hspace{10mm}
    \includegraphics[width=0.23\textwidth]{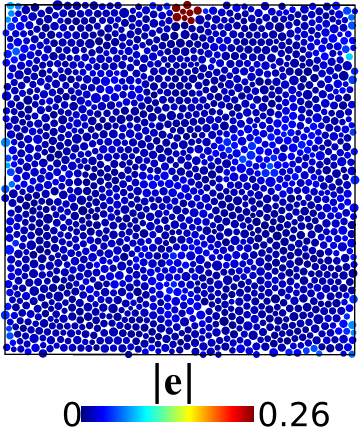}
    \put(-375,13){\textbf{(a) \small{before}}}
    \put(-230,13){\textbf{(b) \small{at} }}
    \put(-105,13){\textbf{(c) \small{after} }}\vspace{5mm}\\
    \includegraphics[width=0.23\textwidth]{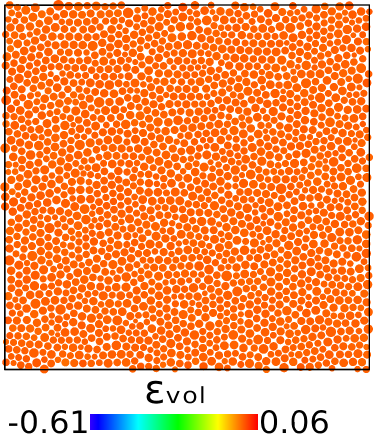}\hspace{10mm}
    \includegraphics[width=0.23\textwidth]{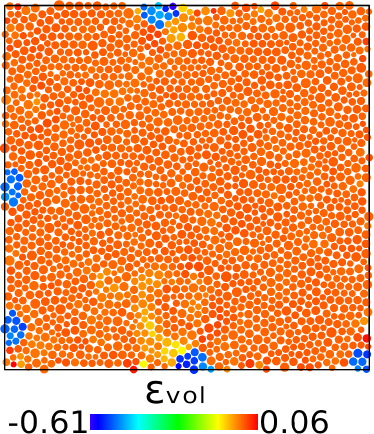}\hspace{10mm}
    \includegraphics[width=0.23\textwidth]{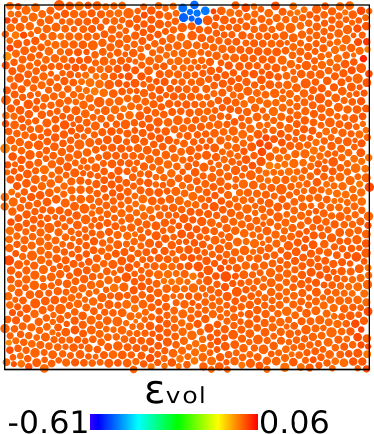}
    \put(-375,10){\textbf{(d) before}}
    \put(-230,10){\textbf{(e) at}}
    \put(-105,10){\textbf{(f) after}}\vspace{5mm}\\
    \includegraphics[width=0.23\textwidth]{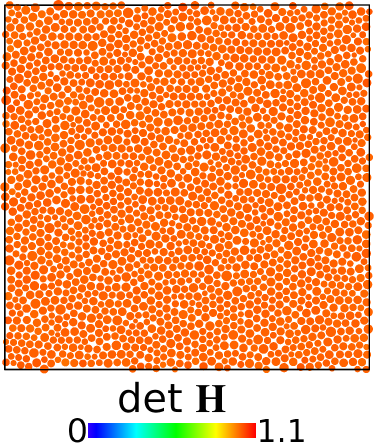}\hspace{10mm}
    \includegraphics[width=0.23\textwidth]{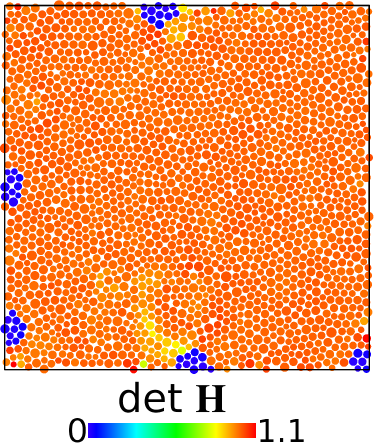}\hspace{10mm}
    \includegraphics[width=0.23\textwidth]{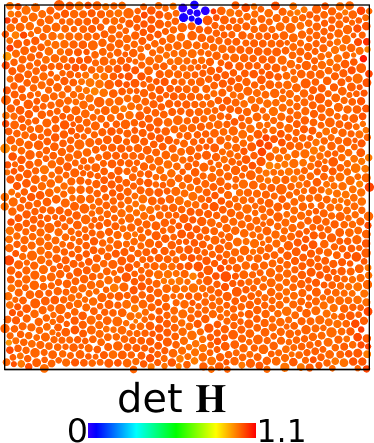}
    \put(-375,13){\textbf{(g) \small{before}}}
    \put(-230,13){\textbf{(h) \small{at}}}
    \put(-110,13){\textbf{(i) \small{after}}}\vspace{5mm}\\
    \includegraphics[width=0.23\textwidth]{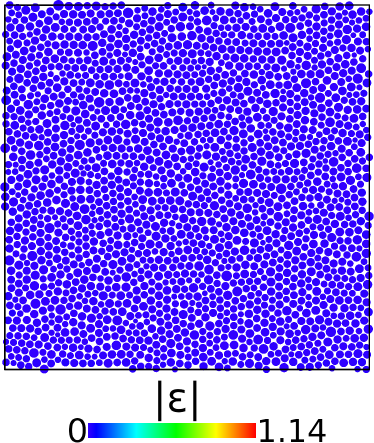}\hspace{10mm}
    \includegraphics[width=0.23\textwidth]{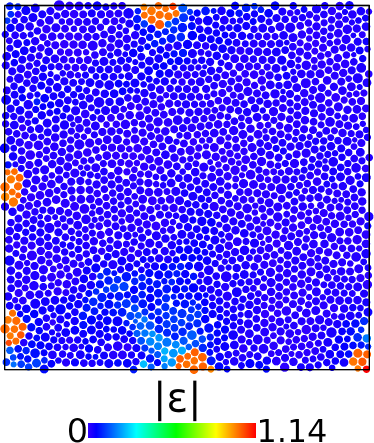}\hspace{10mm}
    \includegraphics[width=0.23\textwidth]{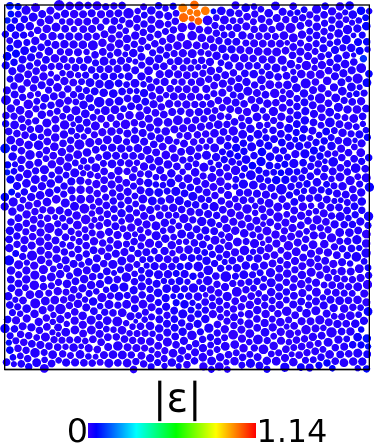}
    \put(-375,13){\textbf{(j) before}}
    \put(-230,13){\textbf{(k) at}}
    \put(-105,13){\textbf{(l) after}}
    \caption{\SSG{
    \emph{Visualization of local strain field measures across an avalanche event.}
    Color‐coded representative configurations taken from MD trajectory before, at and after an avalanche event for \textbf{(a)-(c)} norm $\Vert\mathbf{e}\Vert$ of local deviatoric strain tensor [Eq.~\eqref{22}], \textbf{(d)-(f)} local $\epsilon_{vol}$, \textbf{(g)-(i)} $\det \mathbf{H}$ [Eqs.~\eqref{d2xy_sup3a} and \eqref{25}] at particle level and \textbf{(j)-(l)} norm $\Vert \mathbf{\epsilon}\Vert$ of total local strain tensor [Eq.~\eqref{23ad1}]. 
    In \textbf{(a)-(c)}, red regions denote higher shear‐like deformations, which intensify at the avalanche and remain scattered throughout the system afterward, underscoring the persistent impact of the structural relaxation event. 
    In \textbf{(d)-(f)}, blue regions represent local compression and red regions denote local dilation. Blue patches increase dramatically at avalanche and persist after the avalanche. Note that here $\epsilon_{vol}$ has sign, to distinguish between local compression and dilation, see Eq.~\eqref{23_vol}. 
    In \textbf{(g)-(i)}, blue patches highlight regions of local compression, which intensify during the avalanche and remain scattered in the post‐avalanche configuration, revealing a bias toward compressive rearrangements.
    In \textbf{(j)-(l)} patches of orange color indicate regions of high strain that persist after the avalanche, revealing that the norm of strain is enhanced due to the avalanche. 
    }}
    \label{fig:av5}
\end{figure*}

\paragraph{Temperature dependence of average strains:}
In Fig. \ref{fig:av3}, \SSG{we present the ensemble-averaged distributions of (a) the norm  $\Vert\mathbf{e}\Vert$ of the local deviatoric strain tensor, (b) the norm $\Vert \epsilon_{vol}\Vert$ of the local volumetric strain tensor, (c) the determinant $\det \mathbf{H}$ of the optimal deformation gradient tensor, and (d) the norm $\Vert \mathbf{\epsilon}\Vert$ of the total strain tensor over a range of low temperatures to demonstrate} how they depend on temperature. Each curve represents an average over 12 independent trajectories at a given temperature. 

\SSG{The trend is same for all the measures: lowering the temperature reduces the peak height, shifts it toward smaller values, and narrows the width of the distribution.} This parallels the behavior observed in $D^2_{min}$ [Fig. \ref{fig:av2}(h)]. Overall they suggest that decrease in thermal \SSG{fluctuations leads to} more quiescent, less heterogeneous strain environment. \SSG{We however emphasize that values of $\Vert \epsilon_{vol}\Vert$ are comparable to those of $\Vert\mathbf{e}\Vert$ indicating that both volumetric and deviatoric strains play significant roles in thermal-mediated local rearrangements.} Also in Fig. \ref{fig:av3}(c), although the distributions exhibit tails for both ``compression'' and ``dilation'', the compression side ($\det \mathbf{H}<1$) is more pronounced at all temperatures. This finding suggests that thermal fluctuations often manifest as compressive rearrangements rather than expansions when viewed at the particle level. \SSG{Note that the net global volume change must be zero, as the MD is performed at constant volume.}

\paragraph{Evolution of strain across an avalanche event:}
\SSG{In Fig.\ref{fig:av4} we compare the distribution of local strain measures {\it viz.} (a) $\Vert\mathbf{e}\Vert$, (b) $\Vert \epsilon_{vol}\Vert$, (c) $\det \mathbf{H}$, and (d) $\Vert \mathbf{\epsilon}\Vert$ before and after a representative thermally mediated avalanche event. To assess statistical significance of our results, the same data for four more avalanche events are included in SI Figs. \ref{fig:av14nw10}, \ref{fig:av14nw8}, \ref{fig:av14nw9} and \ref{fig:av14nw11} respectively.}

\SSG{For \emph{all} local strain measures, there is a clear increase in the height of the peak of the distribution \emph{after} the avalanche. This suggests that thermal-mediated avalanche also increases the propensity to further deformation - both deviatoric and volumetric - similar to the case of the shear deformation response. Further,}  Fig.\ref{fig:av4}(c) shows that while both compression and dilation intensify after avalanche, the compression tail $\det \mathbf{H}<1$ becomes more prominent.

These trends are further supported by color-coded, particle-level visualizations in Fig. \ref{fig:av5} at three \SSG{representative} key moments: immediately before,  at and after the avalanche. Thus Fig. \ref{fig:av5}(a)-(c) show $\Vert\mathbf{e}\Vert$ field, Fig. \ref{fig:av5}(d)-(f) display signed $\epsilon_{vol}$ [Eqn. \ref{23_vol}], Fig. \ref{fig:av5}(g)-(i) exhibit $\det \mathbf{H}$ and Fig. \ref{fig:av5}(j)-(l) visualize $\Vert \mathbf{\epsilon}\Vert$ field respectively along the same MD trajectory. In these images, local compression appears in blue, while local dilation appears in \SSG{yellow or red} depending on the \SSG{magnitude}.

Several \SSG{interesting} observations emerge: the configuration before the avalanche display \SSG{uniform and relatively} low deviatoric as well as volumetric strains indicative of a relatively \SSG{homogeneous}, stable arrangement. \SSG{Further, $\det \mathbf{H} > 1$ indicates local dilation.} The configuration after the avalanche also exhibit mostly uniform strain and $\det \mathbf{H}$ fields. However, after the avalanche, there are clear patches with very distinct deviatoric strain and elevated compressive strains. Thus Fig. \ref{fig:av5} visually demonstrates that the system goes to a new configuration due to the avalanche.

\SSG{More interestingly,} the configuration at the avalanche - Fig.  \ref{fig:av5}(b),(e),(h),(k) - shows distinct \SSG{localized shear deformation event} in deviatoric as well as in volumetric strains at the same region where $D^2_{min}$ field is also large [Fig. \ref{fig:av2}(e)]. In particular, Fig. \ref{fig:av5}(e) shows that blue patches indicative of local compression appear in the regime \SSG{where the avalanche is triggered. Thus,} during the avalanche, specific regions exhibit noticeable compression (blue) coinciding with the zones of elevated $D^2_{min}$ and deviatoric strain. \SSG{Thus \emph{volumetric strain is significant in thermal-mediated avalanche events.} In other words, thermally mediated plasticity cannot be fully described by shear alone. As shown in Figs. \ref{fig:av3}, \ref{fig:av4}, and \ref{fig:av5}, thermal fluctuations trigger rearrangements that combine both deviatoric (volume-preserving) and volumetric (compressive or dilative) components, leading to a broader class of local transformations than is typically encountered in shear deformation response. This is a central result of the present work.}

\section{Softness Evolution in Thermal Structural Relaxation}\label{sec:soft}
\begin{figure}[ht!]
    \centering
    \includegraphics[width=0.48\textwidth]{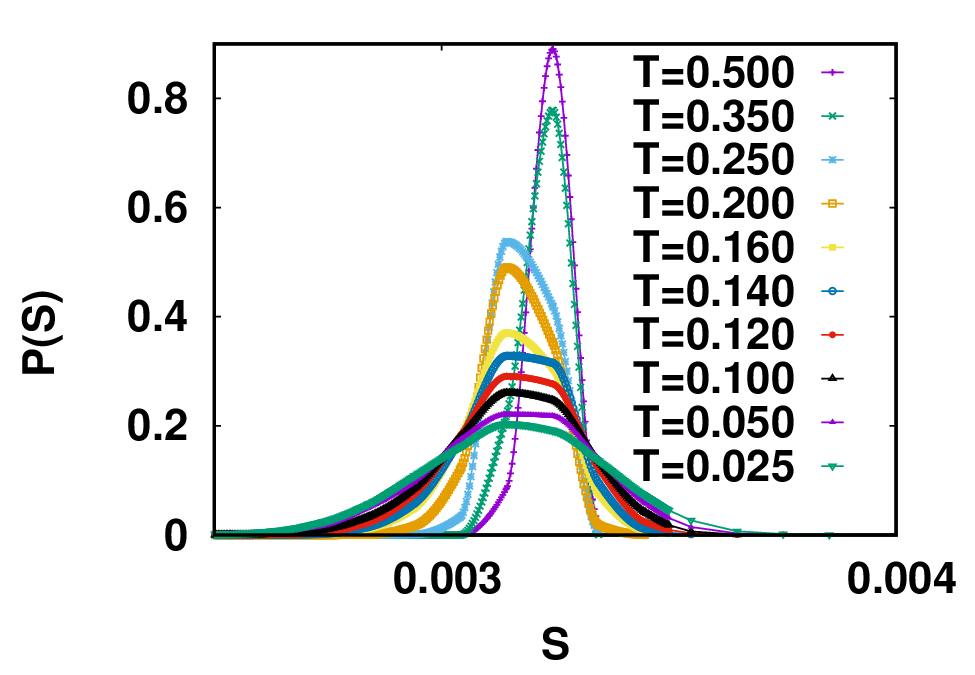}
    \put(-150,80){\textbf{(a)}}\\
    \vspace{5mm}
    \includegraphics[width=0.48\textwidth]{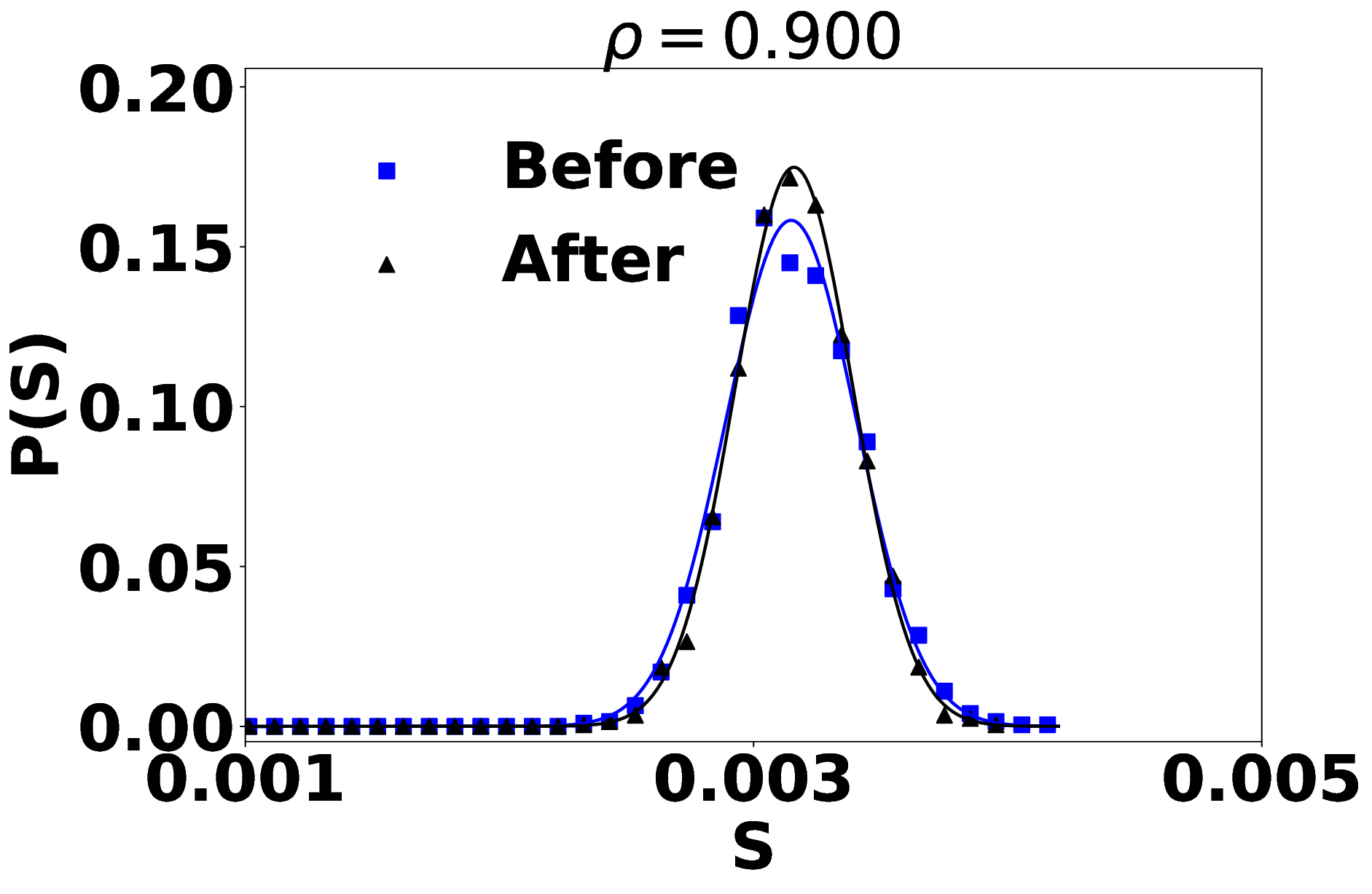}
    \put(-150,70){\textbf{(b)}}
    \caption{
    \textbf{(a)} Distribution of softness at various temperatures averaged over multiple trajectories. The distributions shift toward lower softness as temperature decreases. 
    \textbf{(b)} Comparison of softness before (blue) and after (black) a thermally mediated avalanche indicating that the system becomes softer despite a seemingly similar mean square displacement plateaus.
    }
    \label{fig:av8}
\end{figure}

\SSG{So far we have characterized thermally mediated avalanche event by stress and strain response, vibrational characteristic, and (non-Affine) particle displacement fields. For each measure, our analysis suggests that the system ``softens'' due to the avalanche event. In this section we directly test this hypothesis by analyzing a structural order parameter known as ``softness'' ($S$).}  Derived from a mean-field approximation using the Ramakrishnan–Yussouff free energy functional \cite{ramakrishnan1979first} softness captures how a tagged particle interacts with a “frozen” background, reflecting the depth of the local potential well or “cage” in which it resides. If a region of the system is “softer,” it indicates that the particles there experience a shallower caging potential, making them more susceptible to rearrangements. 
\SSG{The softness order parameter is given by} \cite{sharma2022identifying,patel2023dynamic,sahu2024structural}
\begin{align}
    S &= \frac{1}{|\beta(\phi(\Delta r=0)|}
    \label{main11}
\end{align}
Here $\phi$ represents the depth of the local mean caging potential felt by a particle around minima $\Delta r=0$, $\beta=\frac{1}{k_{B}T}$. \SSG{See Appendix \ref{sec:appS} for more details.}

Fig. \ref{fig:av8}(a) describes the \SSG{ensemble-averaged} softness distribution for a range of temperatures \SSG{spanning out of equilibrium glassy regime to equilibrium supercooled liquid regime}. As the temperature increases, the peak of the distribution systematically shifts to higher softness values, mirroring the trends observed in other mechanical and dynamic metrics such as $D^2_{min}$, mobility field $m$, deviatoric ($\Vert\mathbf{e}\Vert$), volumetric ($\Vert \epsilon_{vol}\Vert$), and total strain ($\Vert \mathbf{\epsilon} \Vert$) measures [cf: Figs. \ref{fig:av2}, \ref{fig:av6}, \ref{fig:av3}]. Physically, at higher temperature thermal fluctuations loosens the local cage surrounding each particle, thereby increasing the overall softness of the system. 

Fig. \ref{fig:av8}(b) compares softness distributions before and after a representative thermally mediated avalanche event. Four more independent events are shown in the SI Fig. \ref{fig:av14nw12}. \emph{In all cases, the peak shifts towards higher $S$ value post‐avalanche}. This indicates that the local caging potential wells become shallower after an avalanche event, rendering the system more susceptible to rearrangements. This observation is consistent with our previous findings that the smallest eigenvalue of the Hessian matrix $\lambda_{smallest}$ becomes larger accompanied by a simultaneous increase in dynamic and elastic indicators of instability. \SSG{In other words}, increasing softness makes it easier for the system to access low-energy vibrational modes. 

\section{Relationship among Mechanical, Dynamic, and Structural Metrics}\label{sec:relat}

\begin{figure*}[ht!]
    \centering
    \includegraphics[width=0.32\textwidth]{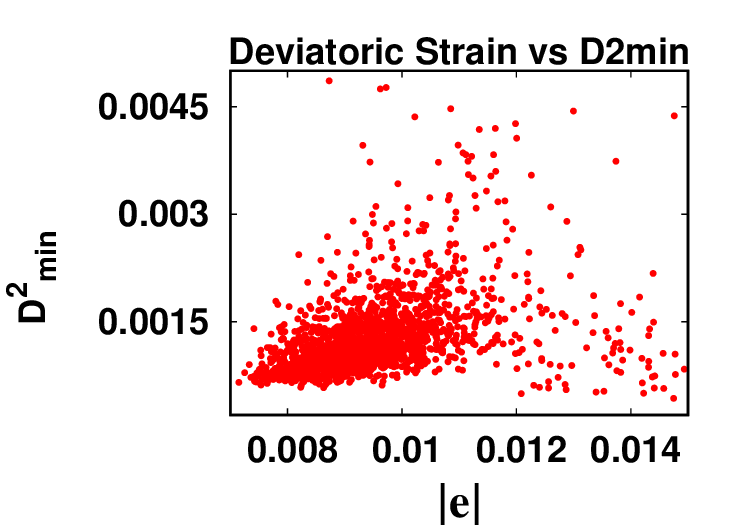}
    \includegraphics[width=0.32\textwidth]{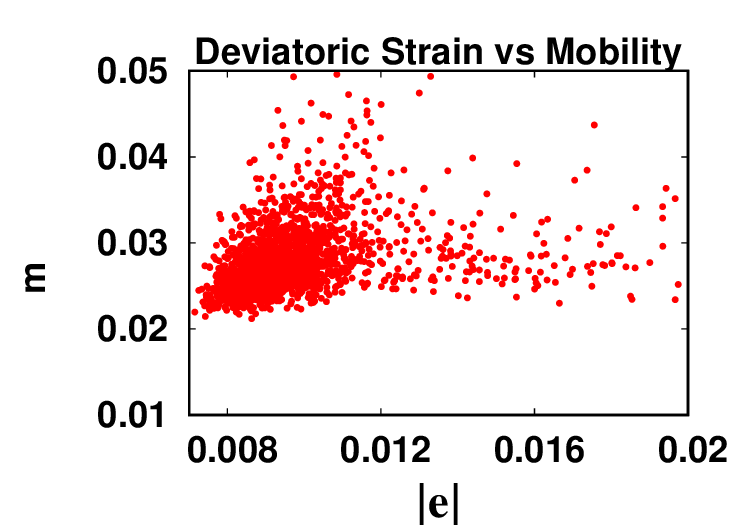}
    \includegraphics[width=0.32\textwidth]{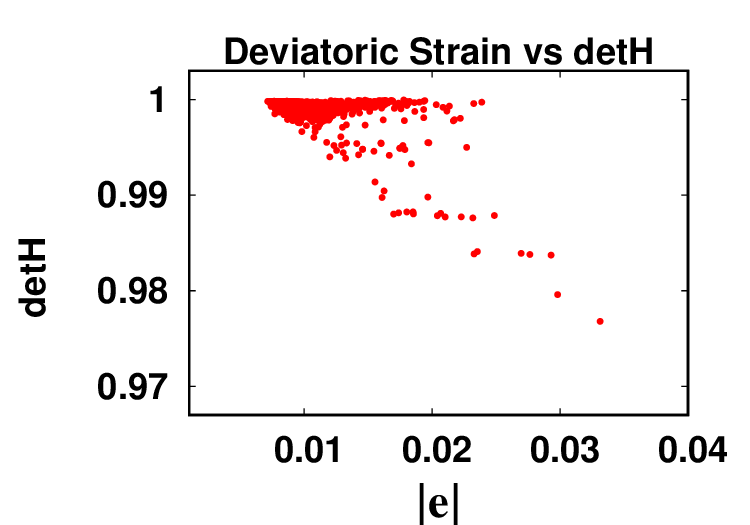}\\
    \medskip
    \includegraphics[width=0.32\textwidth]{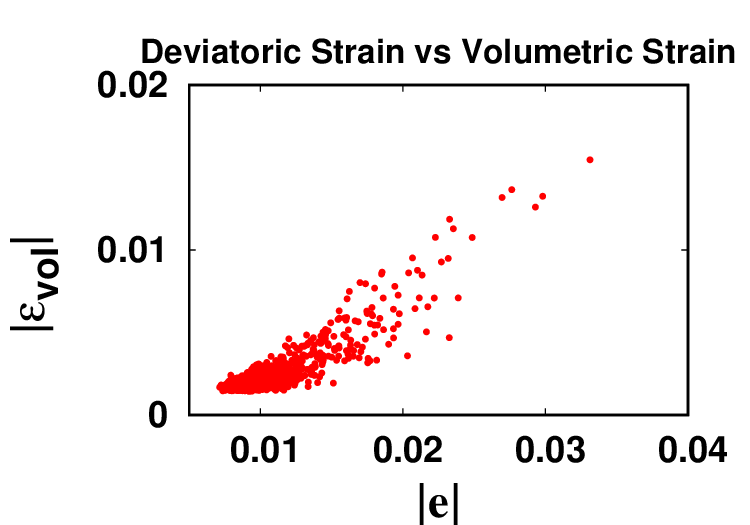}
    \includegraphics[width=0.32\textwidth]{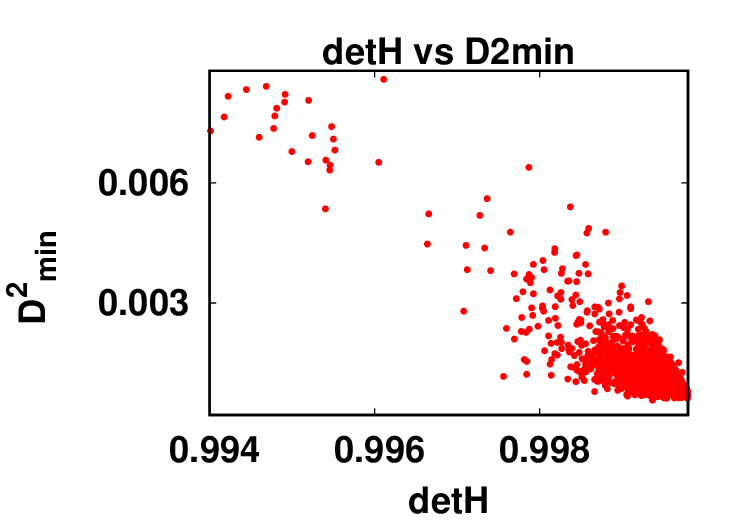}
    \includegraphics[width=0.32\textwidth]{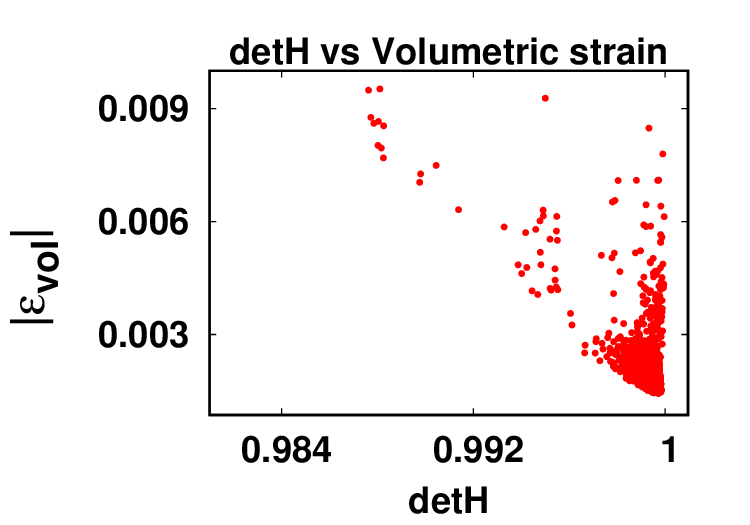}\\
    \medskip
    \includegraphics[width=0.32\textwidth]{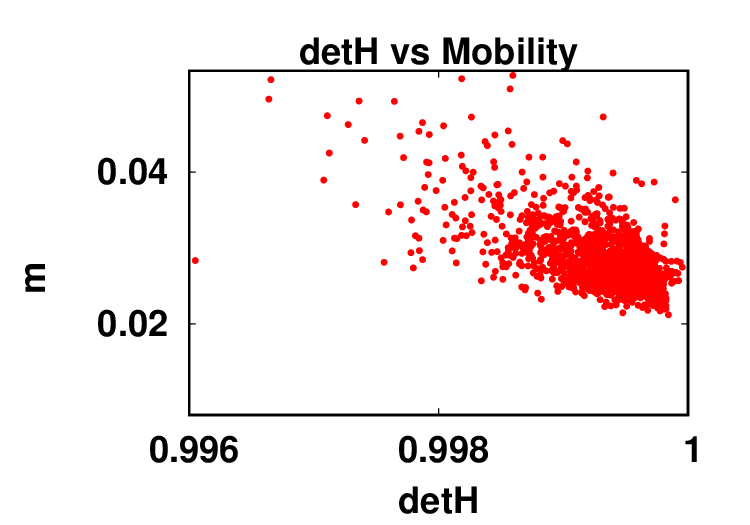}
    \includegraphics[width=0.32\textwidth]{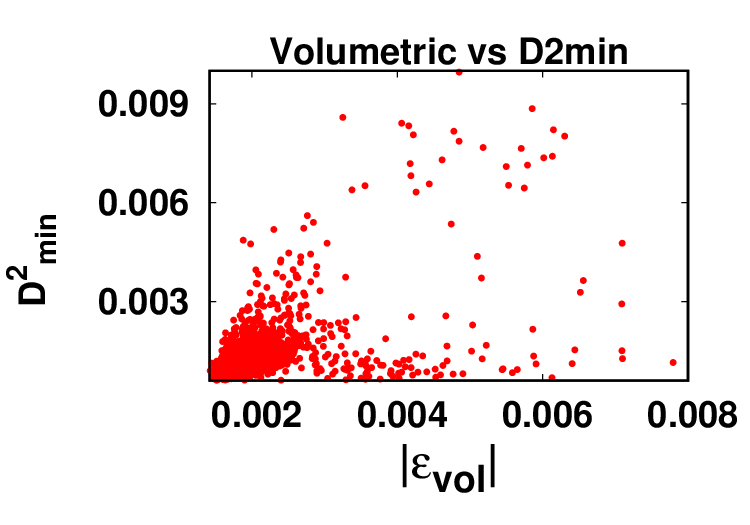}
    \includegraphics[width=0.32\textwidth]{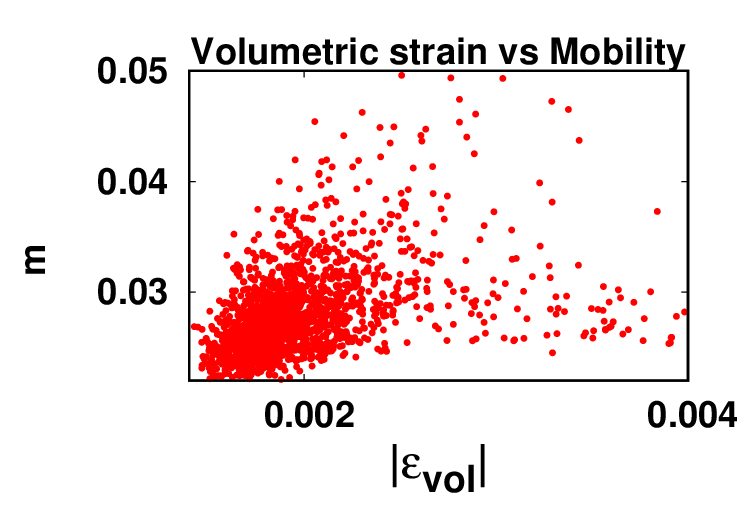} 
    \caption{
    Scatter plots depicting correlations among key characteristics: $D^2_{min}$ (non-affine displacement), Deviatoric strain $\lvert \boldsymbol{e}\rvert$, volumetric strain $\lvert\boldsymbol{\varepsilon}_{\mathrm{vol}}\rvert$, the determinant of the deformation gradient tensor ($\det\mathbf{H}$) and mobility $m$. These plots illustrate how elastic and dynamic properties interrelate in the context of thermally driven plasticity.
    }
    \label{fig:av9}
\end{figure*}

\begin{figure*}[ht!]
    \centering
    \includegraphics[width=0.48\textwidth]{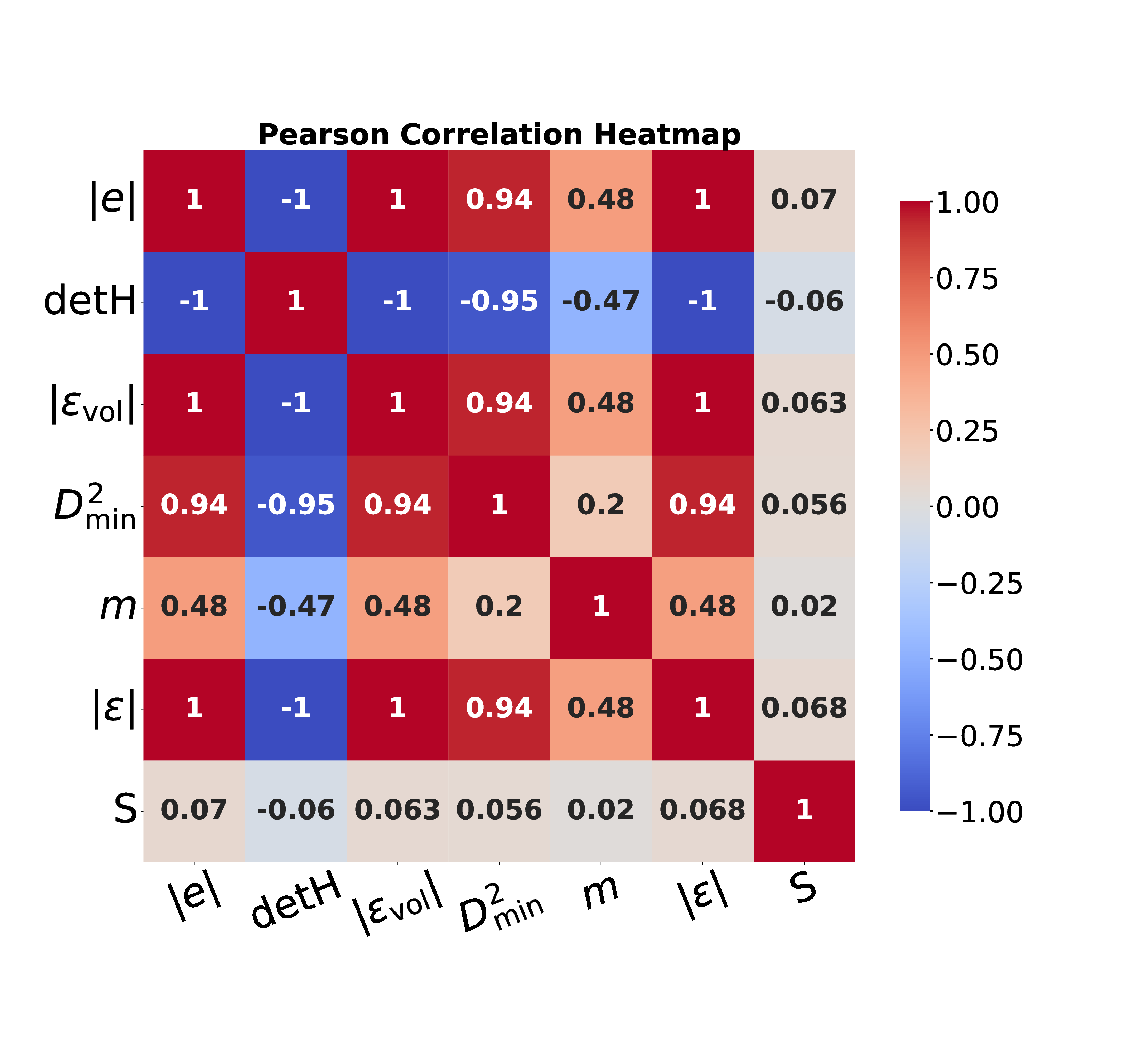}
    \includegraphics[width=0.48\textwidth]{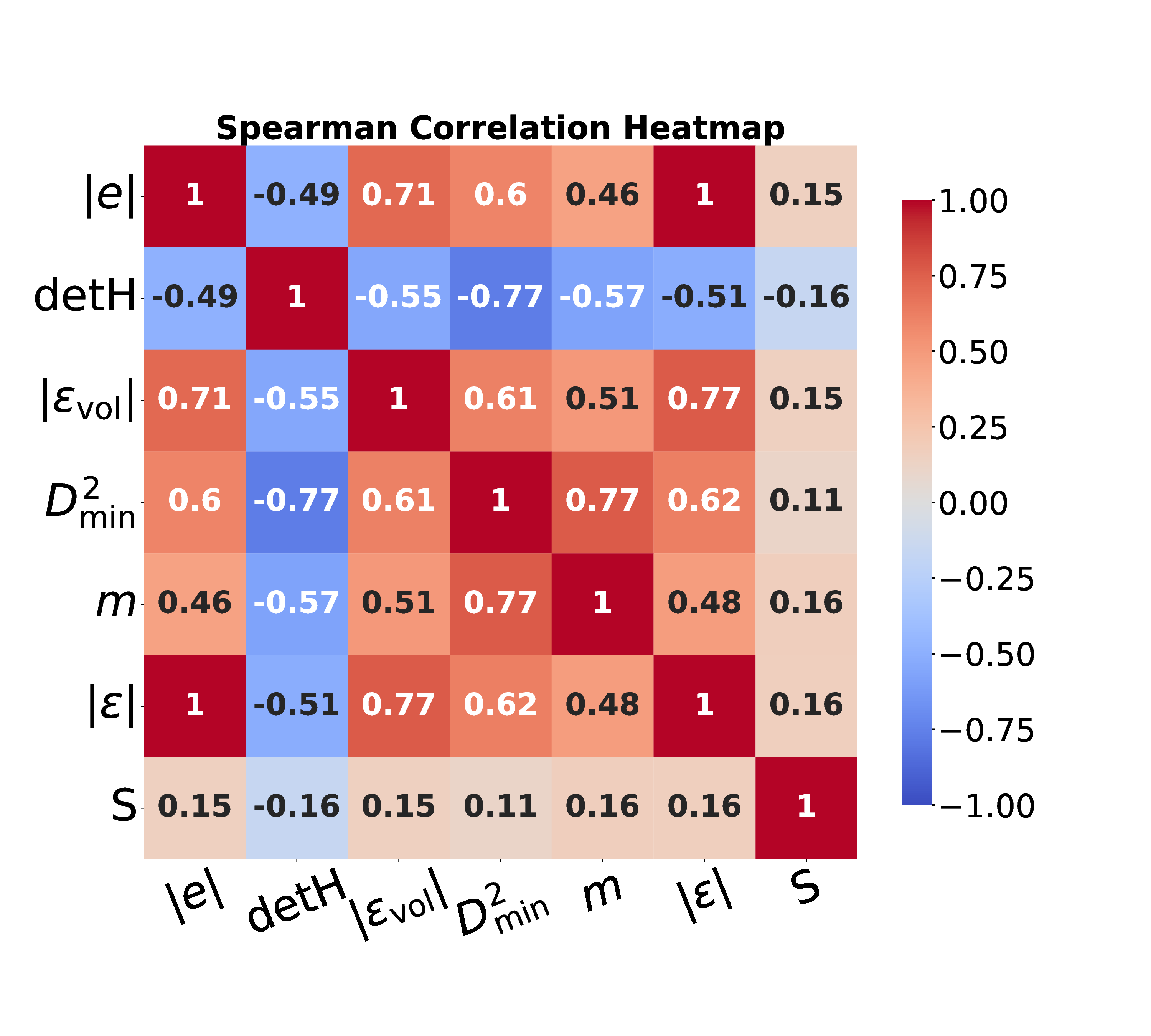}
    \put(-450,160){\textbf{(a)}}
    \put(-200,160){\textbf{(b)}}
    \caption{
    \textbf{(a):} Pearson correlation heatmap between various order parameters: $\lvert \boldsymbol{e}\rvert$ (norm of deviatoric strain), $\det \boldsymbol{H}$ (determinant of the deformation gradient tensor), $\lvert\boldsymbol{\varepsilon}_{\mathrm{vol}}\rvert$ (norm of volumetric strain), $D^{2}_{\min}$ (non-affine displacement), $m$ (dynamic mobility), $\lvert\boldsymbol{\epsilon}\rvert$ (norm of total strain) and  $S$ (softness). 
    \textbf{(b):} Spearman correlation heatmap for the same parameters. Notably, the Spearman correlations for softness are higher than those from the Pearson correlation coefficients. This suggests a nonlinear coupling between structural (softness) and elastic, dynamic responses.
    }
    \label{fig:av10}
\end{figure*}

Having introduced a variety of measures to capture different aspects of thermally mediated avalanche, the next logical step is to investigate \SSG{the relationship among} these observables. 
In this section, we employ two complementary approaches to assess correlations among different metrics. 

\paragraph{Scatter plots:}
In Fig.\ref{fig:av9} we present several scatter plots among various pairs of metrics. They clearly show that elastic and dynamic metrics are correlated, as the data points cluster along distinct trends instead of being randomly dispersed. For example, the relationship between  $\det \mathbf{H}$ and $D^2_{min}$ follows an ordered pattern:
\SSG{lower $\det \mathbf{H} < 1$ indicates higher $D^2_{min}$.} It implies that areas undergoing significant compression also experience more pronounced non-affine displacements. This finding supports the idea that thermally induced plastic events often lead to locally compressed regions. Similarly, the scatter plot comparing deviatoric strain with $\det\mathbf{H}$ exhibits a negative correlation, suggesting that regions with high compression also display elevated deviatoric strain. \SSG{Such correlations reveal} that the elastic and dynamic responses of the system are intricately linked to each other. 

\paragraph{Correlation Heat Maps:}
To capture relationships more \SSG{quantitatively}, we construct Pearson and Spearman correlation matrices \cite{spearman1961proof, sedgwick2012pearson, richard2020predicting}. Pearson’s method quantifies linear correlations, whereas Spearman’s approach also captures non-linear monotonic trends.
The Pearson correlation matrix in Fig. \ref{fig:av10}(a) reveals remarkably strong linear relationships among the primary elastic metrics. For example deviatoric strain $\lvert \boldsymbol{e}\rvert$, the determinant of the deformation gradient tensor $\det\mathbf{H}$ and volumetric strain $\lvert\boldsymbol{\varepsilon}_{\mathrm{vol}}\rvert$ are strongly correlated - with \SSG{magnitude of} correlation coefficients \SSG{$>0.9$}, 
$D^2_{min}$ also shows a high positive correlation with these elastic measures suggesting that regions with significant elastic deformation are prone to large non-affine displacements. Mobility $m$, although correlated with the elastic metrics, exhibits a more moderate relationship.  
For the Spearman correlation matrix [Fig.(\ref{fig:av10}(b)],
\SSG{qualitative} trends remain \SSG{the same}, with pronounced monotonic correlations observed among the elastic parameters as well as between $D^2_{min}$ and mobility $m$. \SSG{However, the magnitude of correlation coefficients are lower.} 
Both the Pearson and Spearman correlation matrices reveal that the norm of total strain $\lvert\boldsymbol{\epsilon}\rvert$ exhibits a strong positive correlation with both deviatoric and volumetric strain components. This indicates that $\lvert\boldsymbol{\epsilon}\rvert$ is an effective metric for capturing the combined magnitude of shear and volumetric deformations. Additionally, \SSG{we note that} $\lvert\boldsymbol{\epsilon}\rvert$ is negatively correlated with $\det\mathbf{H}$.
\SSG{Finally,} the Spearman correlation coefficients between softness $S$ and the other metrics are \SSG{significantly} higher than \SSG{the corresponding} Pearson coefficients. This 
suggests that the relationship between softness and these other order parameters \SSG{might be} nonlinear.

\section{Summary and Conclusions}\label{sec:disc}
\SSG{While the elasto-plastic response of amorphous solids under shear is more well-known, the nature of avalanches and plasticity driven by thermal noise has been less explored.} In this work, we have conducted an extensive \SSG{computer simulation} investigation of thermally mediated plasticity by employing a comprehensive array of interdependent metrics and a 2D model soft glass. \SSG{To understand the nature of the thermally-mediated avalanches events, we analyze the system from five different aspects, namely, (1) \emph{softness} probing \emph{structural} evolution; (2) several \emph{dynamical} observables such as vector displacement field, scalar mobility field as well as $D^2_{min}$ providing non-Affine contribution to displacement field, and distance travelled along the underlying potential energy landscape; (3) \emph{mechanical} response by computing local strain tensor and its volumetric and deviatoric components as well as the stress tensor; (4) \emph{vibrational} properties by probing the lowest eigenvalues and (5) \emph{topological} change by measuring Betti number ($\beta_1$). We also measure correlations among different observables to understand the relationship among them.}

\SSG{For each property we present two kinds of analyses: (i) temperature evolution of ensemble-averaged distributions of observables showing how the system properties change from near equilibrium at $\sim T_g$ to deep inside glassy regime and (ii) evolution of observables across avalanche events describing the change in system properties due to an thermally mediated plastic event.}   

\SSG{We highlight the three main results of the study. 
First, based on linear elasticity, we develop a formalism to compute non-Affine displacement and local strain measures for a system \emph{with no externally applied shear}, {\it i.e.} only driven by thermal fluctuations. 
Second, we show that as a result of an avalanche event there are clear changes in (a) vibrational spectrum, (b) shear stress, (c) topology of particle network, (d) local strain and non-Affine displacements and (e) softness. Overall they indicate that structural reorganization during avalanche leads to \emph{softening} of the material, even for thermally driven plasticity. 
Third, we identify both similarities and differences with the more well-known shear deformation response. 
Our analysis of non-Affine deformation through $D^2_{min}$ and both deviatoric and volumetric contribution to local elastic strains reveals the plastic events are localized in space with characteristics reminiscent of Eshelby-like strain field typical of shear induced avalanches. However, our results show that unlike the shear induced avalanches, local volume change can not be ignored in thermal mediated plasticity, challenging the conventional view. To provide a comparison between shear induced and thermally mediated plasticity, we generalize the idea of shear transformation zone to propose the emergence of Generalized Strain Transformation Zones (GSTZ) where both shear and volumetric effects combine to drive local rearrangements, giving a broader spectrum of local rearrangement events. Visualization of vector (non‐Affine) displacement field reveals a rich pattern of inward translations, swirling motion and rotation around GSTZ that is reminiscent of Eshelby-like response but shows richer behaviour, presumably due to the contributions from volumetric deformations.  }

In summary, our multi-faceted analysis provides a comprehensive picture of thermally mediated avalanche events \SSG{which is driven purely by thermal noise in complete absence of any external force such as shear.} This work not only deepens our understanding of the mechanisms underpinning thermal plasticity in amorphous materials but also establishes a robust framework for understanding the relationship among diverse properties across mechanical, dynamical, and structural domains.

\begin{acknowledgments}
A.J. acknowledges junior and senior research fellowships from the University Grants Commission, India. S.S. acknowledges the Indian
Institute of Technology Roorkee for providing support via the Faculty Initiation Grant (No. PHY/FIG/100804). 
\end{acknowledgments}

\appendix

\section{\SSG{Extracting elastic measures at zero applied strain }} \label{sec:appStrain}
\SSG{In this section, we discuss the definitions of different strains and other quantifiers of elasticity and our method of extracting them from configurations at zero applied strain. For simplicity, we consider only the Euclidean space.}

\subsubsection{\SSG{Optimal deformation gradient tensor $\mathbf{H}$ by minimizing local $D^2_{min}$:}}\label{sec:appStrainH}
To derive \SSG{an expression for the non-affine displacement} $D^2_{min}$ from undeformed configurations, we relate the separations between particles at two different times t and $t+\Delta t$.
Let ${\mathbf{X}(t)}$ denote the position of a particle at time $t$ and ${\mathbf{x}(t+\Delta t)}$ denote its position at a later time $t+\Delta t$. The separation vectors between particles $i$ and $j$ at a reference time $t$ and a later time $t+\Delta t$ are, respectively,
\begin{align}
 \Delta \mathbf{X}_{ij}(t)&=\mathbf{X}_j(t)-\mathbf{X}_i(t) \nonumber\\
 \Delta\mathbf{x}_{ij}(t+\Delta t)&=\mathbf{x}_j(t+\Delta t)-\mathbf{x}_i(t+\Delta t)
 \label{d3}
\end{align}
\SSG{Now consider an Affine mapping from time $t$ to $t+\Delta t$. We relate the separation vectors at the two instants} with the help of the second rank tensor $\mathbf{H}$ as
\begin{align}
\mbox{Affine:}\quad \Delta\mathbf{x}_{ij}(t+\Delta t)=\mathbf{H}\cdot\Delta \mathbf{X}_{ij}(t) 
     \label{d5}
\end{align}
\SSG{where $\mathbf{H}$ is called the deformation gradient tensor. Upto linear order, $\mathbf{H}$ is given by}
\begin{align}
 \mathbf{H}&=\mathbf{I}+\mathbf{\epsilon} \nonumber\\
 H_{\alpha\beta} &= \delta_{\alpha\beta} +\epsilon_{\alpha\beta}
 \label{d1}
\end{align}
\SSG{where $\mathbf{\epsilon}$ is the desired strain tensor and $\alpha, \beta$ denote Cartesian components.}
\SSG{In actual molecular dynamics,} atoms do not move according to affine transformation. So, there will be \SSG{mismatch or ``error'' between} actual displacement and one-to-one Affine mapping. The square of this residual error is called $D^2_{min}$, \SSG{which is} a measure of ``non-Affineness''. Specifically, the error can be calculated between a pair of particles i and j as
\begin{align}
    D^2_{ij}&=(\Delta\mathbf{x}_{ij}-\mathbf{H}_i\cdot\Delta\mathbf{X}_{ij})^T \cdot (\Delta\mathbf{x}_{ij}-\mathbf{H}_i\cdot\Delta\mathbf{X}_{ij})\nonumber\\
    &=\lbrack(\Delta\mathbf{x}_{ij})^T\cdot(\Delta\mathbf{x}_{ij}) \nonumber\\
    &- (\Delta\mathbf{X}_{ij})^T\cdot(\mathbf{H}_i)^T\cdot(\Delta\mathbf{x}_{ij})\nonumber\\
    &- (\Delta\mathbf{x}_{ij})^T\cdot(\mathbf{H}_i)\cdot(\Delta\mathbf{X}_{ij})\nonumber\\
    &+(\Delta\mathbf{X}_{ij})^T\cdot(\mathbf{H}_i)^T\cdot(\mathbf{H}_i).(\Delta\mathbf{X}_{ij})\rbrack
     \label{d6}
\end{align}
\SSG{Here $i,j$ denote particle indices, {\it i.e.} $\mathbf{H}_i$ denotes the deformation gradient tensor defined at particle $i$ which maps a vector from the reference configuration at $t$ to the current configuration at $t+\Delta t$.}
\SSG{Using the identity for any two vectors $\mathbf{u,v}$ and a second rank tensor $\mathbf{T}$,}
\begin{align}
\mathbf{u}\cdot\mathbf{T}\cdot\mathbf{v}=\mathbf{v}\cdot\mathbf{T}^T\cdot\mathbf{u}
\label{d8}
\end{align}
equation \ref{d6} becomes
\begin{align}
D^2_{ij}&=\lbrack (\Delta\mathbf{x}_{ij})^T\cdot(\Delta\mathbf{x}_{ij})\nonumber\\
&-2(\Delta\mathbf{x}_{ij})\cdot\mathbf{H}_i\cdot(\Delta\mathbf{X}_{ij})^T\nonumber\\
&+(\Delta\mathbf{X}_{ij})^T\cdot(\mathbf{H}_i)^T\cdot(\mathbf{H}_i)\cdot(\Delta\mathbf{X}_{ij}) \rbrack
\label{10}
\end{align}

For a given particle $i$ we sum over contributions from all other particles $j$, \SSG{and obtain}
\begin{align}
D^2_{i}&= \sum\limits_{j=1}^N D^2_{ij}
\label{11}
\end{align}
\SSG{Next, to determine the optimal value of $\mathbf{H}_i$, we minimize the ``loss function'' $D^2_{i}$ with respect to the deformation gradient tensor $\mathbf{H}_i$. To this end, we compute} the derivative $\frac{\partial D^2_i}{\mathbf{\partial H}}$.
\begin{align}
\frac{\partial D^2_i}{\mathbf{\partial H}_i}&=\frac{\partial}{ \mathbf{\partial H}_i } \sum_{j=1}^N \,\,\lbrack(\Delta\mathbf{x}_{ij})^T\cdot(\Delta\mathbf{x}_{ij})\nonumber\\
&-2(\Delta\mathbf{x}_{ij})\cdot(\mathbf{H}_i)\cdot(\Delta\mathbf{X}_{ij})^T\nonumber\\
&+ (\Delta\mathbf{X}_{ij})^T\cdot(\mathbf{H}_i)^T\cdot(\mathbf{H}_i)\cdot(\Delta\mathbf{X}_{ij}) \rbrack\nonumber\\
&=\sum_{j=1}^N \,\,\frac{\partial}{\partial \mathbf{H}_i}\lbrack-2(\Delta\mathbf{x}_{ij})\cdot(\mathbf{H}_i)\cdot(\Delta\mathbf{X}_{ij})^T \nonumber\\
&+(\Delta\mathbf{X}_{ij})^T\cdot(\mathbf{H}_i)^T\cdot(\mathbf{H}_i)\cdot(\Delta\mathbf{X}_{ij}) \rbrack\nonumber\\
&=\sum_{j=1}^N \,\,\,\frac{\partial}{\partial H_i^{\delta \psi}}\lbrack -2(\Delta{x}_{ij})^{\alpha}H_i^{\alpha \beta}(\Delta X_{ij})^{\beta} \nonumber\\
&+(\Delta X_{ij})^{\alpha}H_i^{\beta \alpha}H_i^{\beta \gamma}(\Delta X_{ij})^{\gamma} \rbrack\nonumber\\
&=\sum_{j=1}^N \,\,\,\lbrack -2(\Delta x_{ij}^\delta \Delta X_{ij}^\psi)+ \Delta X_{ij}^\psi H_i^{\delta \gamma} \Delta X_{ij}^\gamma \nonumber\\
&+\Delta X_{ij}^\alpha H_i^{\delta \alpha} \Delta X_{ij}^\psi \rbrack
\label{12}
\end{align}
\SSG{Here Greek letters denote Cartesian components.} 
Changing dummy indices we obtain
\begin{align}
\frac{\partial D^2_i}{\mathbf{\partial H}_i}
&=\sum_{j=1}^N \,\,\,\lbrack -2(\Delta x_{ij}^\delta \Delta X_{ij}^\psi) +2 \Delta X_{ij}^\psi H_i^{\delta \alpha}\Delta X_{ij}^\alpha \rbrack
\label{13}
\end{align}
Now we set $\frac{\partial D^2_i}{\mathbf{\partial H}}=0$ and extract the optimal deformation gradient tensor that minimizes the loss function as,
\begin{align}
H^{\delta \alpha}_i &=\lbrack \sum_{j=1}^N \,\,\, (\Delta x_{ij}^\delta \Delta X_{ij}^\psi) \rbrack. \lbrack \sum_{j=1}^N \,\,\,(\Delta X_{ij}^\psi \Delta X_{ij}^\alpha) \rbrack^{-1}\nonumber\\
&=A_i^{\delta \psi} (B_i^{\psi \alpha})^{-1}
\label{d2xy_sup3a}
\end{align}
where $A_i^{\delta \psi} \equiv \sum_{j=1}^N \,\,\,(\Delta x_{ij}^\delta \Delta X_{ij}^\psi) $ and $B_i^{\psi \alpha} \equiv \sum_{j=1}^N \,\,\,(\Delta X_{ij}^\psi \Delta X_{ij}^\alpha)$, \SSG{and Einstein sum convention is implied.}

So, square of loss function is given by following equation 
\begin{align}
D^2_{i}&=\frac{\sum\limits_{j \in N_i} (\Delta\mathbf{x}_{ij}-\mathbf{H}_i.\Delta\mathbf{X}_{ij})^T.(\Delta\mathbf{x}_{ij}-\mathbf{H}_i.\Delta\mathbf{X}_{ij})}{N_i}
\label{d15a}
\end{align}
Where optimal deformation gradient tensor $\mathbf{H}_i$ is obtained from equation \ref{d2xy_sup3a}.
In equation \ref{d15a} we have normalized the loss function by dividing with the second nearest neighbour distance obtained from the pair correlation function to incorporate a sufficient number of neighboring particles for affine mapping. 

Equation \ref{d15a} is a scalar metric that provides information about the magnitude of non-affine displacements, without revealing their directional characteristics. To address this limitation, we further decompose the non-affine displacement into its Cartesian components. Specifically, the non-affine displacement vector can be defined using the product of local deformation gradient tensor $\mathbf{H}$ and separation vector $\Delta\mathbf{X}_{ij}$ as
\begin{align}
   \Delta x^{\text{NA}}_{i} &= \frac{\sum\limits_{j \in N_i} \Bigl[ \Delta x_{ij} - \Bigl( H_{xx}\,\Delta X_{ij} + H_{xy}\,\Delta Y_{ij} \Bigr) \Bigr]}{N_i} \nonumber\\
   \Delta y^{\text{NA}}_{i} &= \frac{\sum\limits_{j \in N_i} \Bigl[ \Delta y_{ij}- \Bigl( H_{yx}\,\Delta X_{ij} + H_{yy}\,\Delta Y_{ij} \Bigr) \Bigr]}{N_i}
\label{d2xy_sup}
\end{align}
The elements of the second rank tensor $H_{xx}$, $H_{xy}$, $H_{yx}$, and $H_{yy}$ can be fetched from equation \ref{d2xy_sup3a}.
From equation \ref{d2xy_sup} we can incorporate both x and y components of non-affines in one single vector equation for $i^{th}$ particle as  as
\begin{align}
    \mathbf{D}_i &=\mathbf{\Delta x}^{\text{NA}}_i + \mathbf{\Delta y}^{\text{NA}}_i
 \label{d2xy_sp2}   
\end{align}

\subsubsection{\SSG{Local elastic strain tensor $\mathbf{\epsilon}$ from optimal deformation gradient tensor $\mathbf{H}$}}\label{sec:appStraine}
\SSG{Combining Eqns. \ref{d1} and \ref{d2xy_sup3a}, we obtain the local strain tensor $\mathbf{\epsilon}_i$ from the optimal $\mathbf{H}_i$ as,}
\begin{align}
\epsilon_i^{\delta \alpha}&=\sum_{\psi=1}^d \,\,\,A_i^{\delta \psi}(B_i^{ \alpha\psi})^{-1}-\delta^{\delta \alpha}
\label{17}
\end{align}
\SSG{where $d$ is the spatial dimension. Note that this strain is locally induced by thermal fluctuation. In general, it is composed of both deviatoric or shear ($e_{\alpha\beta}$) and volumetric local strain:}
\begin{align}
\epsilon_{\alpha\beta} &= e_{\alpha\beta} + \frac{1}{d}\,\epsilon_{\gamma\gamma}\,\delta_{\alpha\beta} \nonumber\\
\epsilon_{\alpha\beta} &= e_{\alpha\beta} + \frac{1}{2}\,\epsilon_{\gamma\gamma}\,\delta_{\alpha\beta} \;\ \mbox{(in 2d)} \nonumber\\
\Rightarrow e_{\alpha\beta} &= \epsilon_{\alpha\beta} - \frac{1}{2}\,\epsilon_{\gamma\gamma}\,\delta_{\alpha\beta}
\label{18}
\end{align}

\SSG{Using Eqn. \ref{18}, we explicitly write down the Cartesian components of the local deviatoric strain in 2D Euclidean space as,}
\begin{subequations}\label{20}
\begin{align}
e_{11}&=\epsilon_{11}-\frac{1}{2}(\epsilon_{11} +\epsilon_{22}) \nonumber\\ &=\frac{1}{2}(\epsilon_{11}-\epsilon_{22})\\
e_{22}&=\epsilon_{22}-\frac{1}{2}(\epsilon_{11} +\epsilon_{22}) \nonumber\\ &=\frac{1}{2}(\epsilon_{22}-\epsilon_{11})\\
e_{12}&=\epsilon_{12}\\
e_{21}&=\epsilon_{21}
\end{align}
\end{subequations}
\SSG{The norm of the local deviatoric strain is given by}
\begin{align}
\Vert \mathbf{e}\Vert &\equiv \sqrt{\mathbf{e:e}} \equiv \sqrt{e_{11}^2+e_{22}^2+e_{12}^2+e_{21}^2} \nonumber\\
&=\sqrt{\frac{1}{2} (\epsilon_{11}-\epsilon_{22})^2 +\epsilon_{12}^2 +\epsilon_{21}^2}
\label{22}
\end{align}
\SSG{Similarly, the norm of the total (local) strain $\mathbf{\epsilon}$ is defined as }
\begin{align}
\Vert \mathbf{\epsilon} \Vert&=\sqrt{\mathbf{\epsilon}:\mathbf{\epsilon}}\nonumber\\
  &=\sqrt{\epsilon_{11}^2+\epsilon_{22}^2 +\epsilon_{12}^2 +\epsilon_{21}^2}
\label{23ad1}
\end{align}
\SSG{where the components $\epsilon_{11}, \epsilon_{22}, \epsilon_{12}, \epsilon_{21},$ are computed for each particle $i$ using Eqn. \ref{17}.}
\AV{The volumetric strain is defined as}
\begin{equation}
\epsilon_{\mathrm{vol}} \;=\; \tfrac{1}{2}\,(\epsilon_{11}+\epsilon_{22})
\label{23_vol}
\end{equation}
\AV{For non-negative scalar quantities (e.g., distributions), we use the magnitude of the volumetric strain, i.e. the absolute value of the strain-tensor trace:}
\begin{align}
\Vert\epsilon_{vol}\Vert =\frac{1}{2}\vert\;(\epsilon_{11}+\epsilon_{22})\;\vert
\label{23}
\end{align}

\subsubsection{\SSG{Indicator $\det \mathbf{H}$ to quantify nature of avalanche event}}\label{sec:appStraindetH}
The trace of strain tensor gives relative change of local area if strain calculated locally as
\begin{align}
\epsilon_{11}+\epsilon_{22} =\frac{\Delta A}{A}
\label{24}
\end{align}
Here \SSG{the area} $A$ is defined locally. \SSG{On one hand, owing to a thermally mediated avalanche event, if the local area remains same ($\Delta A = 0$), it indicates an area preserving plastic event {\it i.e.} local strain at the core of the plastic rearrangement is dominated by deviatoric shear. On the other hand, if the local area decreases ($\Delta A < 0$) or increases ($\Delta A > 0$), then the volumetric strain is important. Thus, the dimensionless quantity $\frac{\Delta A}{A}$ is an indicator of the nature of the thermal plastic event.}

\SSG{We now show that the} same information can be obtained more elegantly in terms of the local deformation gradient tensor. \SSG{To this end,} we relate a local area element before and after an avalanche event. 

Consider a small elemental area in the reference frame $\mathbf{X}$ at time $t$, given by
\begin{align}
d\mathbf{A}&= d\mathbf{X1} \times d\mathbf{X2} = dX1\,dX2\,\,\hat{k}
\label{add1}
\end{align}
Here $d\mathbf{X1}$ and $d\mathbf{X2}$ are differential vectors in the reference configuration and $\hat{k}$ is a unit vector perpendicular to $X1-X2$ plane. Similarly at time $t+\Delta t$ in the current frame $\mathbf{x}$ the deformed area is 
\begin{align}
d\mathbf{A'}= d\mathbf{x1} \times d\mathbf{x2}
\label{add2}
\end{align}
where $d\mathbf{x1}$ and $d\mathbf{x2}$ are differential vectors in the deformed configuration.
\SSG{The coordinate transformation rules between the two frames can be written as}
\begin{align}
dx1&=\hat{i}\, \frac{\partial x1}{\partial X1}\, dX1 \, + \hat{j}\, \frac{\partial x1}{\partial X2}\,dX2 \nonumber\\
dx2&=\hat{i}\, \frac{\partial x2}{\partial X1}\, dX1 \, + \hat{j}\, \frac{\partial x2}{\partial X2}\,dX2
\label{add3}
\end{align}
Note that $\hat{i},\hat{j},\hat{k}$ are Cartesian unit vectors in frame $\mathbf{X}$.
Substituting Eqn. \ref{add3} in Eqn. \ref{add2} \SSG{and using Eqn. \ref{add1},} we get
\begin{align}
 d\mathbf{A'}
&= \begin{vmatrix}
\frac{\partial x_1}{\partial X_1} & \frac{\partial x_1}{\partial X_2} \\
\frac{\partial x_2}{\partial X_1} & \frac{\partial x_2}{\partial X_2}
\end{vmatrix} d\mathbf{A} = \left(\det \mathbf{H} \right) d\mathbf{A}
 \label{add5}
\end{align}
where $\mathbf{H} \equiv\frac{\partial \mathbf{x}}{\partial \mathbf{X}}$ is the local deformation gradient tensor. Hence, for small areas, we can relate these the area $\mathbf{A'}$ at current time to the area $\mathbf{A}$ at reference time by 
\begin{align}
\mathbf{A'} = \left| \det \mathbf{H} \right|\, \mathbf{A}
\label{add9}
\end{align}
Taking the magnitudes we obtain,
\begin{align}
A'&=(\det \mathbf{H})\, A \nonumber\\
\mbox{or,}\;\;\frac{A'}{A} &=(\det\mathbf{H})
\label{25}
\end{align}
So, $|\det\mathbf{H}| < 1$ implies contraction of local region (area),$|\det\mathbf{H}| > 1$ implies expansion of local region, and $|\det\mathbf{H}| = 1$ implies no change of local region.

\section{Topological computation details}\label{sec:appTopo}
\SSG{In this section, we provide details of the computation of the Betti number $\beta_1$. The Betti number $\beta_1$ is the rank of the first homology group and counts the number of independent one-dimensional holes (topological loops) in a point cloud. Physically, $\beta_1$ quantifies the mesoscale connectivity, i.e., the structure formed by rings or loops. Changes in $\beta_1$, therefore, report how the mesoscale network of cavities and links rearranges during an avalanche (loops closing, merging, or appearing). Those rearrangements affect how mechanical and dynamical responses propagate locally.}
We extracted particle positions at each \SSG{configuration of a MD}  trajectory \SSG{showing} avalanche event. Each configuration is represented as a point cloud in a 2D space. We computed the Euclidean distance between every pair of particles to form a distance matrix. \SSG{We then generate a Vietoris–Rips filtration of the Ripser complex over a range of radii (or distances) using the GUDHI library} \citesupp{edelsbrunner2010computational, maria2014gudhi}. A 1D loop ($\beta_1$ feature) is said to be “born” at the smallest distance scale $d_1$ at which it appears in the filtration and “dies” at a distance $d_2$ where it merges or closes off. For each frame, every loop that is born at a distance $d_1 \leq \alpha$ counted once, even if it subsequently dies at a distance $d_2 < \alpha$. In other words, as soon as a loop appears in the range $[0, \alpha]$, it contributes to the $\beta_1$ count for that configuration. To focus on physically relevant loops, we set a filtration threshold $\alpha =(\sigma_{max1}+\sigma_{max2})\times 3.0$. Here $\sigma_{max1}$ and $\sigma_{max2}$ are the two largest particle diameters in the system. Numerically, this yielded a value of approximately $8.19 \langle \sigma \rangle$. By comparing this cutoff with the radial distribution function $g(r)$ at the same density, we observe that $g(r)$ peaks well below $r \approx 8 \langle \sigma \rangle$  and levels off near unity around that scale. Consequently, setting the threshold to $8.19 \langle \sigma \rangle$ captures all significant loops forming within the first few coordination shells where topologically relevant features arise while excluding unphysically large distances where $g(r)$ essentially flattens out. \SSG{In Figs. \ref{fig:av1} and \ref{fig:av14nw5}, $\beta_1$ values denote the the count of 1D holes up to the chosen threshold.}

\begin{figure}[htbp!]
    \centering
    \begin{overpic}[width=0.36\textwidth]{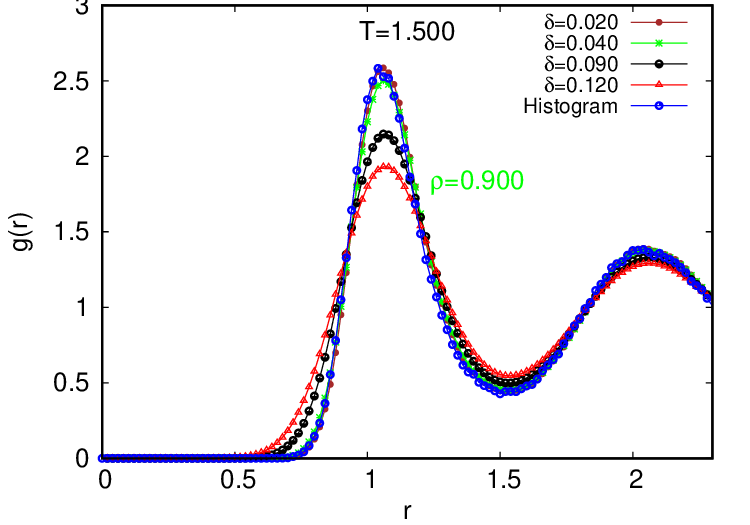}
    \put(30,50){\large\bfseries (a)}
    \end{overpic}
    \begin{overpic}[width=0.36\textwidth]{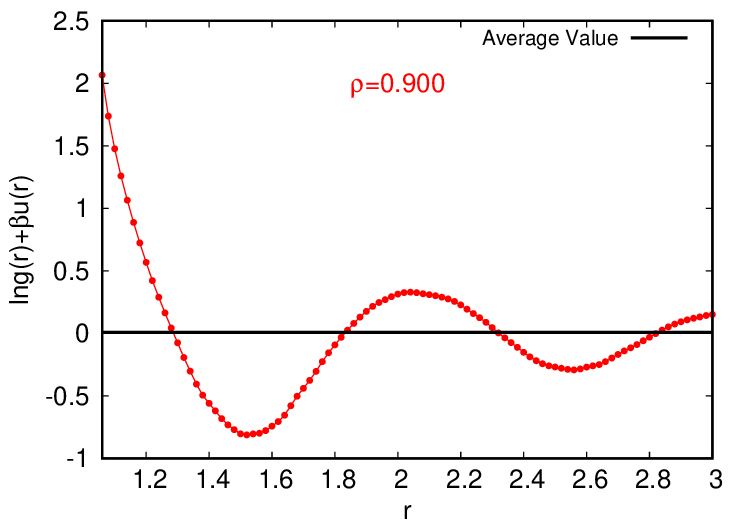}
    \put(32,55){\large\bfseries (b)}
    \end{overpic}
\caption{\textbf{(a)} Pair distribution function \(g(r)\) computed using Equation \ref{eq:gr} at the particle level for various variance values \(\delta\). As \(\delta\) decreases, the particle-level \(g(r)\) converges toward the  \(g(r)\), with optimal agreement at \(\delta = 0.020\). Here, \(r_{\text{min}}\) denotes the distance at which \(g(r) = 0\) (i.e., \(r = 0\)), and \(r_{\text{max}}\) is the distance where \(g(r)\) approaches 1. \textbf{(b)} \(\ln g(r) + \beta u(r)\) as a function of \(r\), where \(u(r)\) is the pair potential. The first peak at \(r = 1.00\) corresponds to the average particle diameter \(\langle \sigma \rangle = 1.0\); hence, the \(r\)-axis begins at \(r = 1.00\).}
\label{fig:av3sup}
\end{figure}

\section{Details of computing softness order parameter}\label{sec:appS}
\SSG{In this section we describe details about the procedure of computation of the ``softness'' order parameter $S$ introduced in Refs. \citesupp{sahu2024structural,patel2023dynamic}. For more background about the underlying physics, we refer to these works.}

\paragraph{Definition of softness:} \SSG{First,} we define pair correlation function for \SSG{individual} particle $i$ in 2D as
\begin{align}
    g_i(r)&=\frac{1}{2\pi r\rho}\sum_{j}\, \frac{1}{\sqrt{2\pi\delta^2}} e^{-\frac{(r-r_{ij})^2}{2\delta^2}}.
    \label{eq:gr}
\end{align} 
Here $\delta$ is variance or width parameter of Gaussian function, $\rho$ is the bulk density of the material, \SSG{and $r_{ij}$ is the distance between particles $i,j$ in a given configuration. From $g(r)$ we compute the direct correlation function $c(r)$ {\it via} the Ornstein- Zernike relation,}
\begin{align}
    g(r)-1 &=c(r) +\rho \int dr' c(|\Vec{r}-\Vec{r'}|) [g(r')-1]
\end{align}
and closure relation of the hypernetted chain approximation
\begin{align}
    c(r) &=g(r)-1- \ln[g(r)]-\beta u(r)
\end{align} 
Following Ref. \citesupp{patel2023dynamic} for polydisperse and repulsive systems, we approximate $c(r)$ as
\begin{align}
    c(r) \approx g(r)-1
\end{align}
Next we define the depth of the caging potential for the polydisperse system as
\begin{align}
    \beta(\phi(\Delta r=0)) &= -2 \pi \rho \int_{r_{min}}^{r_{max}} dr \, r \, c(r) \, g(r)
    \label{eq:dep}
\end{align}
Here $\Delta r$ is the displacement of particle from its minima of caging potential \SSG{and $\beta=\frac{1}{k_BT}$.} \SSG{Note that in the above equations $c(r)$ is also computed for individual particles.} Now softness $S$ is given by \citesupp{sharma2022identifying, patel2023dynamic, sahu2024structural},
\begin{align}
    S &= \frac{1}{\beta |(\phi(\Delta r=0))|}
\end{align}

\paragraph{Choice of parameters $\delta$, $r_{min}, r_{max}$:}
\SSG{In Fig. \ref{fig:av3sup} we show details of choice of the variance $\delta$ in Eqn. \ref{eq:gr} and the values of limits $r_{min}, r_{max}$ in Eqn. \ref{eq:dep}. }

\bibliographystyle{unsrtnat}
\bibliography{main}

\clearpage
\newpage

\renewcommand{\thesection}{S\Roman{section}}
\renewcommand{\thefigure}{S\arabic{figure}}
\renewcommand{\theequation}{S\arabic{equation}}
\setcounter{section}{0}
\setcounter{figure}{0}
\setcounter{equation}{0}

\newpage
\onecolumngrid
\section*{Supplementary Information: Analysis of four additional avalanche events}

\begin{figure*}[b!]
 \centering
    \includegraphics[width=0.45\textwidth]{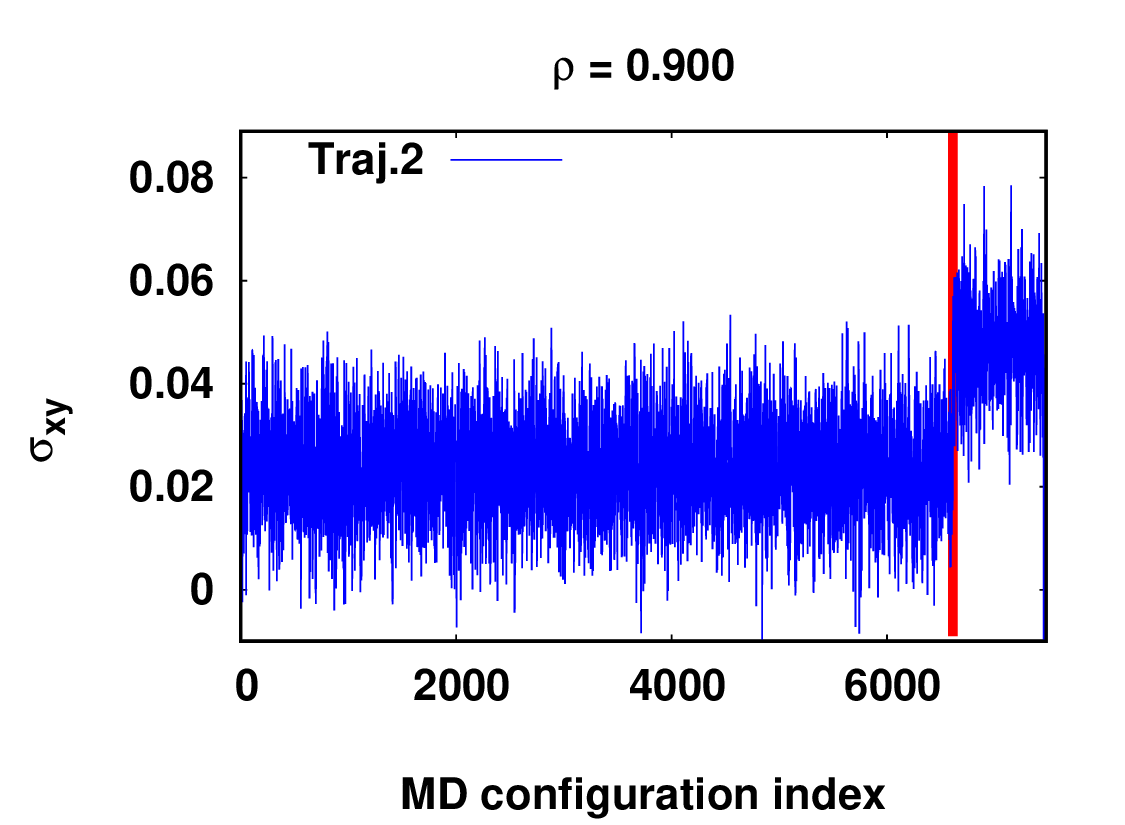}
    \includegraphics[width=0.45\textwidth]{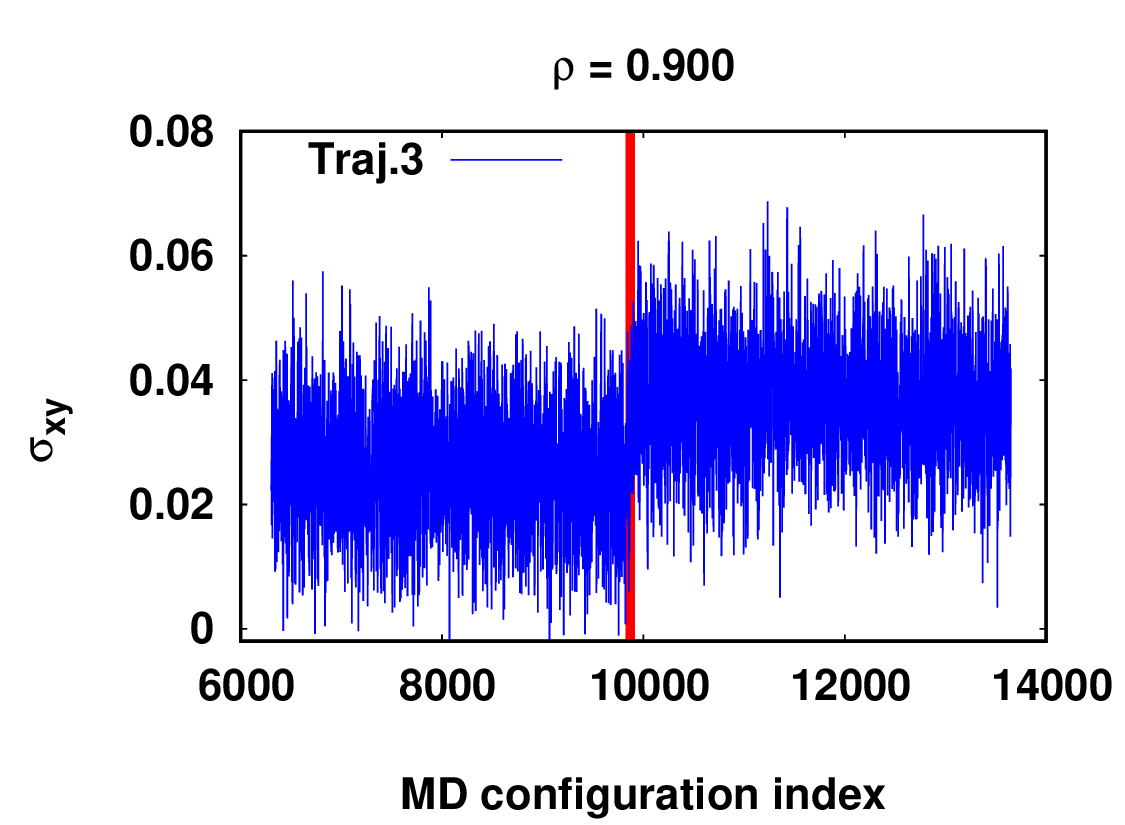}\\
    \includegraphics[width=0.45\textwidth]{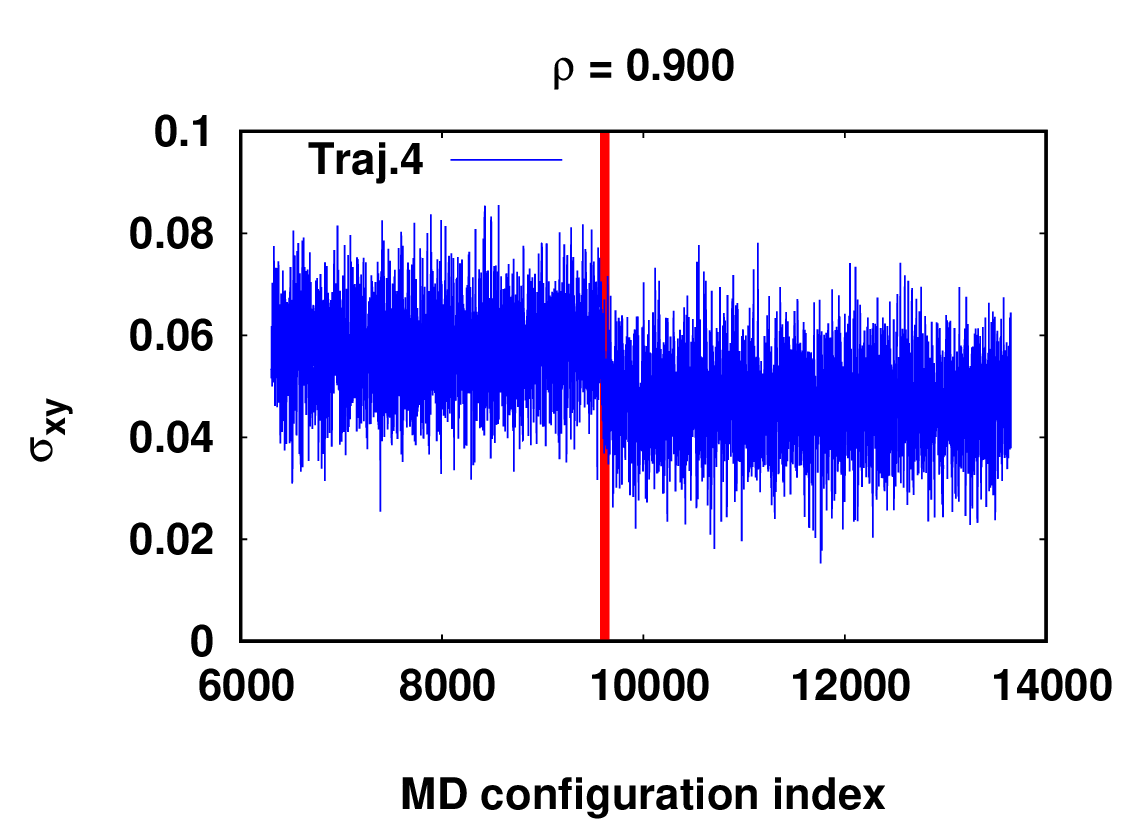}
    \includegraphics[width=0.45\textwidth]{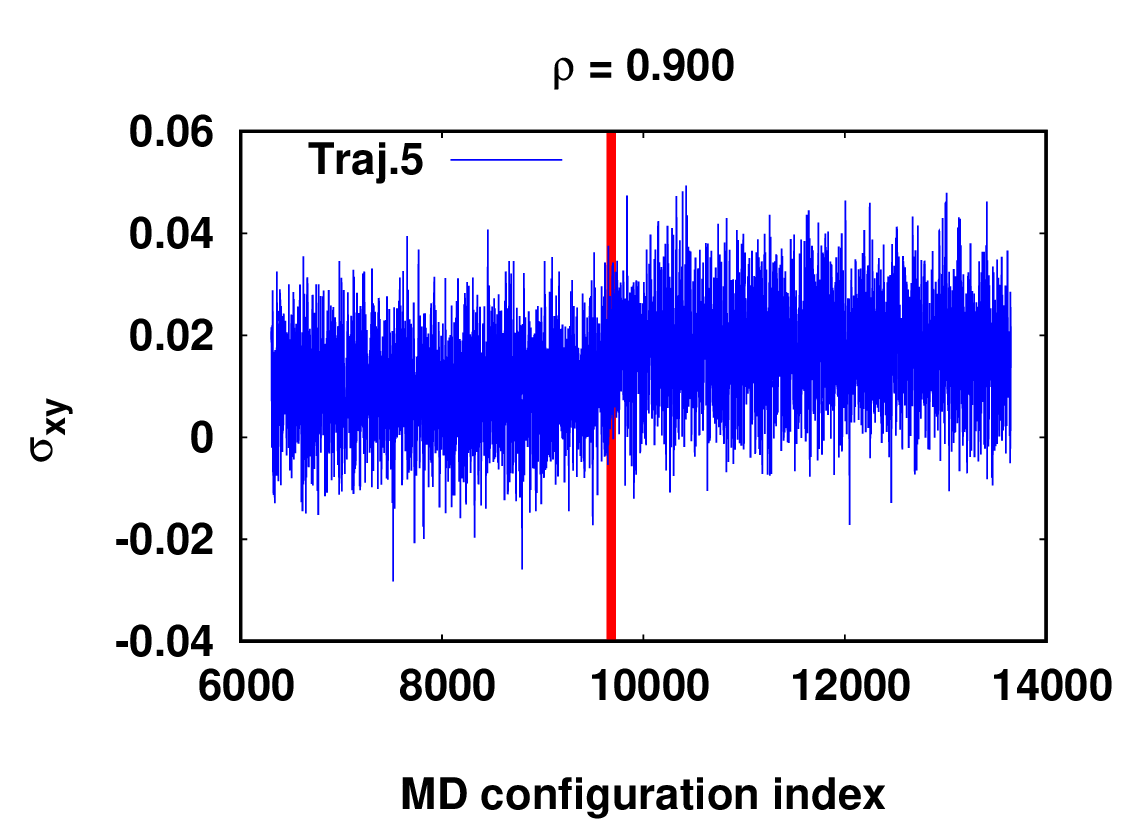}
    \caption{Evolution of shear stress with for four additional MD trajectories at density $\rho=0.900$ showing avalanches. Shear stress is computed at finite temperature (without energy minimization), and shows abrupt jumps during avalanche events. 
    In all panels, the vertical line denotes the configuration at which the thermally mediated plastic event (avalanche) is triggered.}
    \label{fig:av14nw2}
\end{figure*}

\begin{figure*}[h!]
 \centering
    \includegraphics[width=0.32\textwidth]{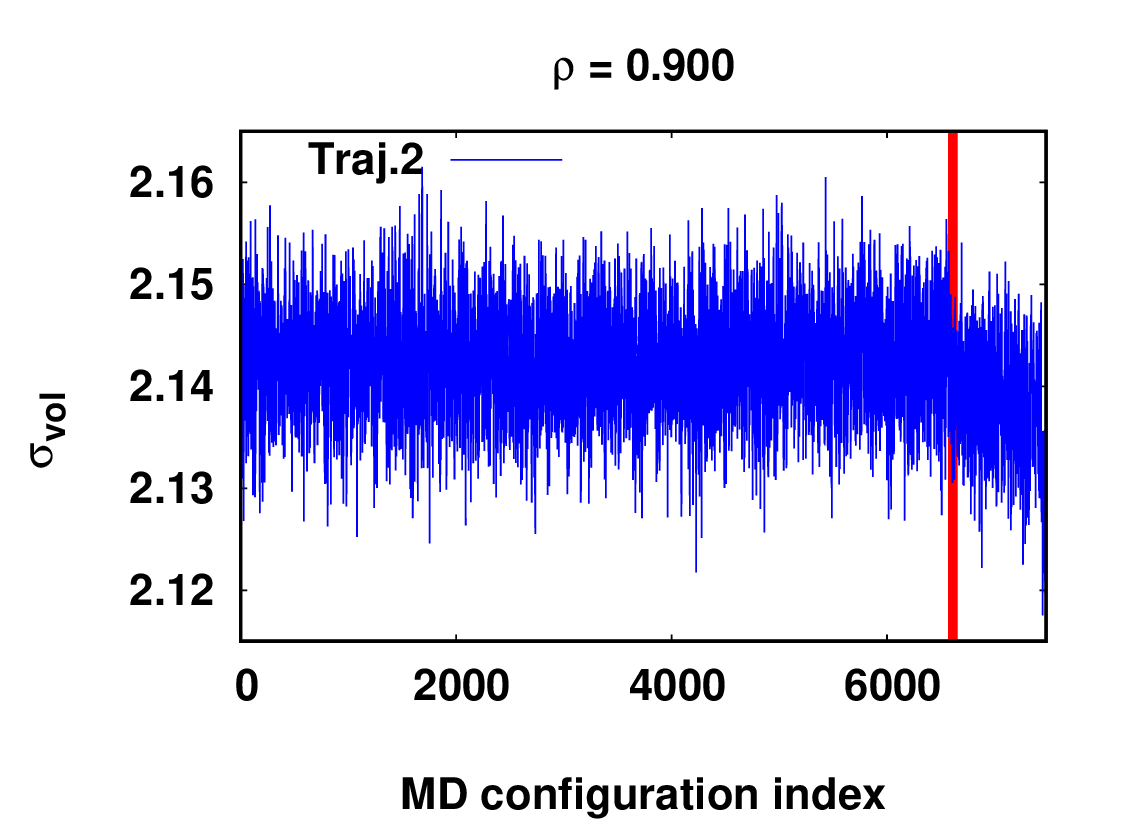}
    \includegraphics[width=0.32\textwidth]{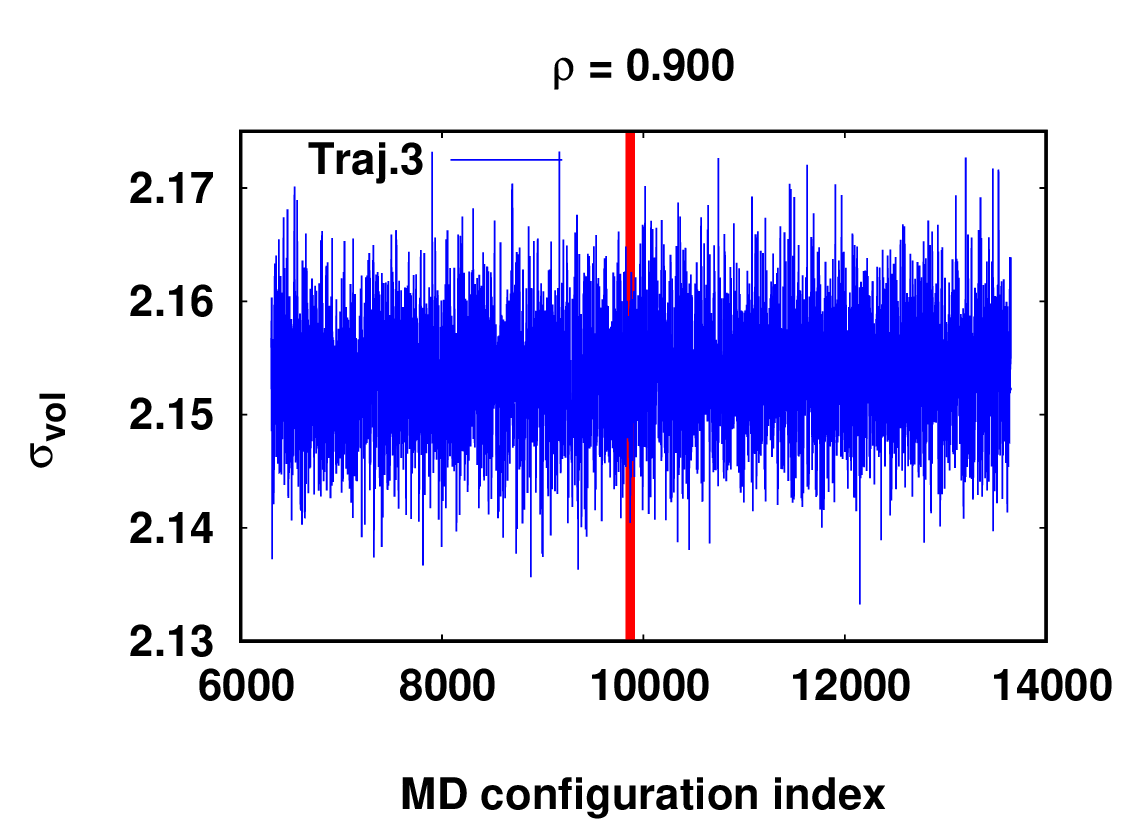}\\
    \includegraphics[width=0.32\textwidth]{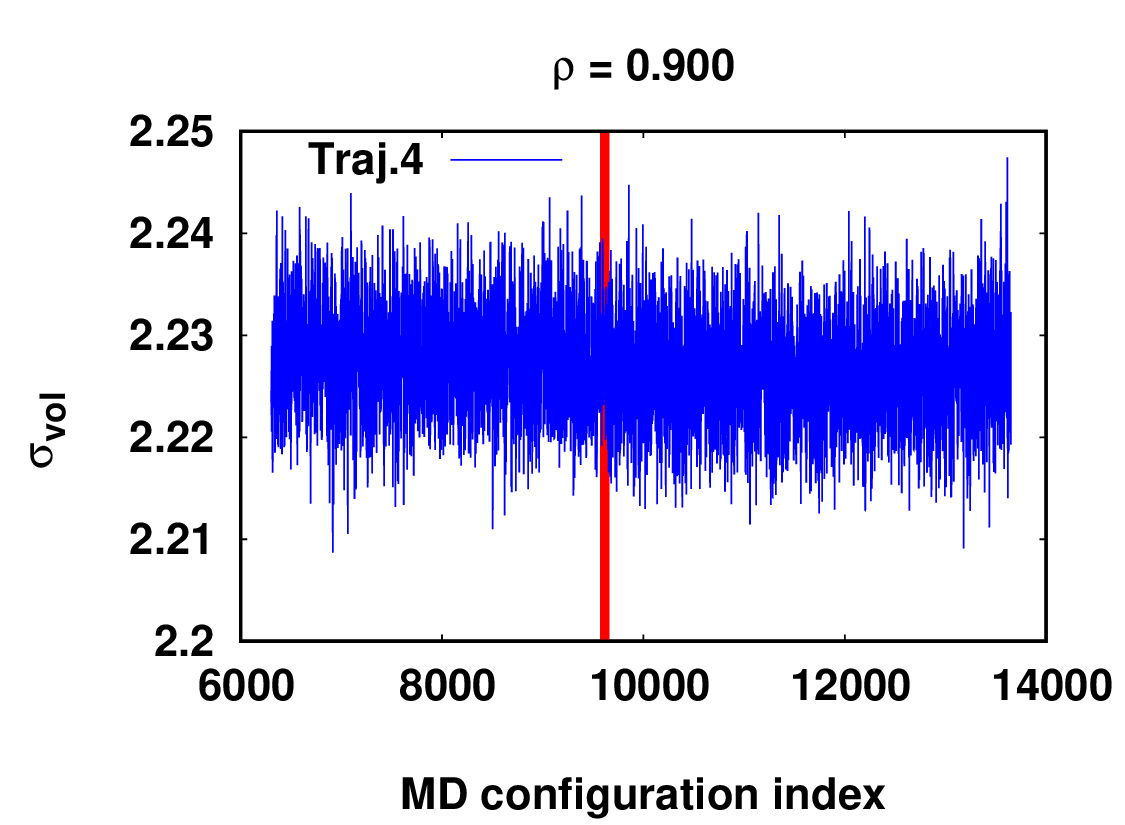}
    \includegraphics[width=0.32\textwidth]{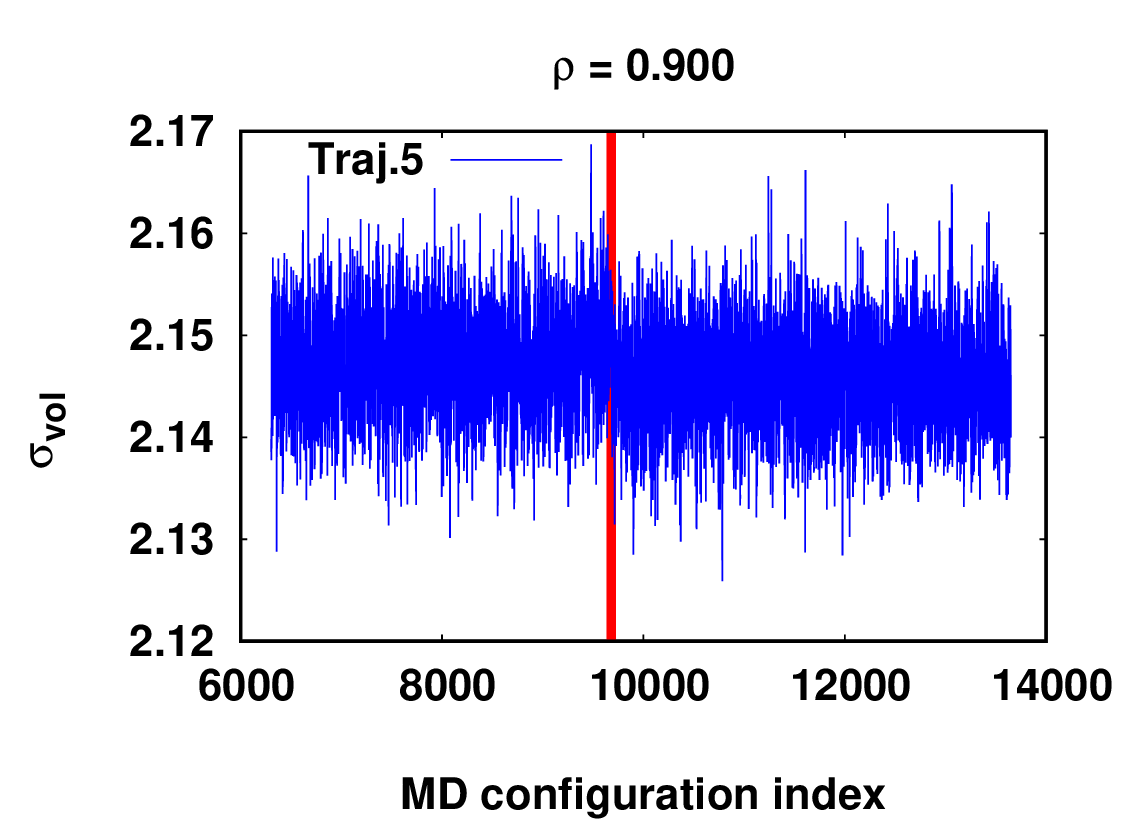}
    \caption{Evolution of volumetric stress for four additional avalanche events at density $\rho=0.900$. Volumetric stress is computed at finite temperature. It shows no abrupt jump during the avalanche events. 
    In all panels, the vertical line denotes the configuration at which the avalanche is triggered.}
    \label{fig:av14nw3}
\end{figure*}

\begin{figure*}[h!]
 \centering
    \includegraphics[width=0.34\textwidth]{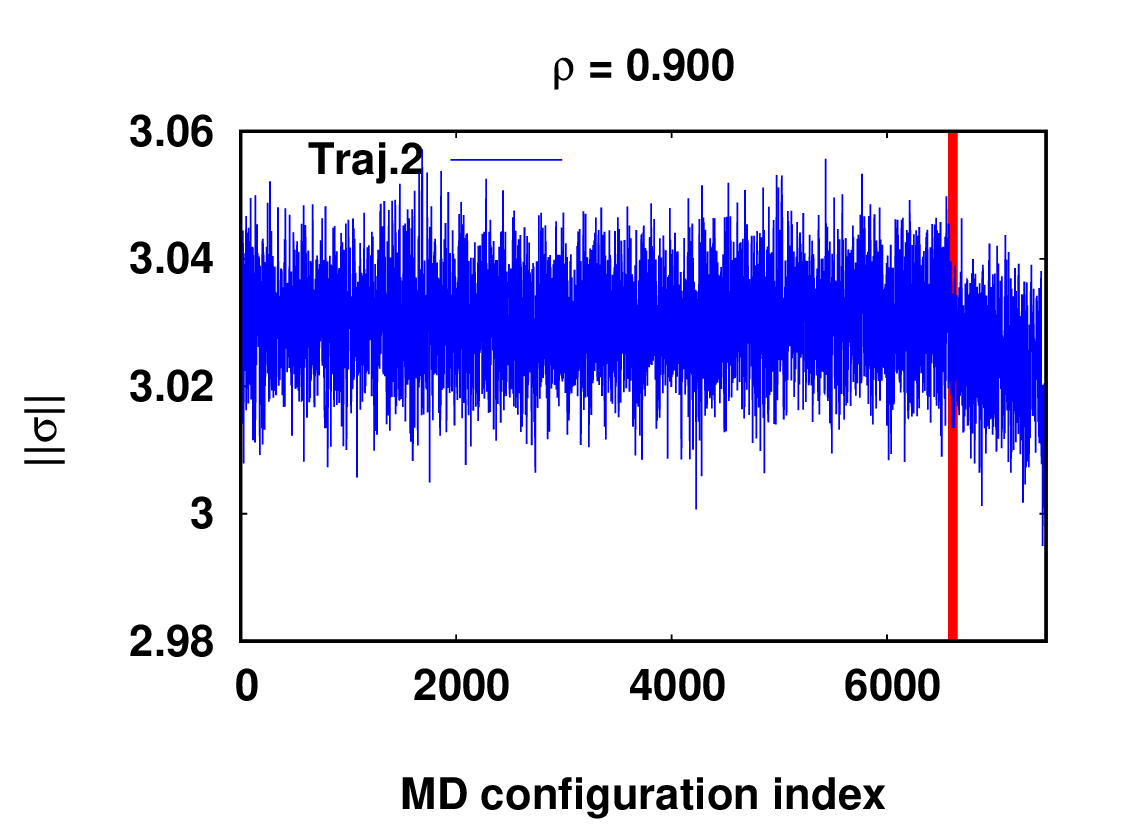}
    \includegraphics[width=0.34\textwidth]{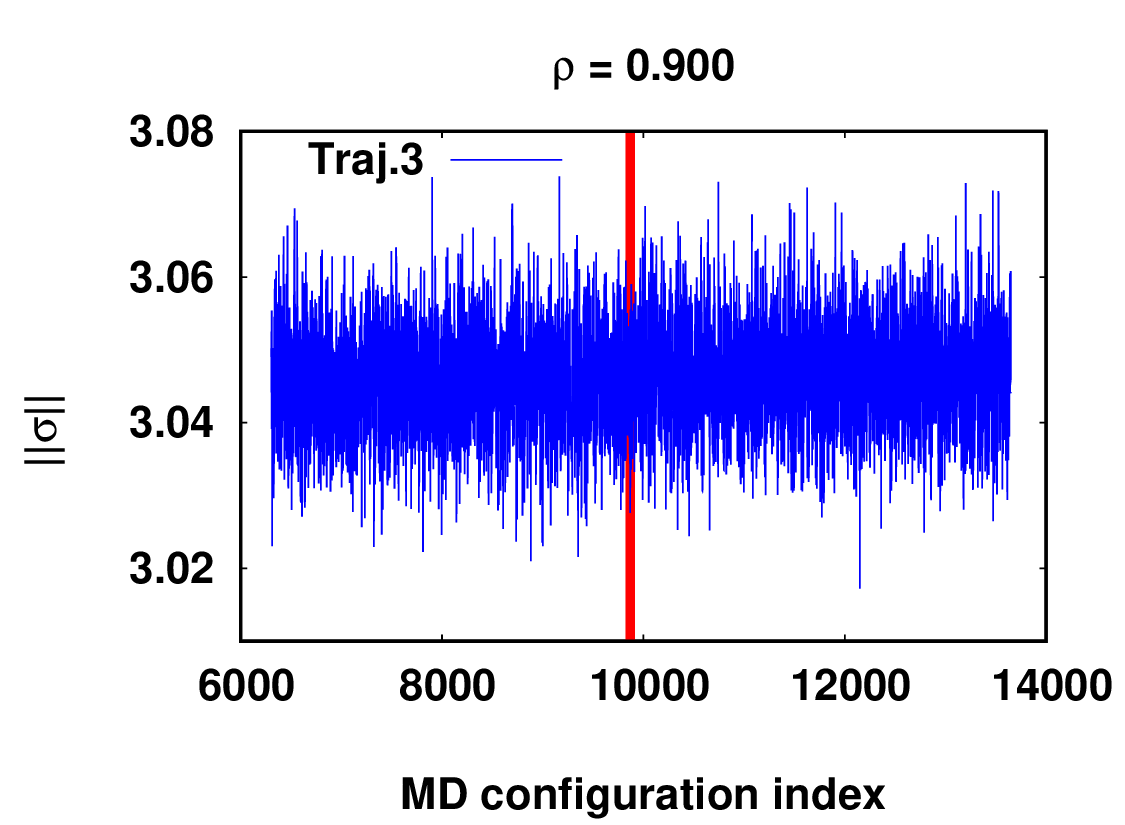}\\
    \includegraphics[width=0.34\textwidth]{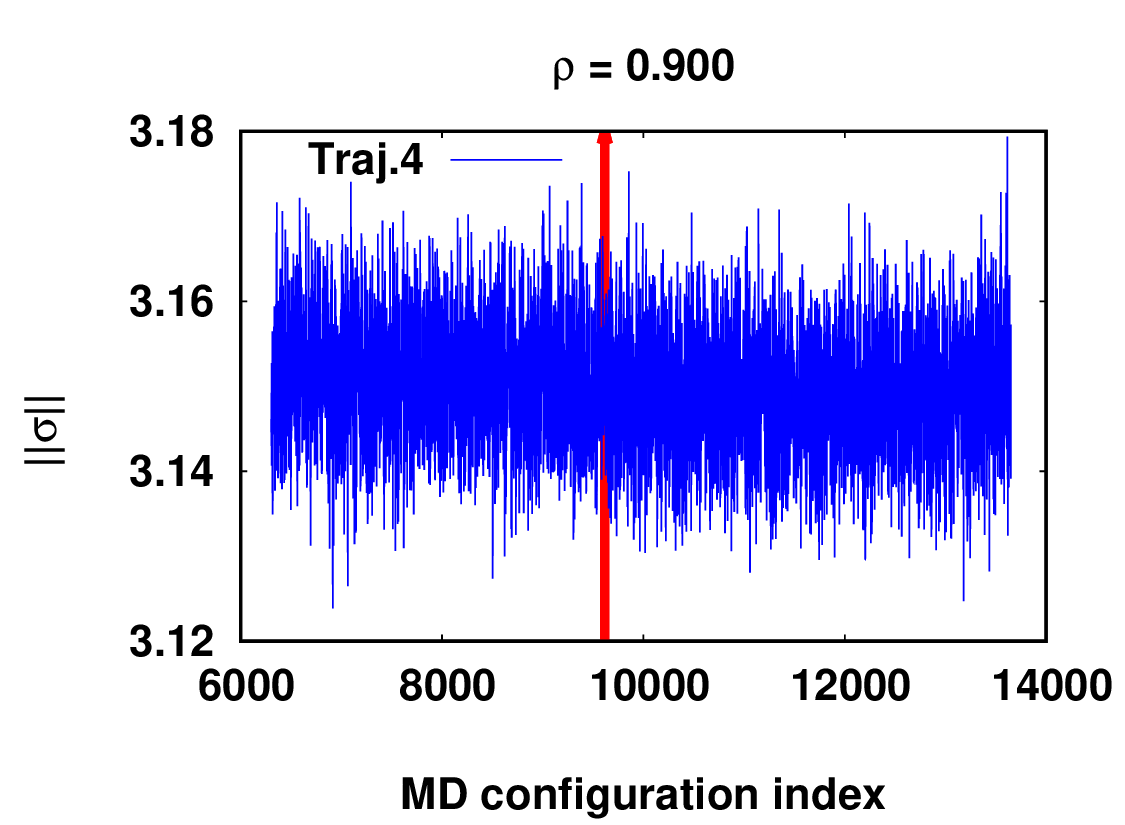}
    \includegraphics[width=0.34\textwidth]{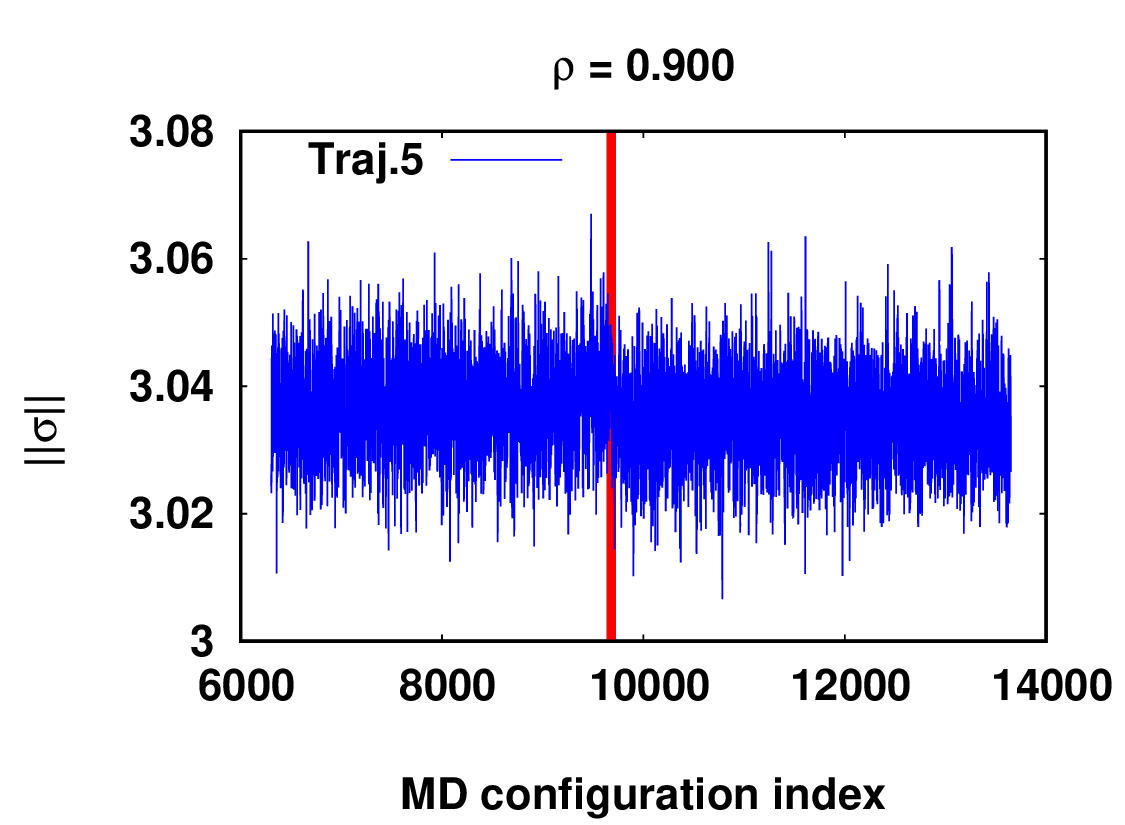}
    \caption{Evolution of the norm of stress for four additional avalanche trajectories at density $\rho=0.900$. The norm of stress is computed at a finite temperature (without energy minimization) and shows no discontinuous change during the avalanches. This contrasts with the abrupt change observed in the shear stress component. This supports the finding that while shear stress plays a critical role in triggering thermally mediated plasticity, the overall stress magnitude remains largely unchanged. In all panels, the vertical line denotes the configuration at which the thermally mediated plastic event (avalanche) is triggered.}
    \label{fig:av14nw4}
\end{figure*}

\begin{figure*}[h!]
 \centering
    \includegraphics[width=0.33\textwidth]{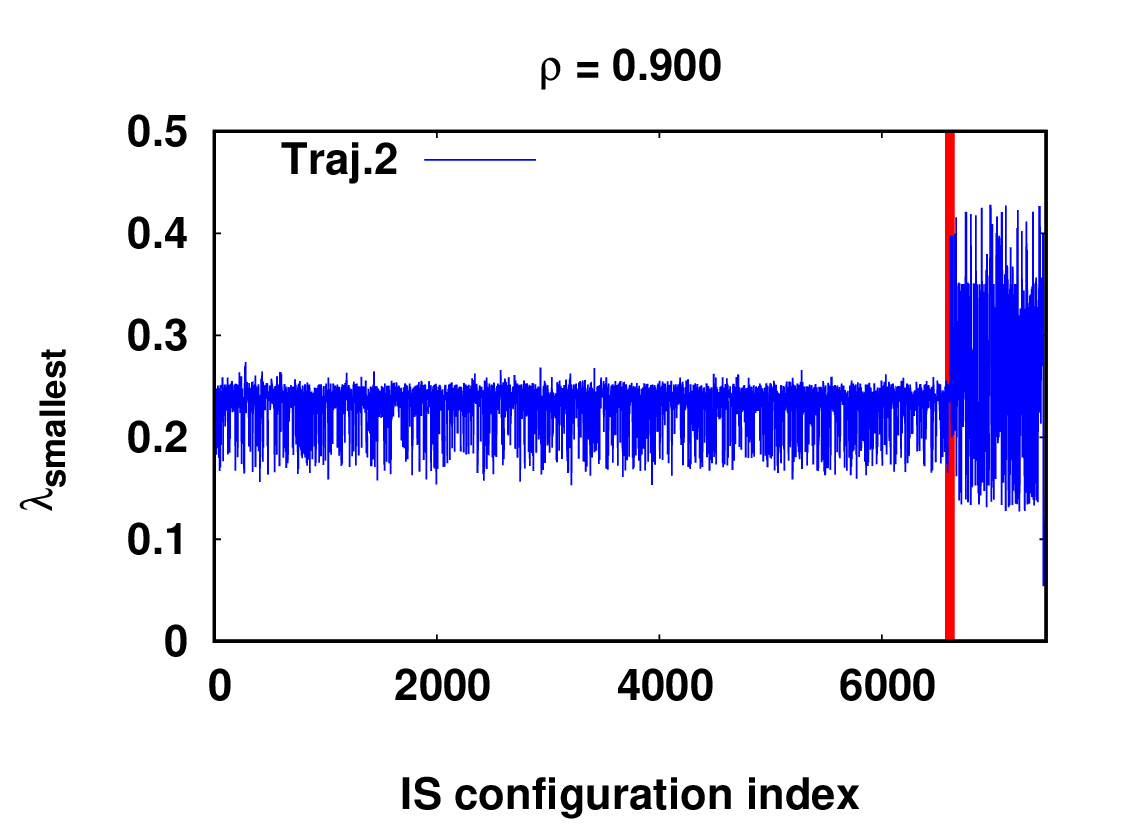}
    \includegraphics[width=0.33\textwidth]{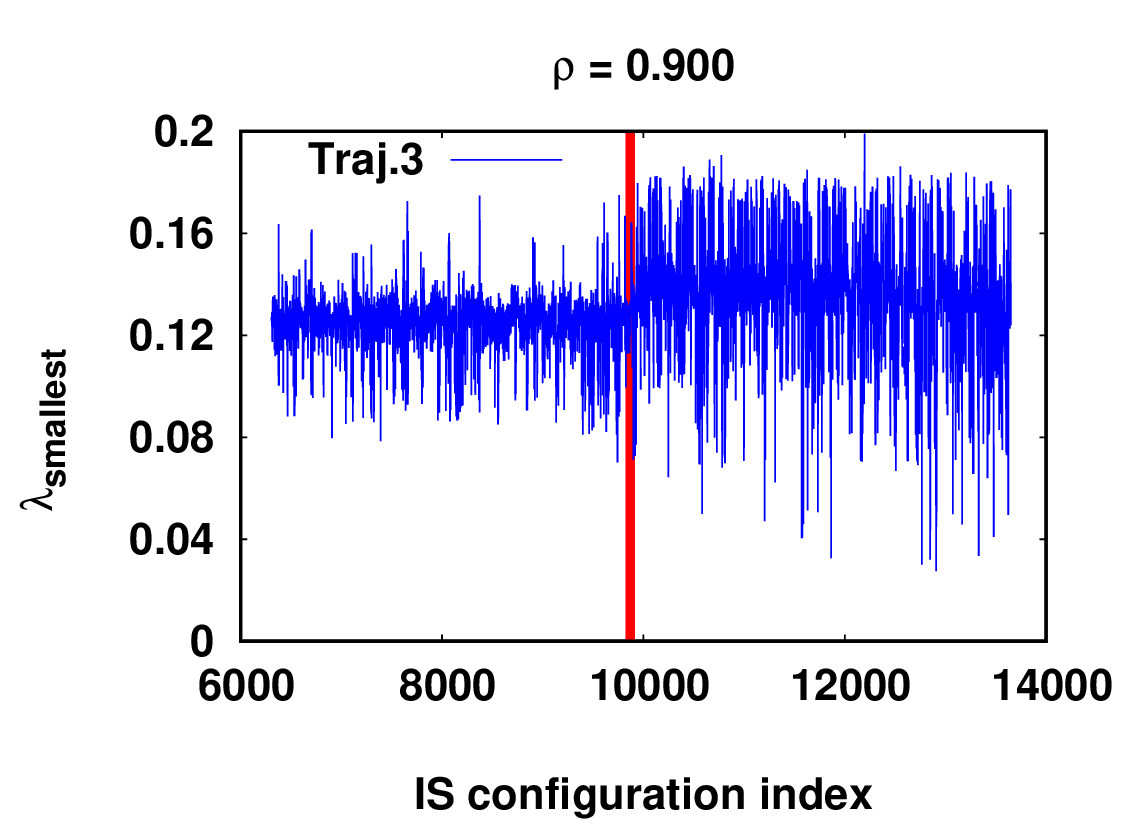}\\
    \includegraphics[width=0.33\textwidth]{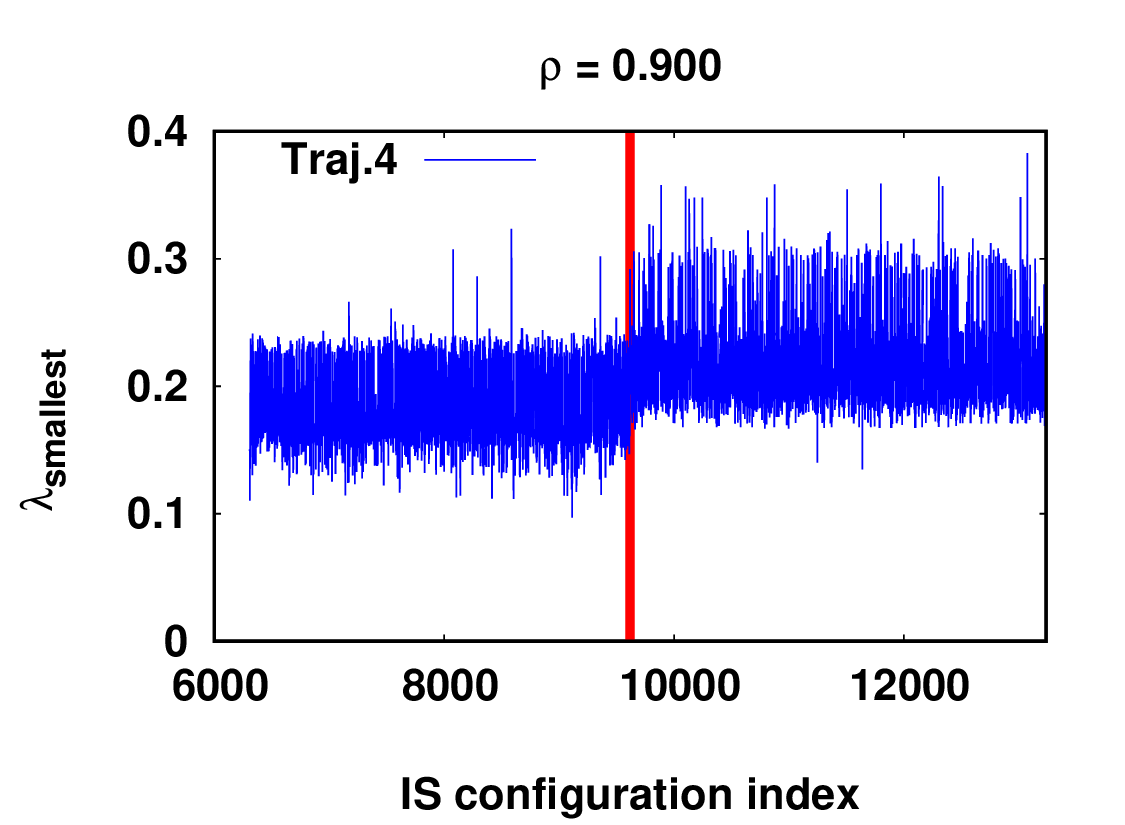}
    \includegraphics[width=0.33\textwidth]{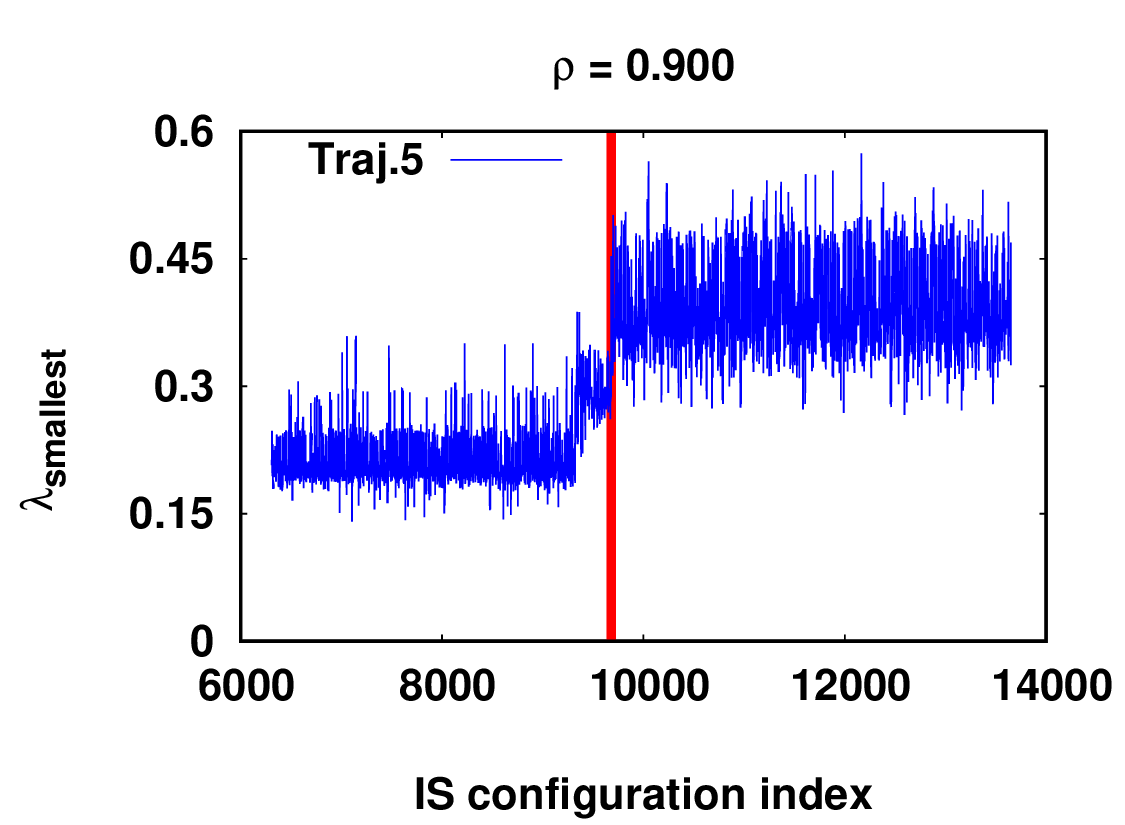}
    \caption{Evolution of the smallest eigenvalue $\lambda_{smallest}$ of the Hessian of the inherent structure (IS) for four additional IS trajectories at density $\rho=0.900$. These results complement the detailed analysis presented in the main text. In all panels, the vertical line denotes the configuration at which the thermally mediated plastic event (avalanche) is triggered.}
    \label{fig:av14nw1}
\end{figure*}

\begin{figure*}[htb!]
 \centering
    \includegraphics[width=0.33\textwidth]{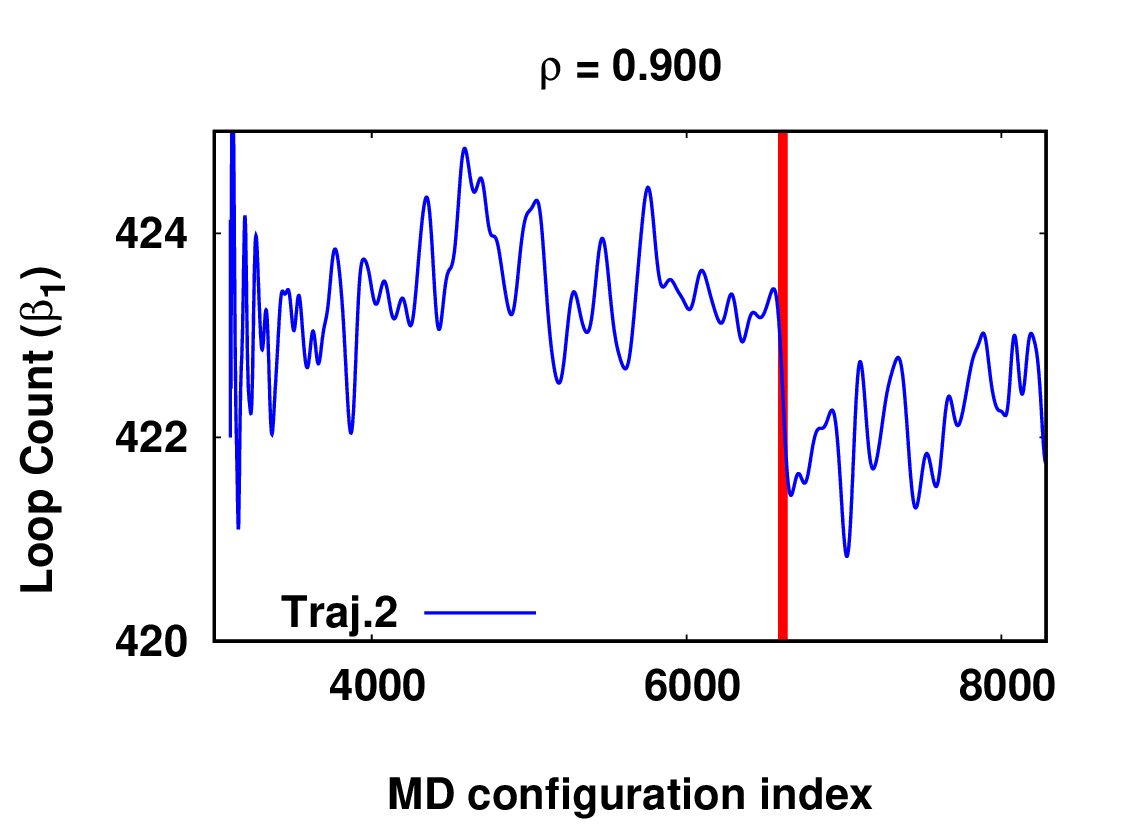}
    \includegraphics[width=0.33\textwidth]{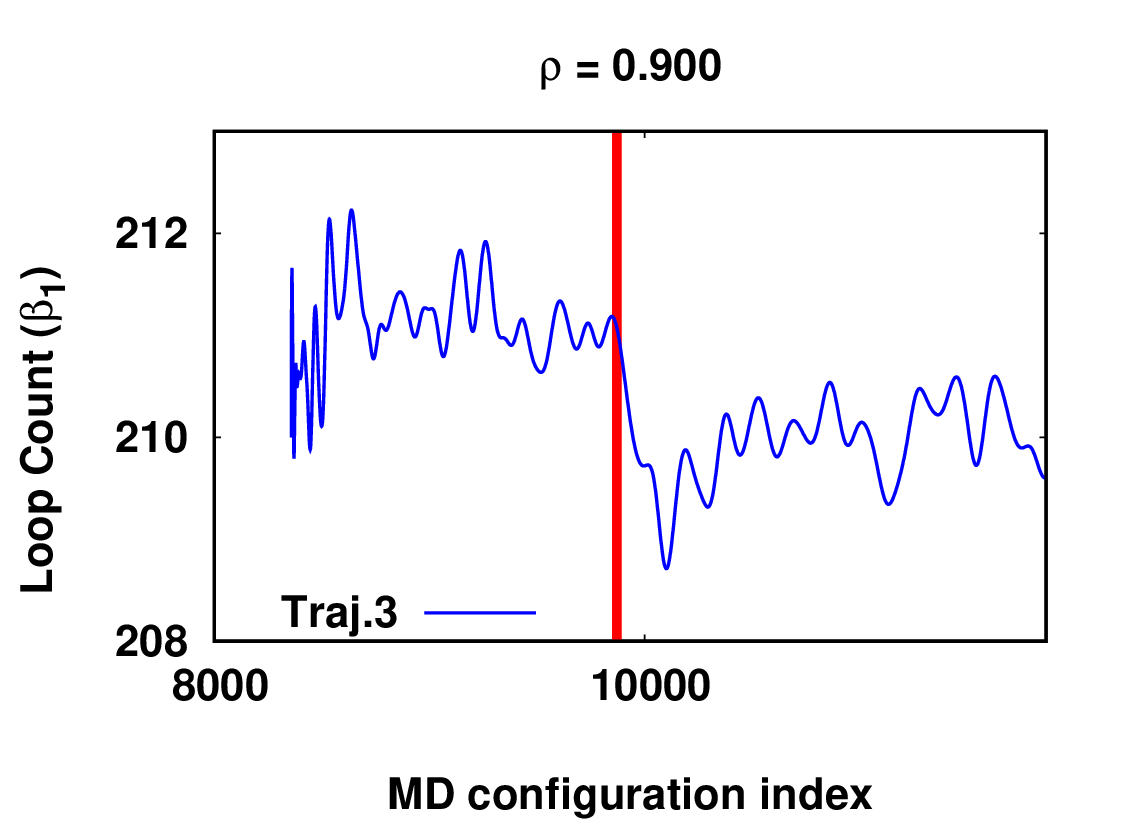}\\
    \includegraphics[width=0.33\textwidth]{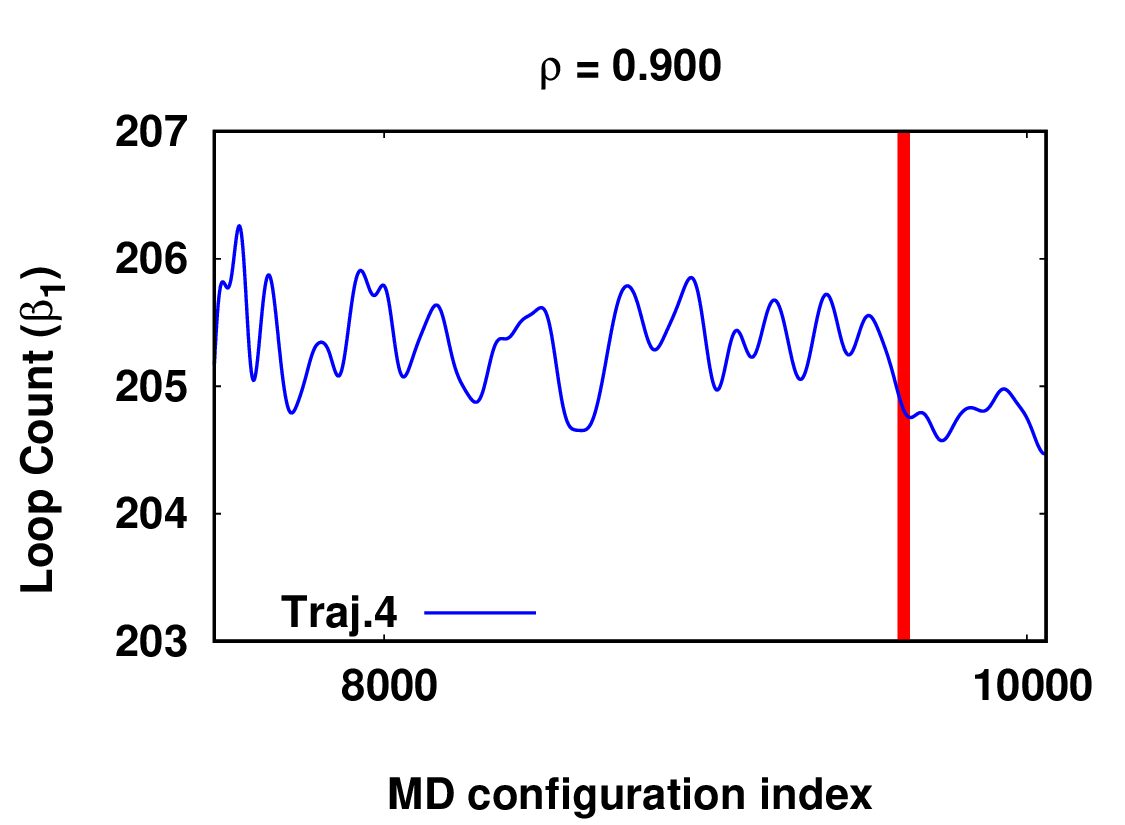}
    \includegraphics[width=0.33\textwidth]{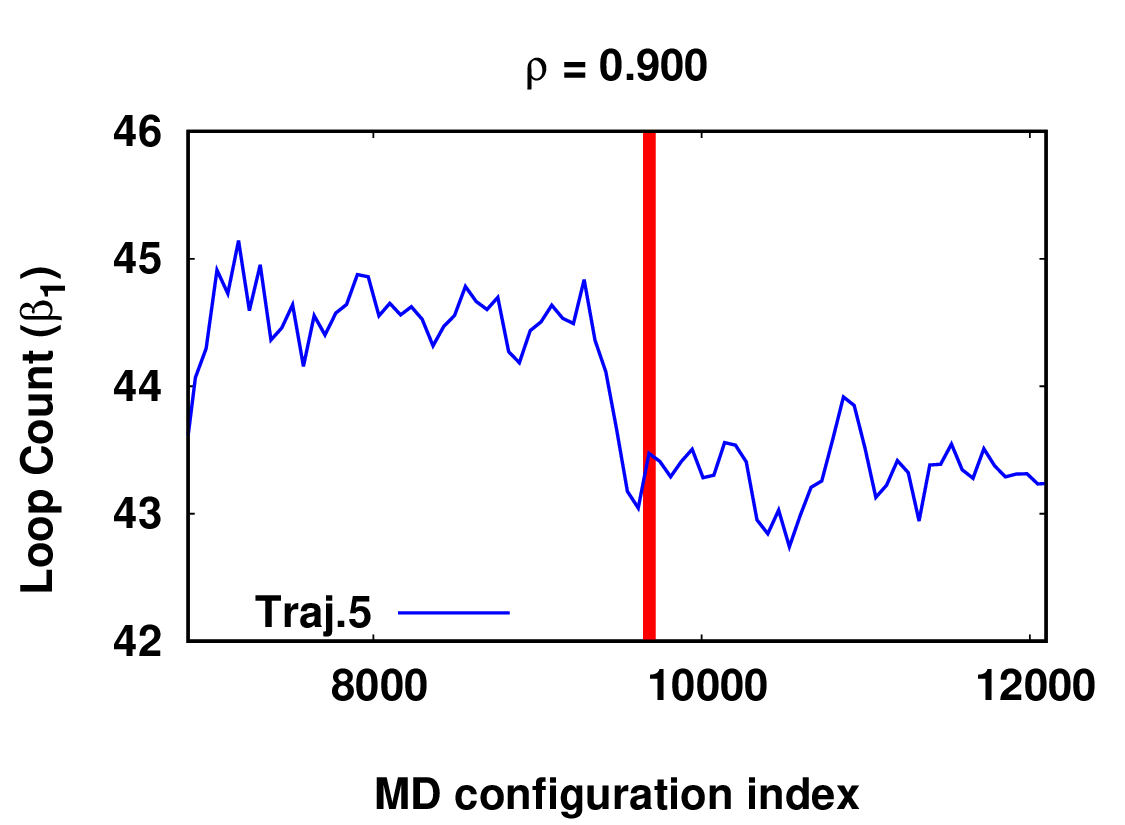}
    \caption{Evolution of the persistent 1D hole (ring) count (\SSG{Betti number} $\beta_1$) along four additional \SSG{MD} trajectories at density $\rho=0.900$. Red vertical lines mark the onset of an avalanche. \SSG{In each case, $\beta_1$ markedly decreases due the avalanche events with some trajectories showing two distinct ``plateaus" before and after avalanche.} It suggests that an avalanche leads to a more uniform and distributed network of smaller, stable holes.
    }
    \label{fig:av14nw5}
\end{figure*}

\begin{figure*}[htb!]
 \centering
    \includegraphics[width=0.45\textwidth]{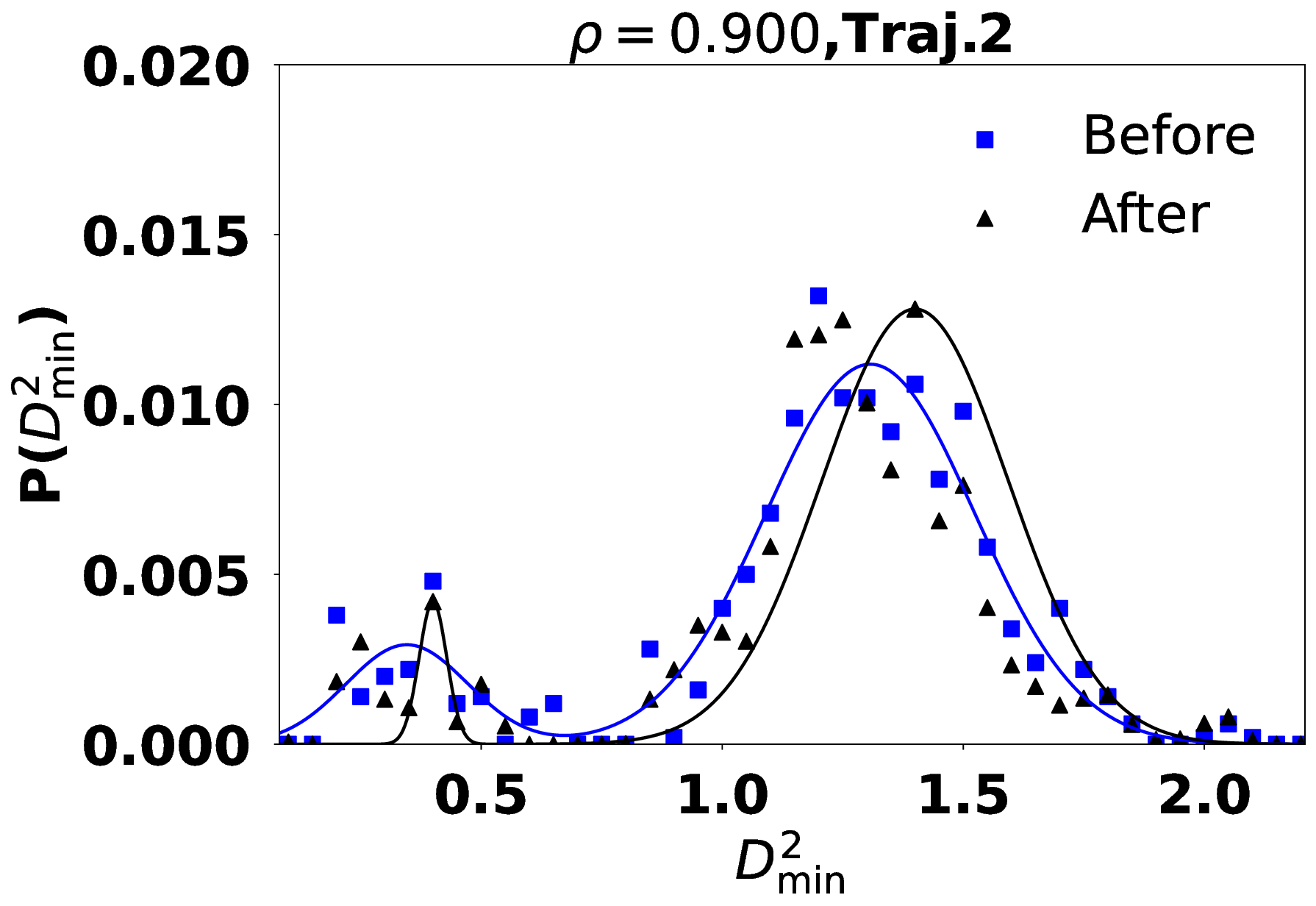}
    \includegraphics[width=0.45\textwidth]{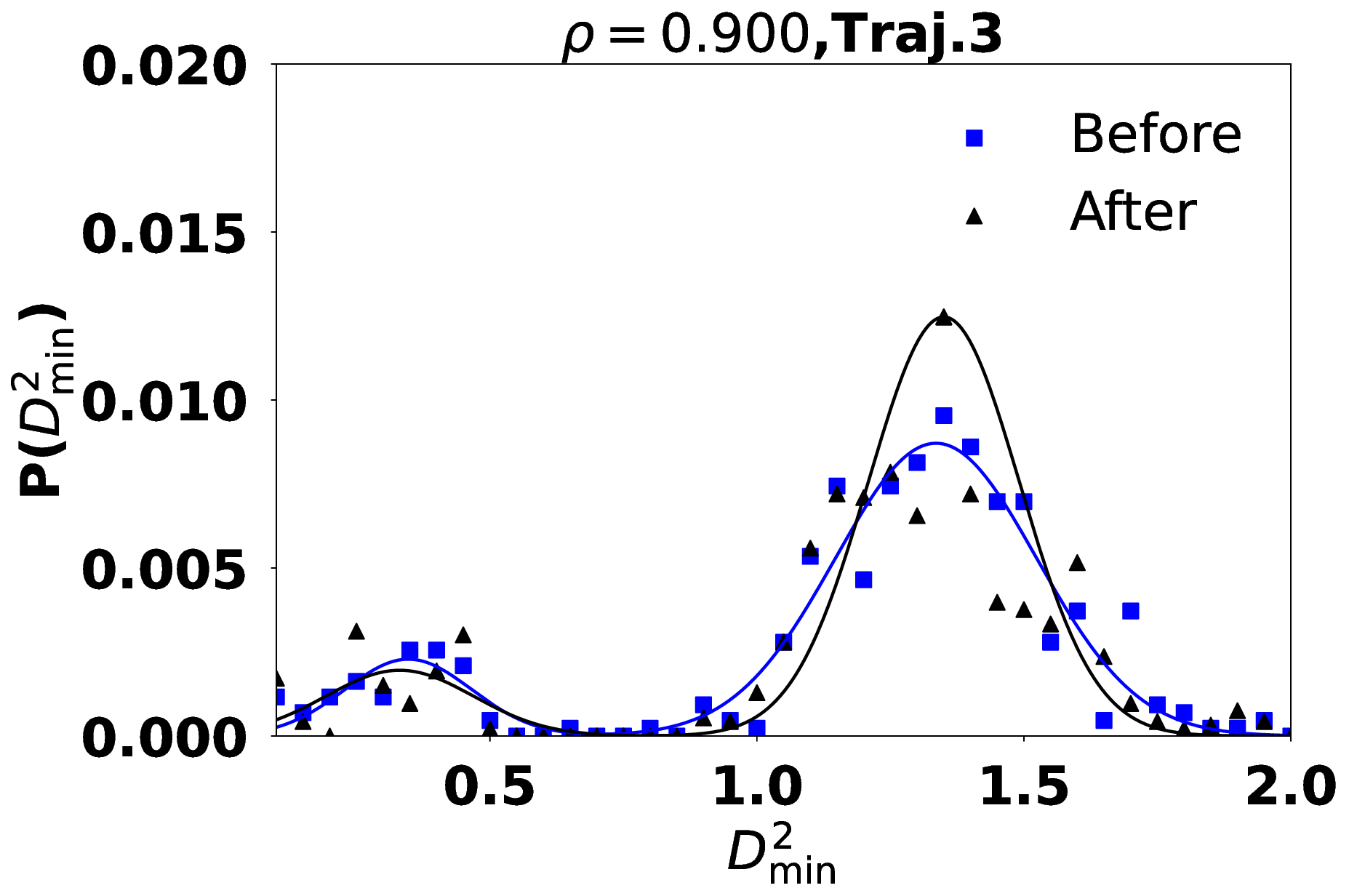}\\
    \includegraphics[width=0.45\textwidth]{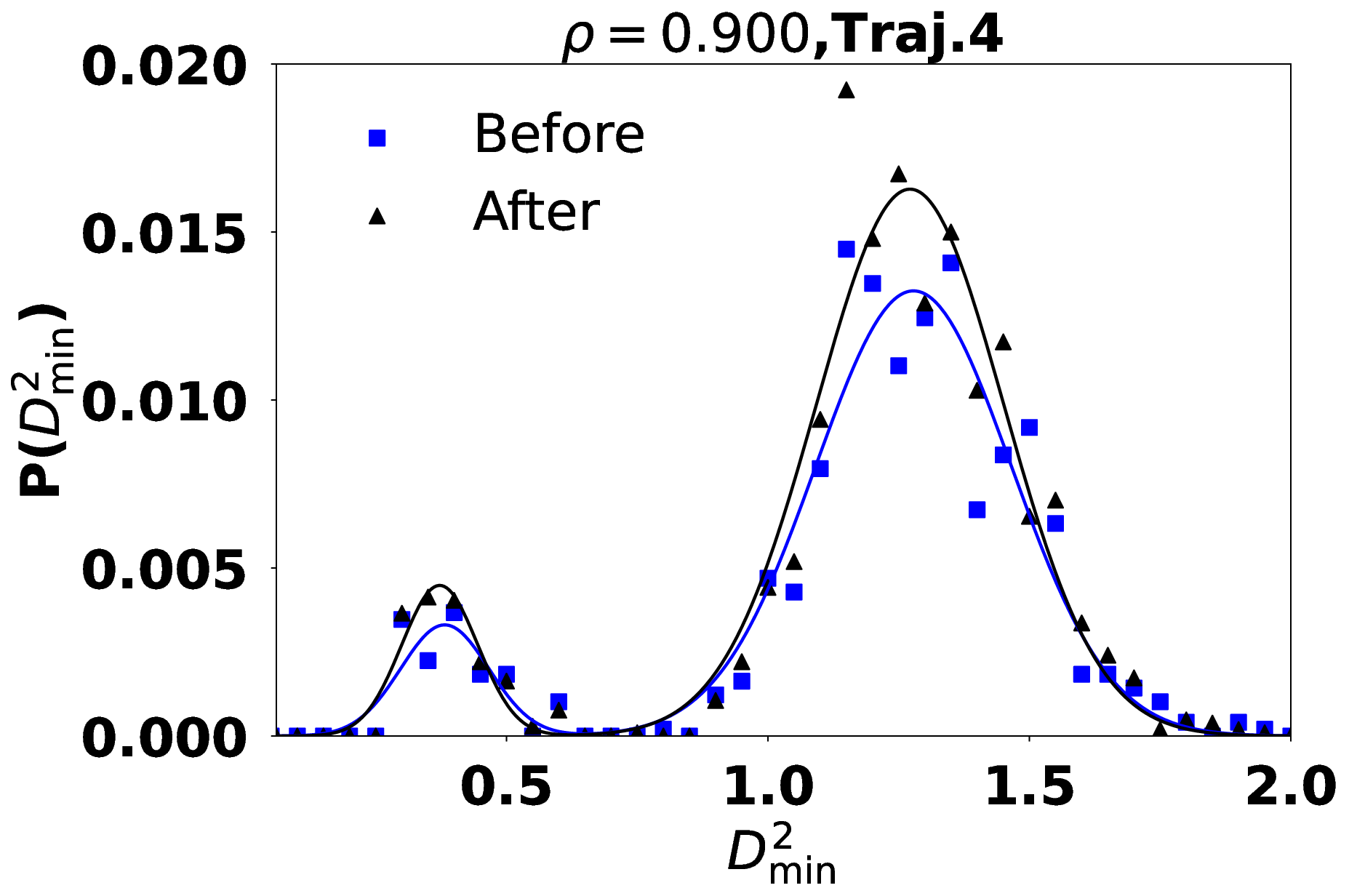}
    \includegraphics[width=0.45\textwidth]{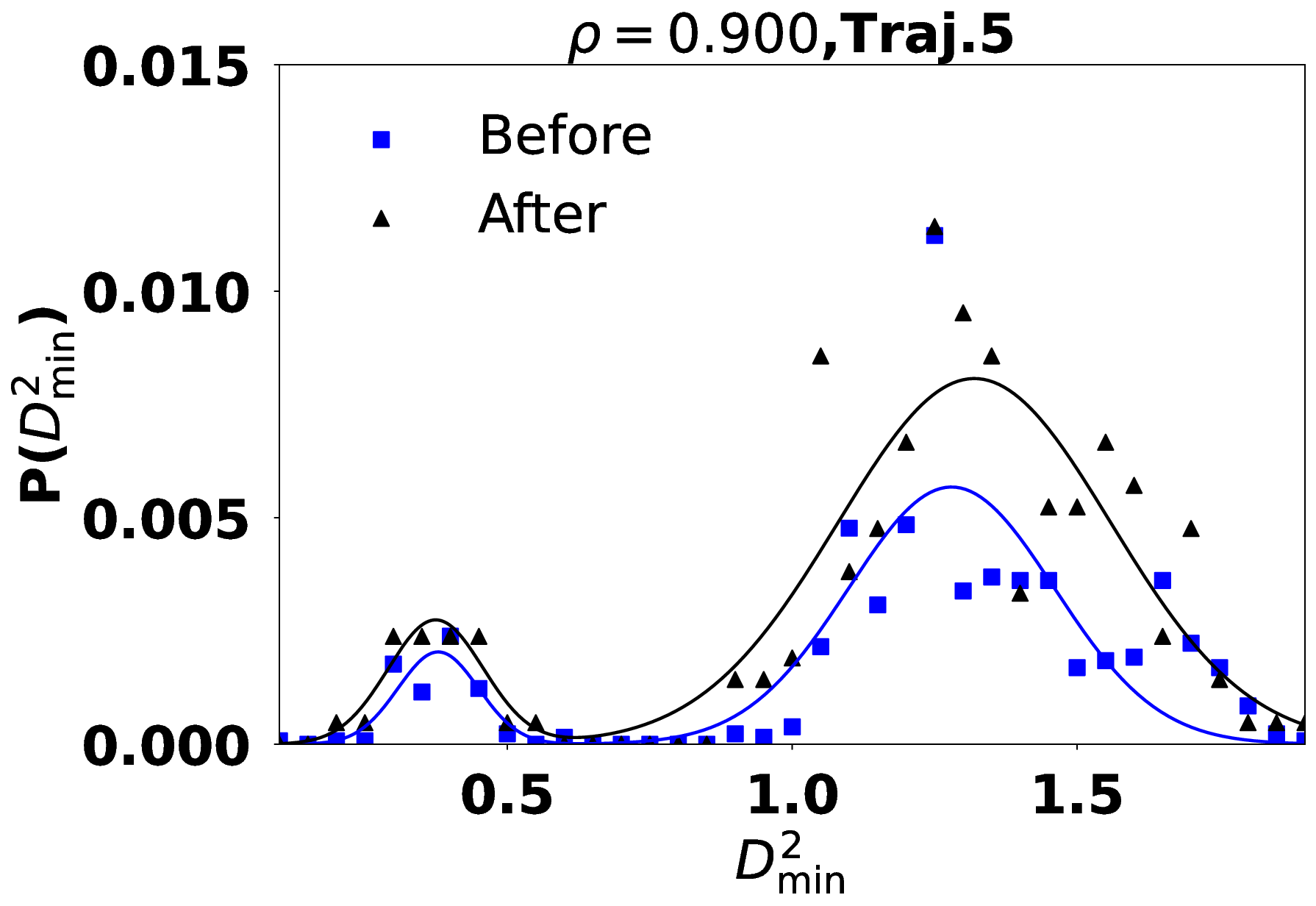}
    \caption{Distribution of $D^2_{min}$ values before and after the avalanche for the additional trajectories at density $\rho=0.900$. The histograms clearly show that $D^2_{min}$ shifts to higher values and increases in peaks at the post-avalanche states, reflecting increased non-affine displacements and reduced mechanical stability. This enhancement in $D^2_{min}$ across multiple trajectories corroborates our main text findings that the post-avalanche configuration is dynamically and mechanically distinct from the pre-avalanche state.}
      \label{fig:av14nw7}
\end{figure*}

\begin{figure*}[htb!]
 \centering
    \includegraphics[width=0.45\textwidth]{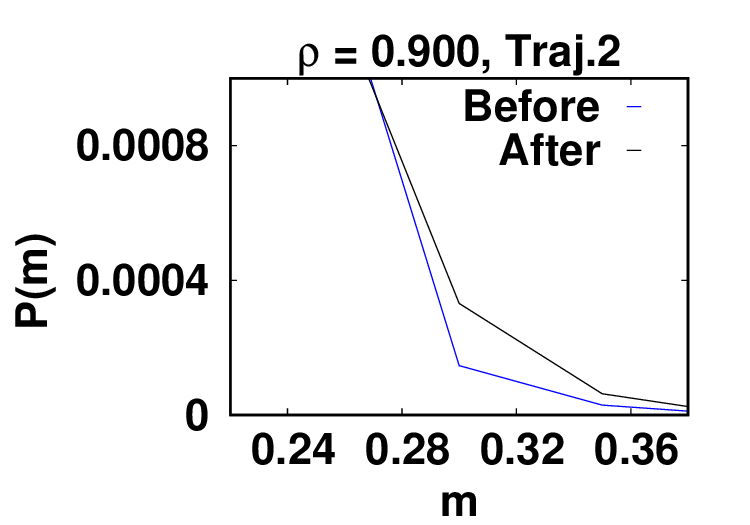}
    \includegraphics[width=0.45\textwidth]{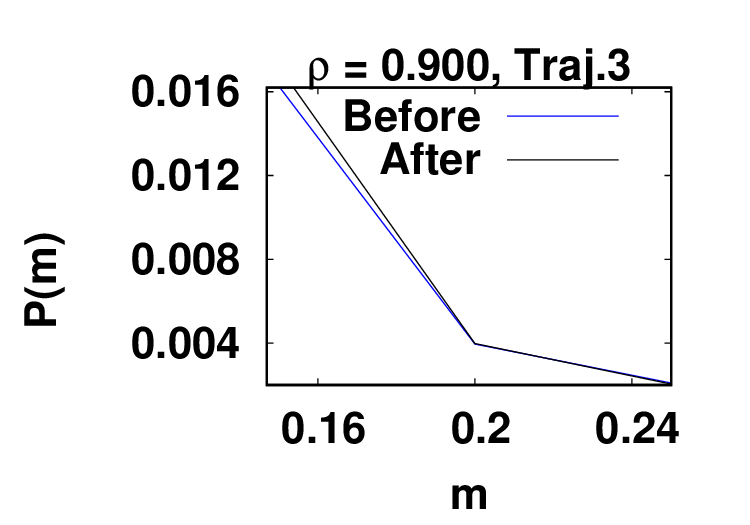}\\
    \includegraphics[width=0.45\textwidth]{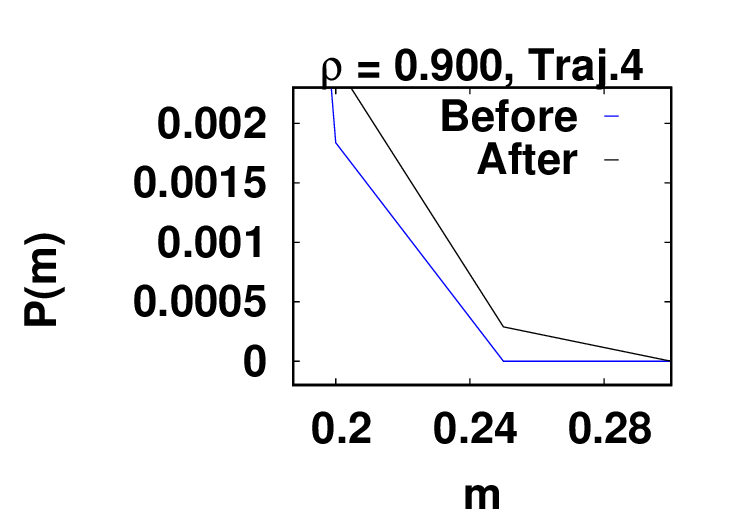}
    \includegraphics[width=0.45\textwidth]{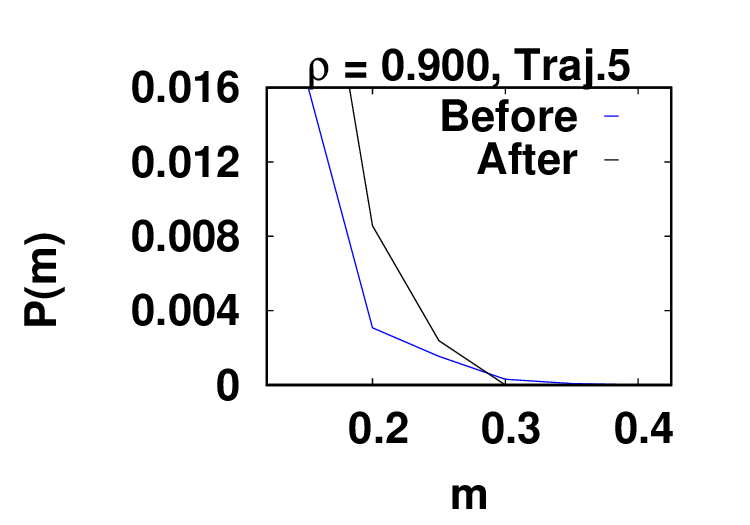}
    \caption{Distribution of dynamic mobility computed using a dynamically updated reference frame for the four additional avalanche trajectories at density $\rho=0.900$. In this analysis, displacements are measured relative to the immediately preceding configuration, thereby preventing the cumulative effects of avalanche-induced jumps. Our dynamic mobility analysis reveals that the states before and after an avalanche event are dynamically distinct and mobility increases due to the avalanche. These supplementary results further corroborate the findings presented in the main text.}
    \label{fig:av14nw6}
\end{figure*}

\begin{figure*}[htb!]
 \centering
    \includegraphics[width=0.45\textwidth]{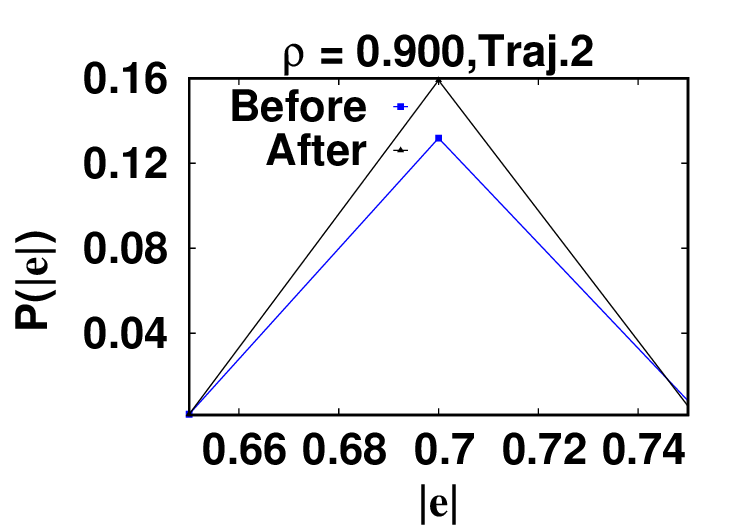}
    \includegraphics[width=0.45\textwidth]{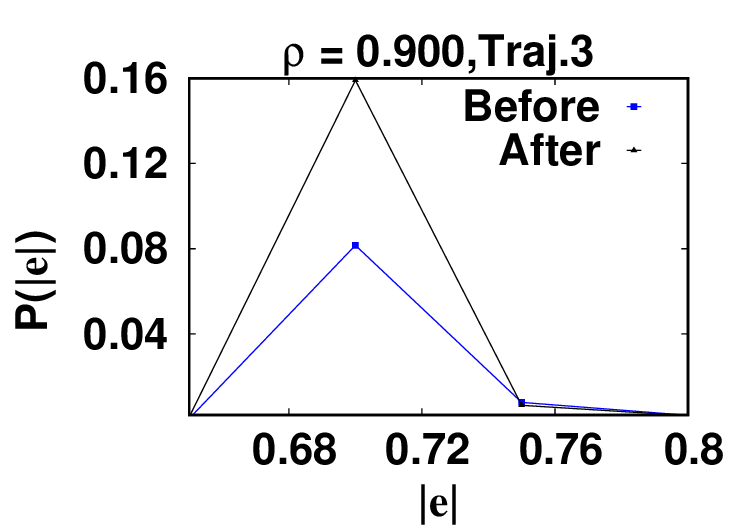}\\
    \includegraphics[width=0.45\textwidth]{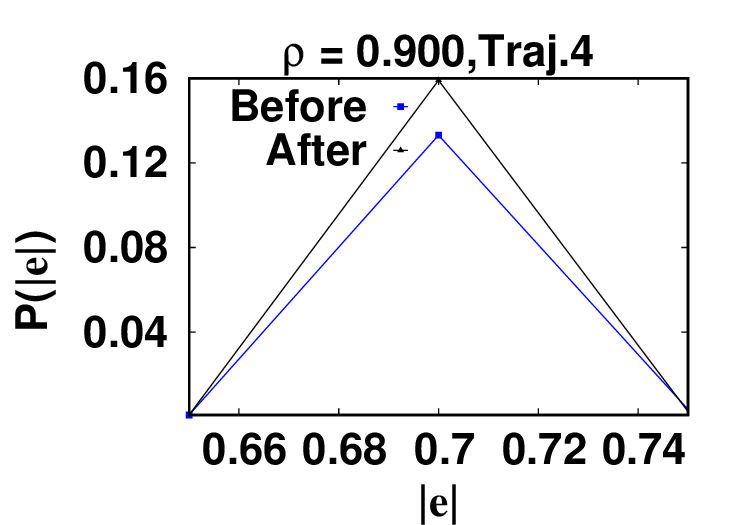}
    \includegraphics[width=0.45\textwidth]{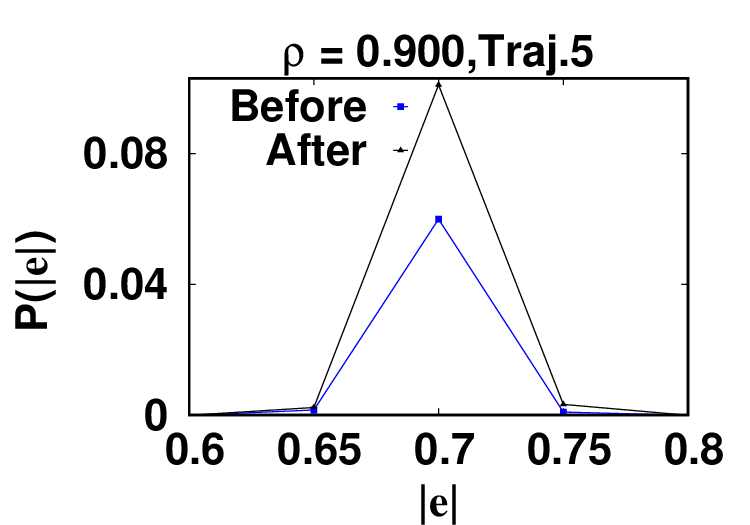}
    \caption{Distribution of deviatoric strain for the additional trajectories at density $\rho=0.900$, comparing the pre and post-avalanche states. The distribution shows significantly higher peaks in the post-avalanches states, indicating an increase in volume-preserving shear strain following the avalanche. This enhancement in deviatoric strain further supports the findings presented in the main text.}
    \label{fig:av14nw10}
\end{figure*}

\begin{figure*}[htb!]
 \centering
    \includegraphics[width=0.45\textwidth]{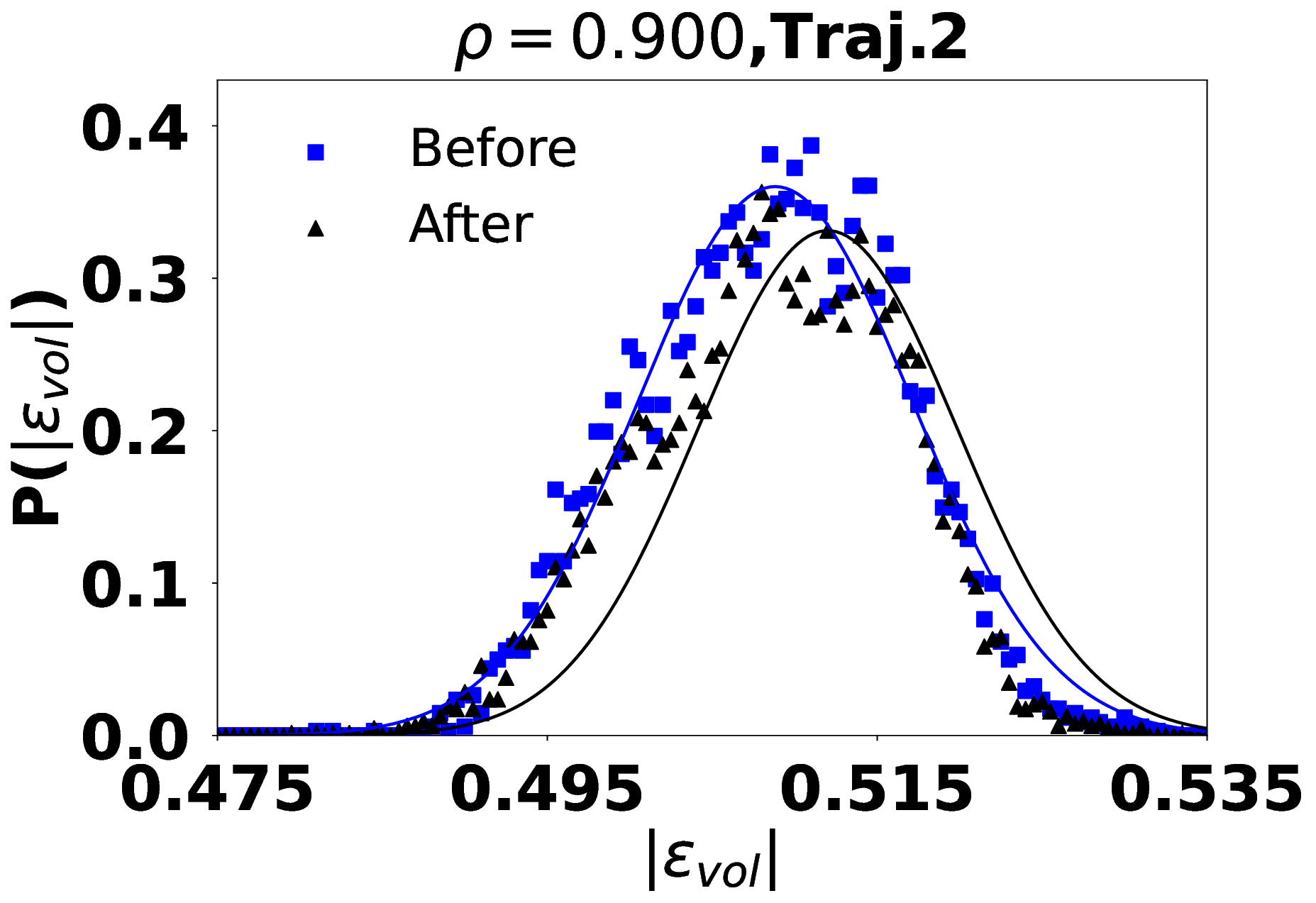}
    \includegraphics[width=0.45\textwidth]{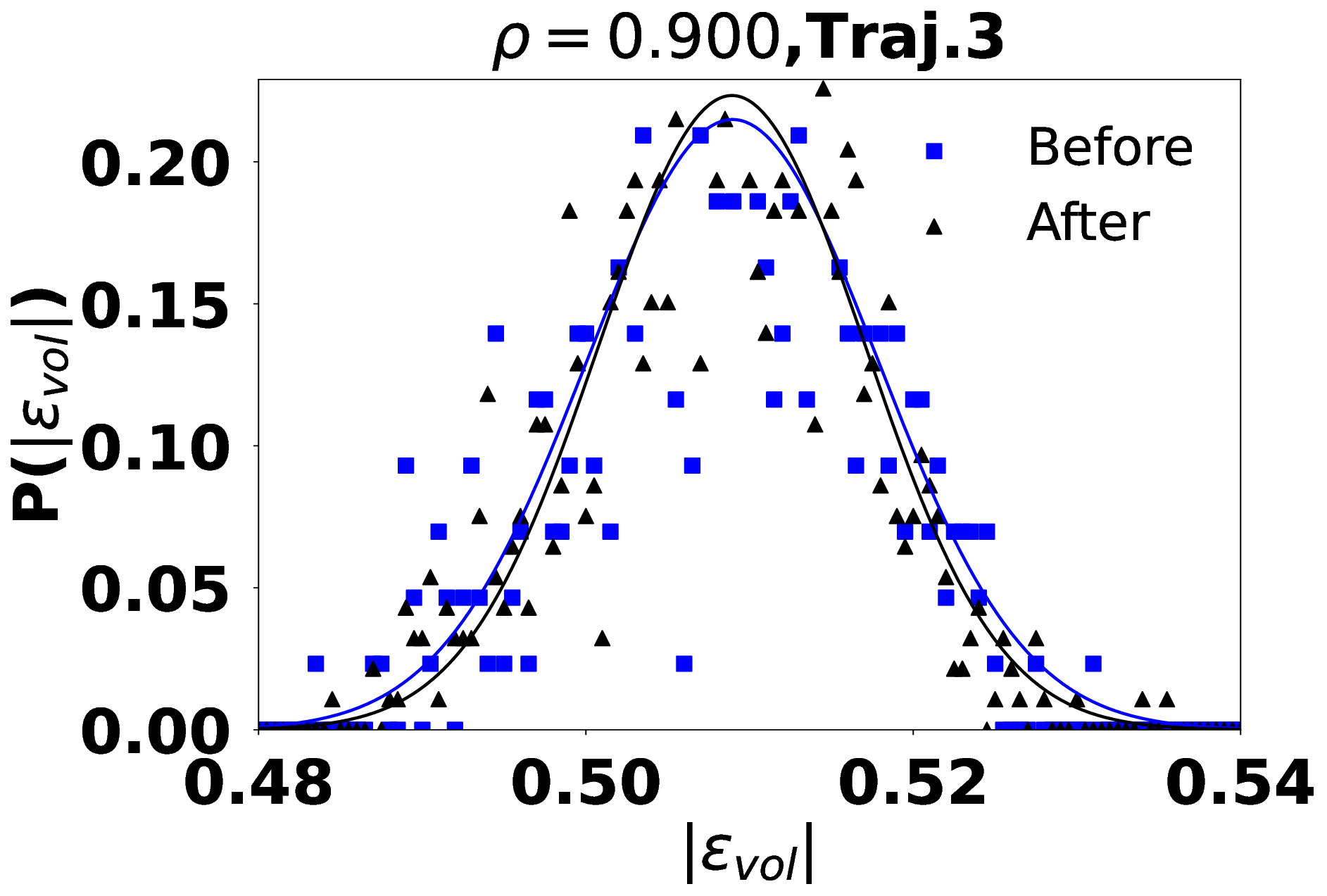}\\
    \includegraphics[width=0.45\textwidth]{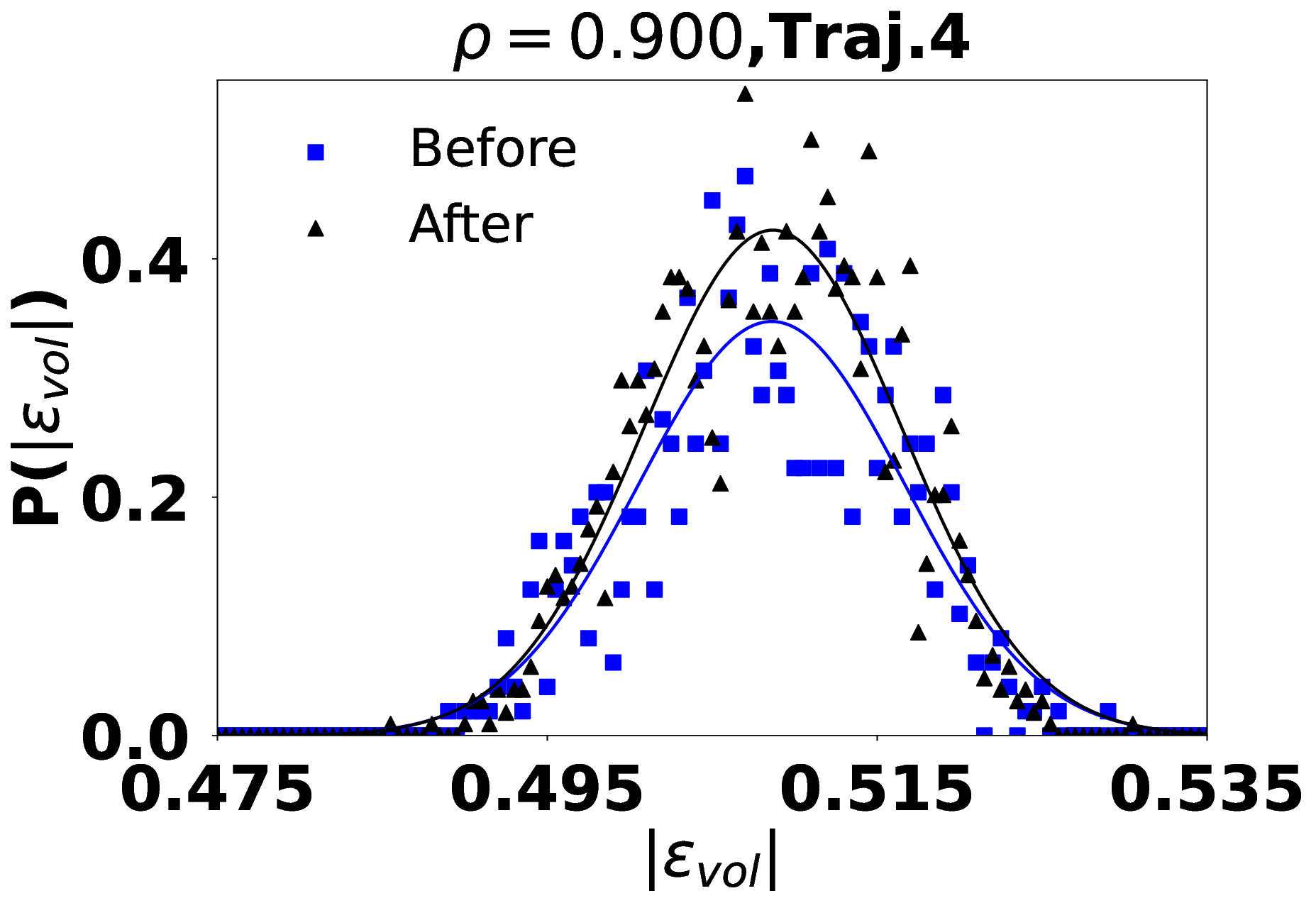}
    \includegraphics[width=0.45\textwidth]{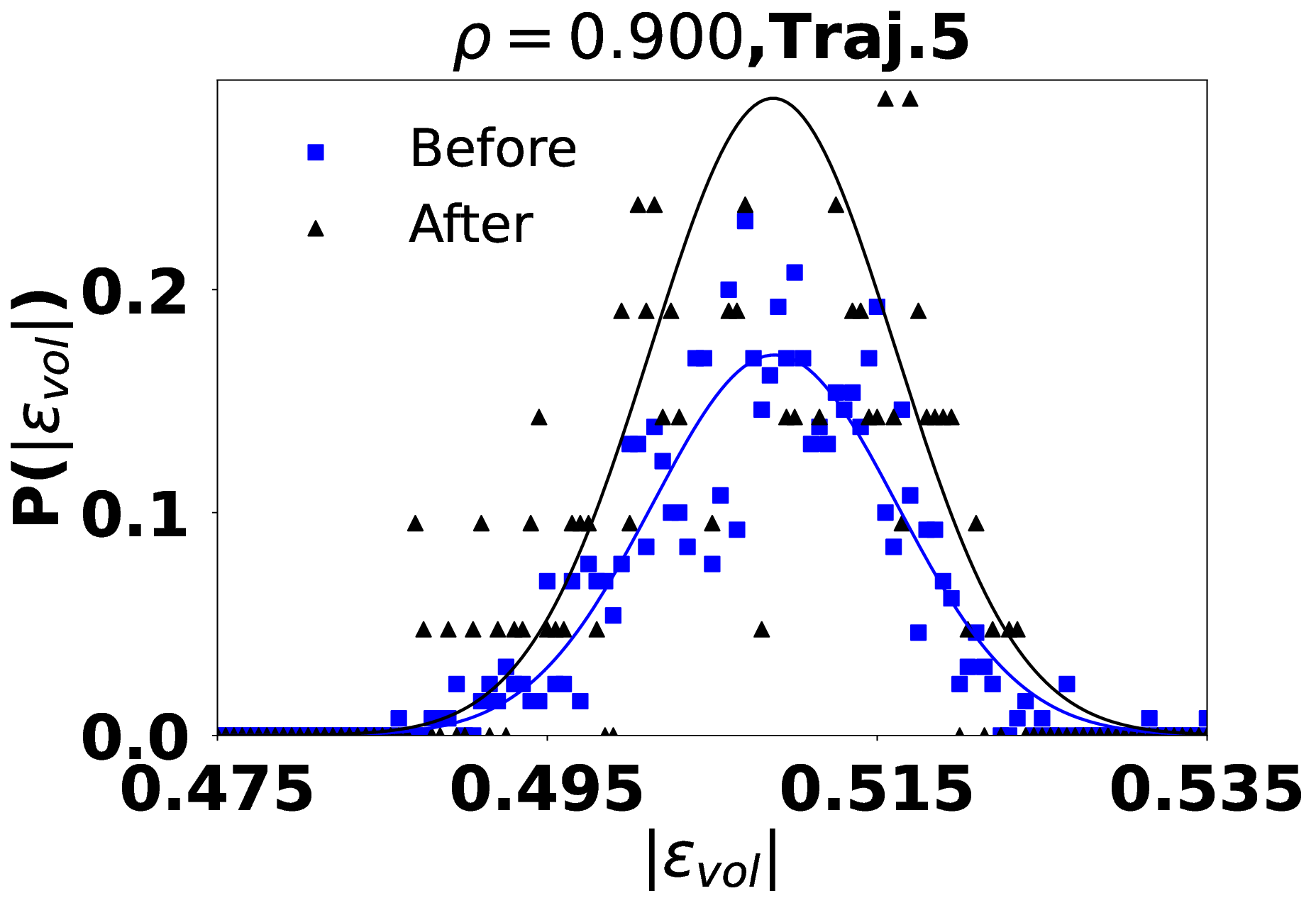}
    \caption{Distribution of the magnitude of volumetric strain $\vert \epsilon_{vol}\vert$ for the four additional avalanche trajectories at density $\rho=0.900$. The distributions compare the volumetric strain before and after the avalanche, showing clear heightening peaks in post-avalanches. This enhancement in volumetric strain indicates that structural reorganization during the avalanche leads to greater local volume changes, further supporting the findings presented in the main text. }
    \label{fig:av14nw8}
\end{figure*}

\begin{figure*}[htb!]
 \centering
    \includegraphics[width=0.45\textwidth]{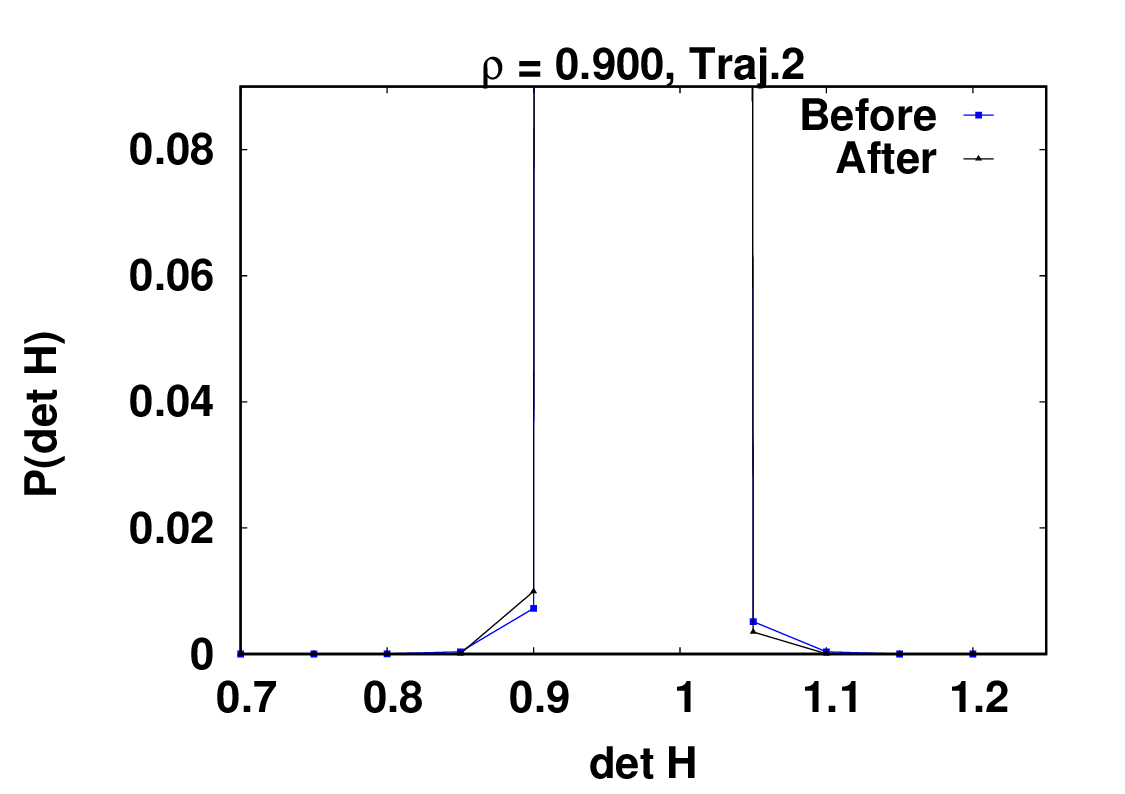}
    \includegraphics[width=0.45\textwidth]{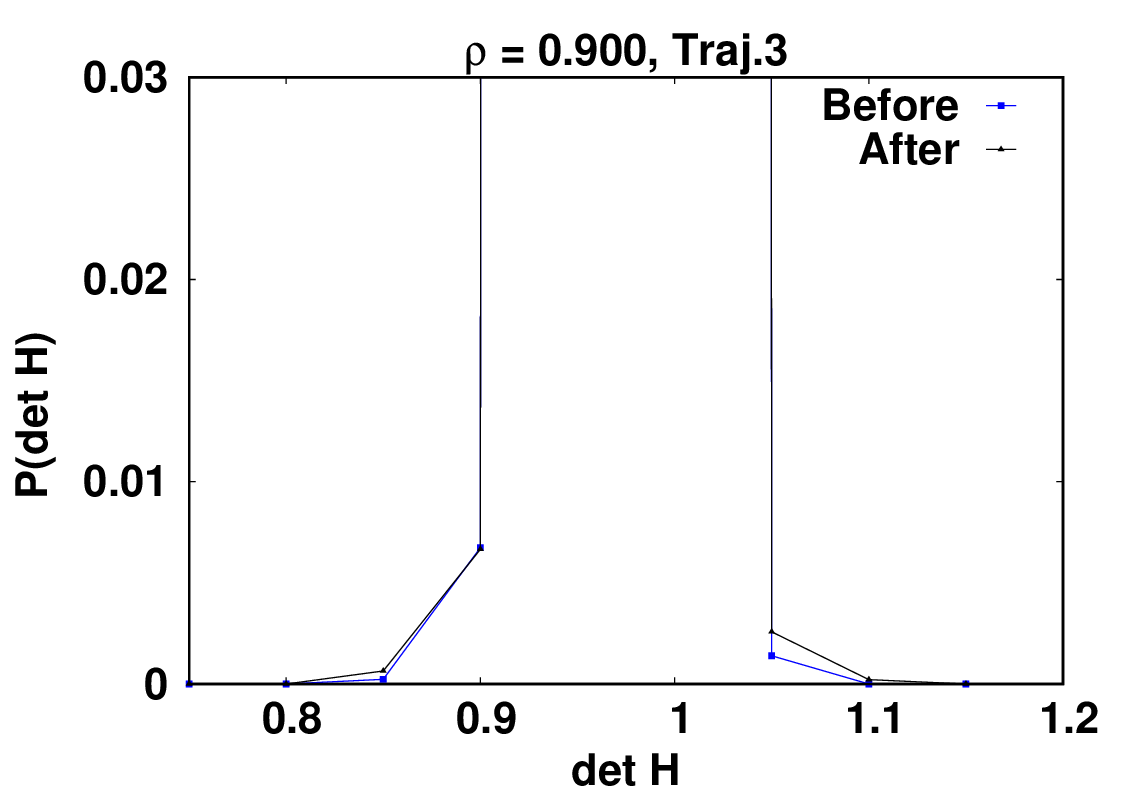}\\
    \includegraphics[width=0.45\textwidth]{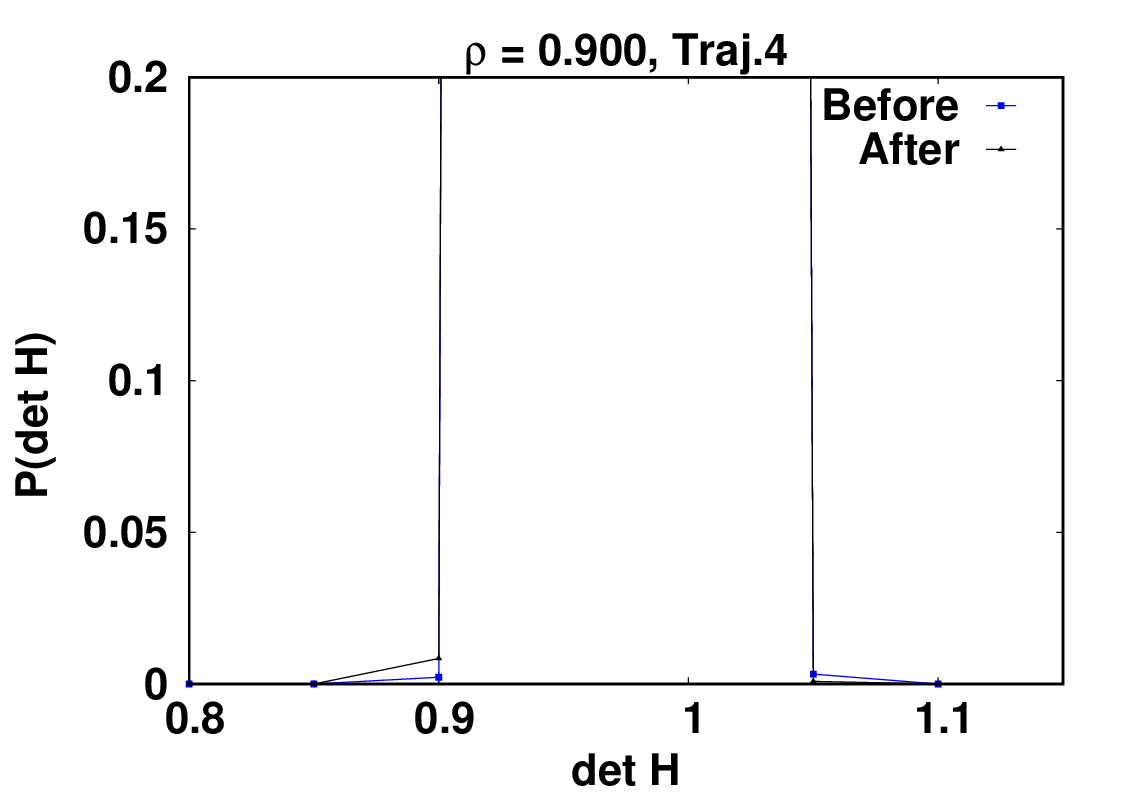}
    \includegraphics[width=0.45\textwidth]{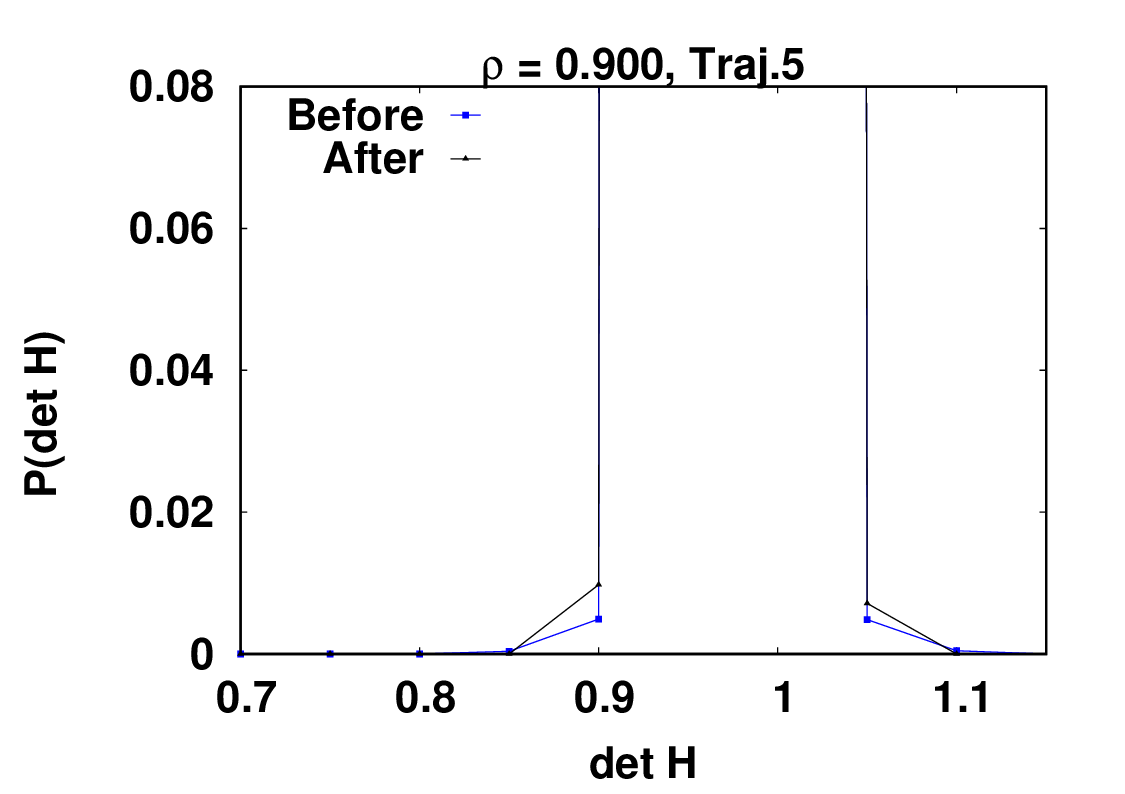}
    \caption{Distribution of $\vert \det \mathbf{H} \vert$ before and after the avalanche for the additional trajectories at density $\rho=0.900$. The histograms reveal that the high-compression tails $\vert \det \mathbf{H} \vert \,\, <1$  are significantly enhanced in the post-avalanche state, indicating that local contraction, rather than expansion, predominates during thermal-mediated plasticity. This observation supports the findings presented in the main text.}
    \label{fig:av14nw9}
\end{figure*}

\begin{figure*}[htb!]
 \centering
    \includegraphics[width=0.45\textwidth]{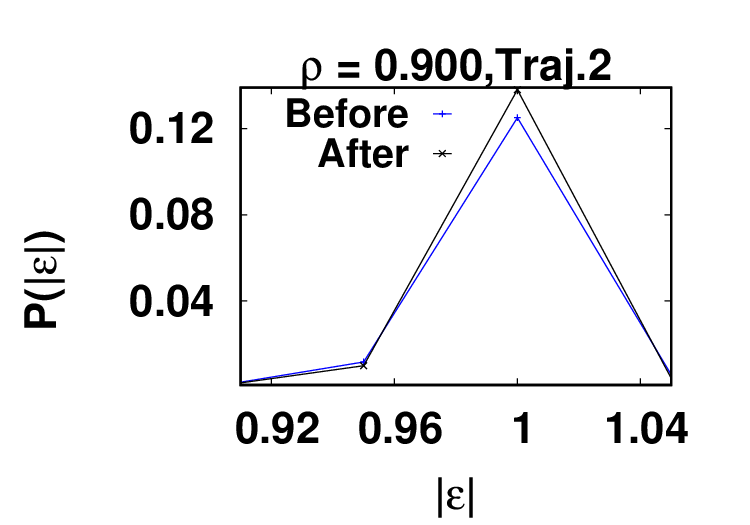}
    \includegraphics[width=0.45\textwidth]{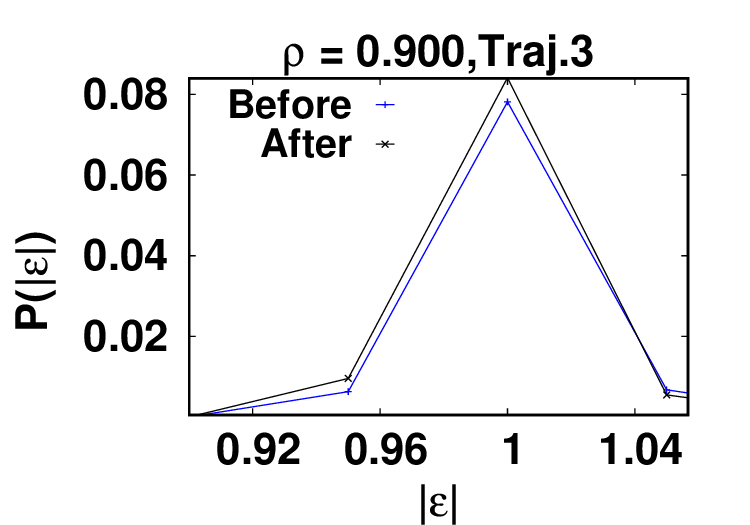}\\
    \includegraphics[width=0.45\textwidth]{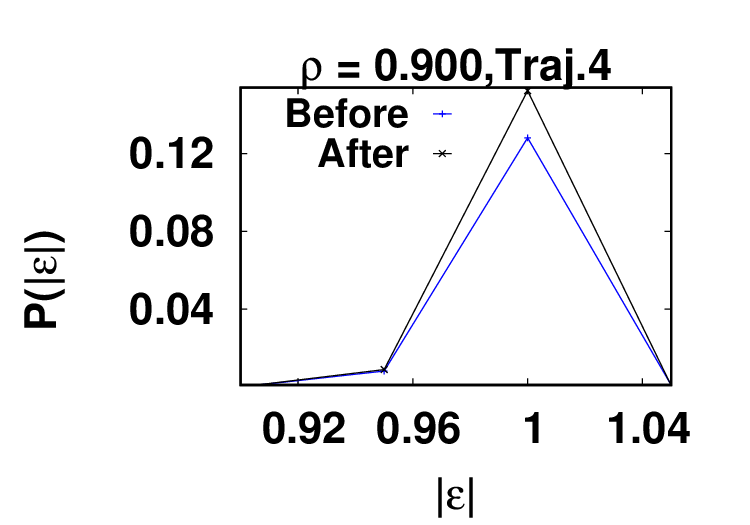}
    \includegraphics[width=0.45\textwidth]{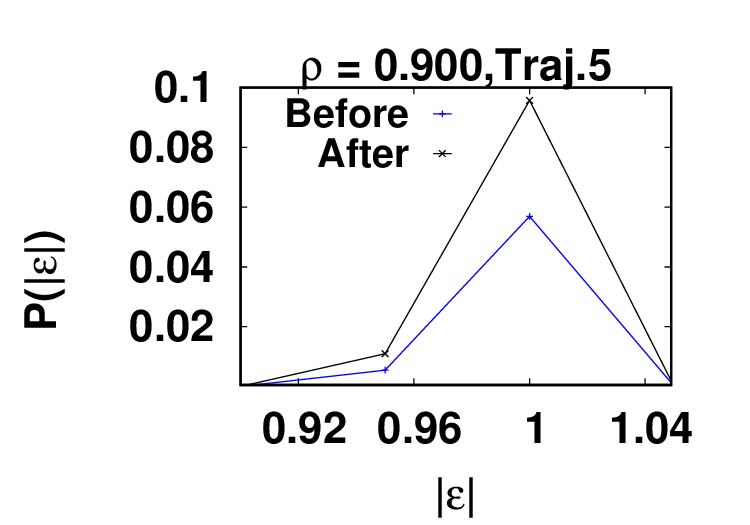}
    \caption{Distribution of the norm of strain for the additional avalanche trajectories at density $\rho=0.900$, comparing the pre- and post-avalanche states. In these histograms, the norm of strain which is calculated as $\Vert \mathbf{\epsilon} \Vert=\sqrt{\mathbf{\epsilon}:\mathbf{\epsilon}}$ quantifies the overall magnitude of local deformation by integrating both shear and volumetric contributions. The distributions show that the post-avalanche state exhibits significantly higher peaks, indicating a marked increase in total strain following the avalanche. This enhanced strain magnitude reflects a more pronounced deformation that is associated with increased local rearrangements, and it further corroborates our main text findings regarding the emergence of a mechanically unstable state.}
      \label{fig:av14nw11}
\end{figure*}

\begin{figure*}[htb!]
 \centering
    \includegraphics[width=0.45\textwidth]{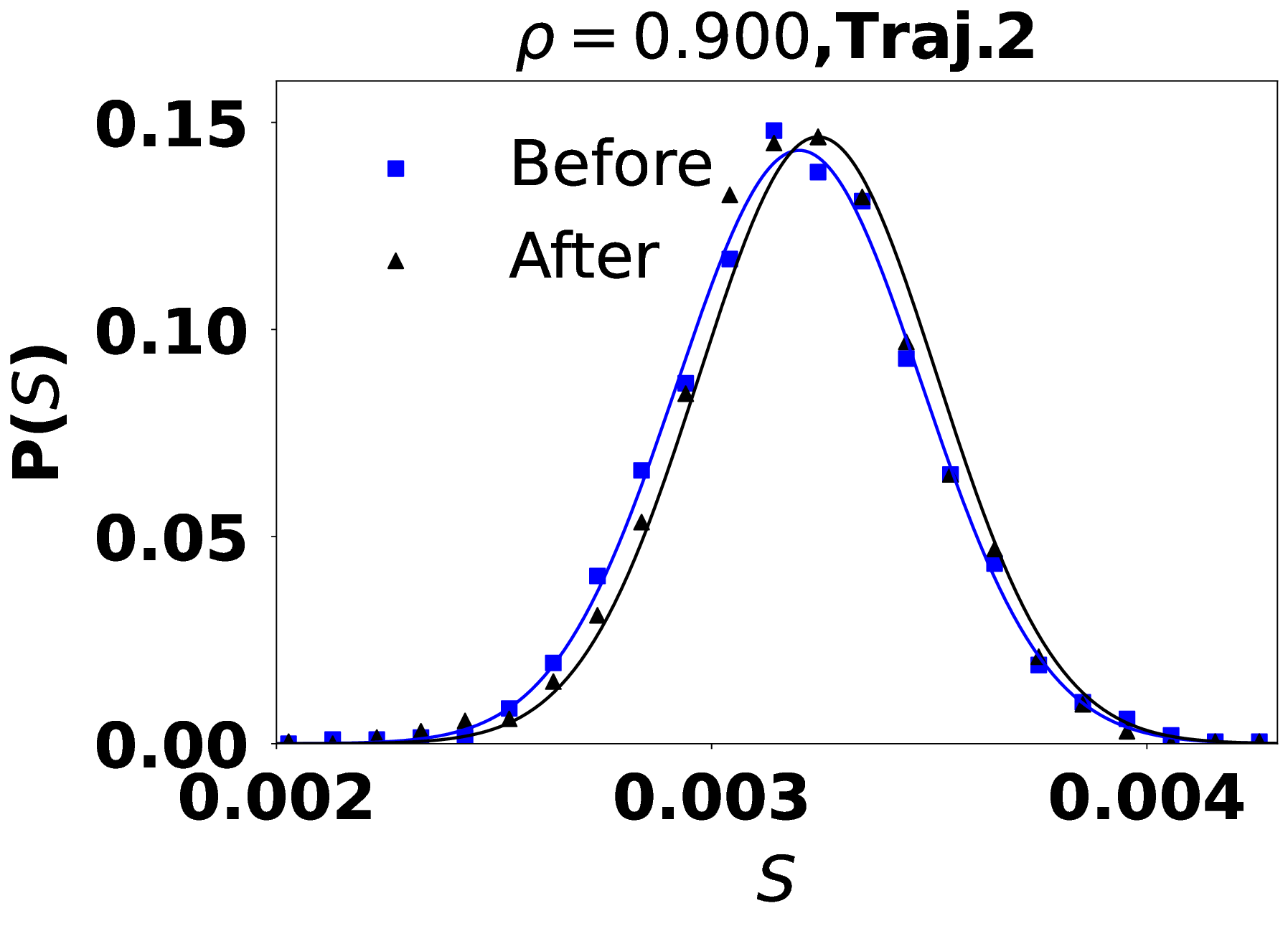}
    \includegraphics[width=0.45\textwidth]{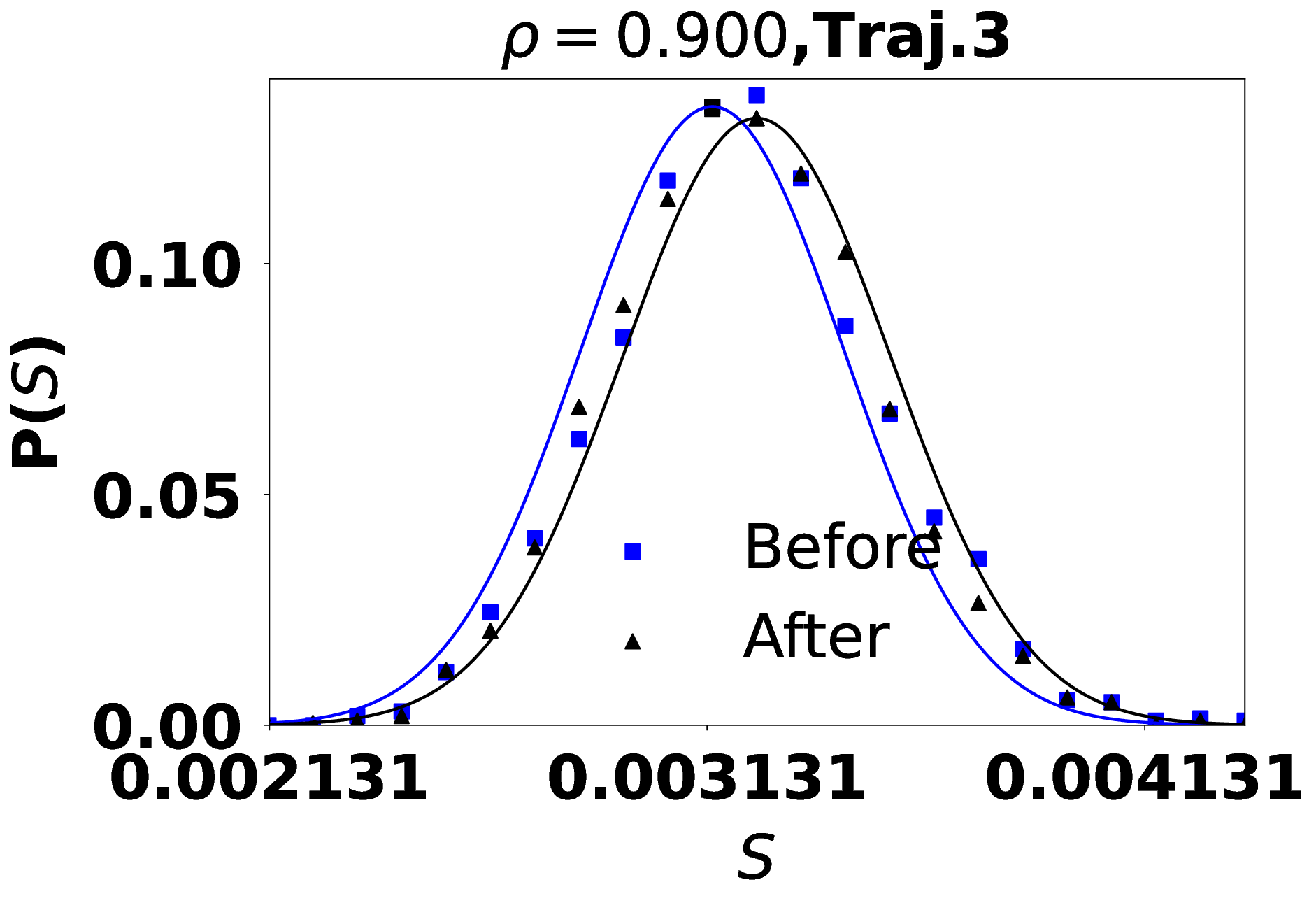}\\
    \includegraphics[width=0.45\textwidth]{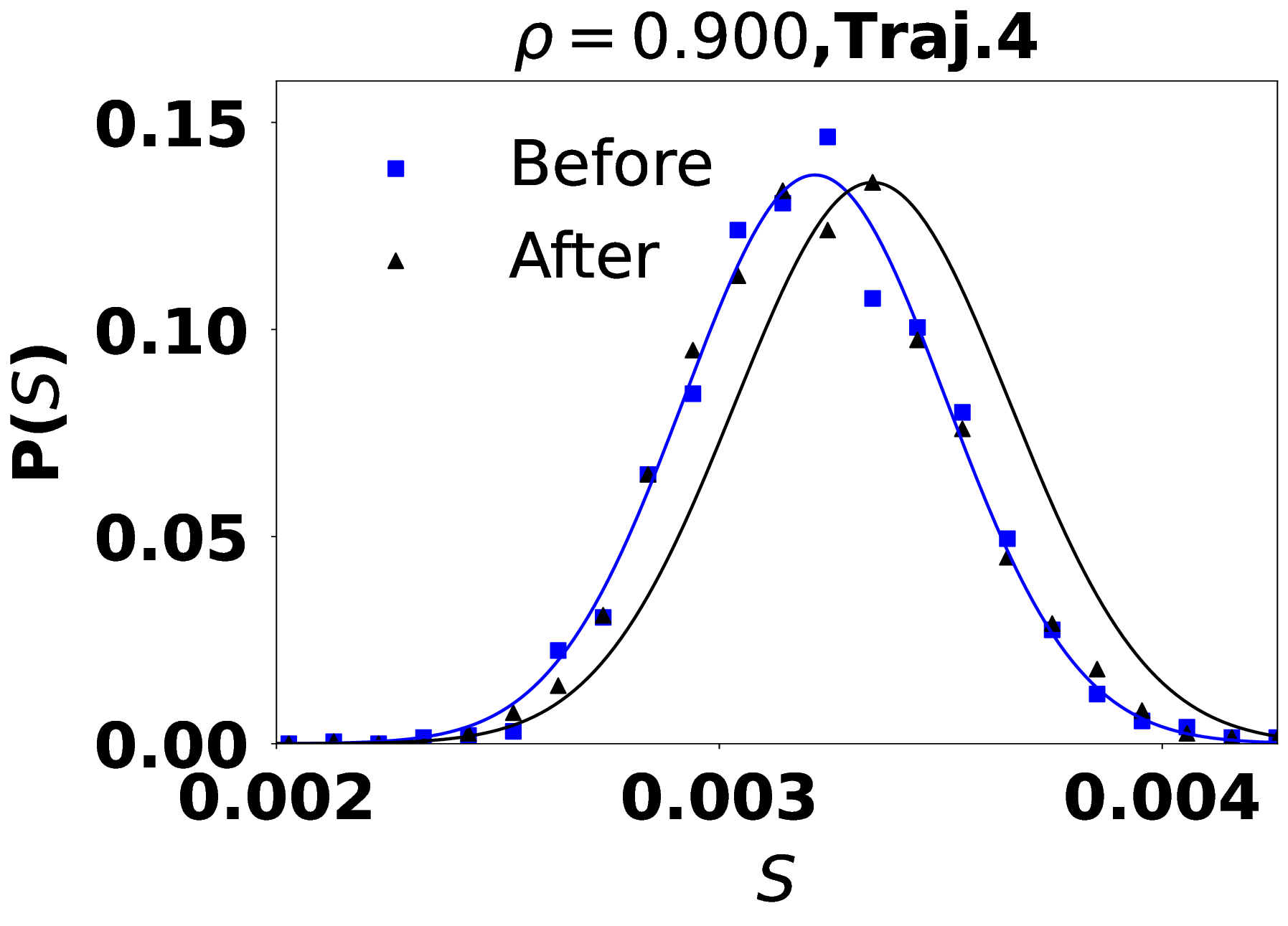}
    \includegraphics[width=0.45\textwidth]{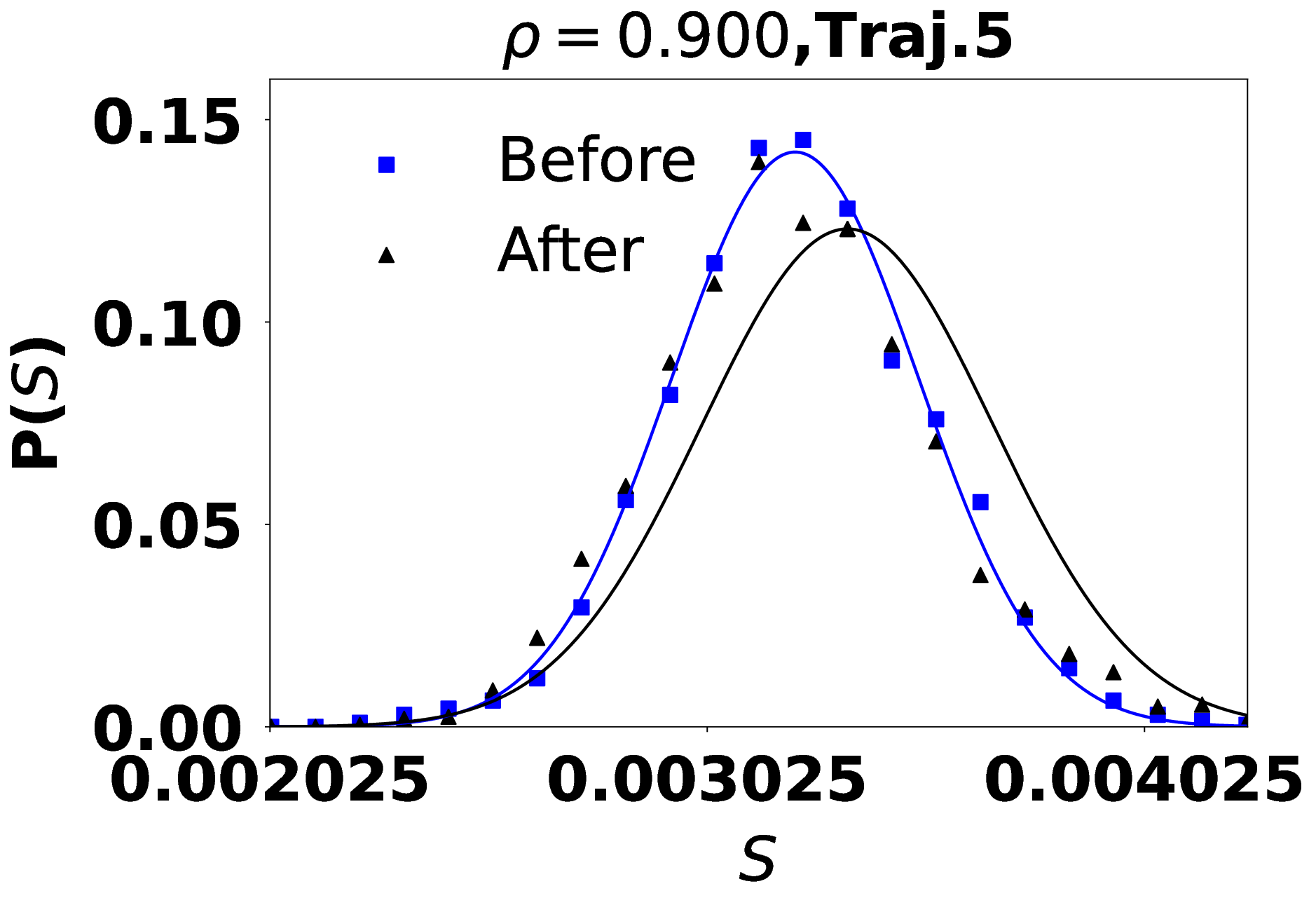}
    \caption{Distribution of softness for the additional avalanche trajectories at density $\rho=0.900$, comparing the pre and post avalanche states. The softness peak shifts toward higher values after the avalanche, indicating an overall increase in local softening that supports the findings presented in the main text.}
      \label{fig:av14nw12}
\end{figure*}

\end{document}